\documentclass[twocolumn,twocolappendix]{aastex631}
\pdfoutput=1 
\usepackage{amsmath,amstext}
\usepackage[T1]{fontenc}
\usepackage[figure,figure*]{hypcap}
\usepackage{mwe}
\usepackage{lipsum}
\usepackage{graphicx} 
\usepackage{gensymb}
\graphicspath{{figures/}}
\usepackage{multirow}
\usepackage{xcolor,colortbl}
\usepackage{tipa}
\usepackage{natbib}
\usepackage{hyperref}
\usepackage{ulem}
\usepackage{float}
\usepackage{enumitem}

\newcommand{\gaia}{\textit{Gaia}\xspace}
\newcommand\HI{$\textrm{H}\scriptstyle\mathrm{I}$\xspace}
\newcommand{\Hpx}{HEALPix\xspace}
\newcommand{\nside}[1]{$N_{\rm{side}} = {#1}$\xspace}
\newcommand{\perch}{{\tt perch}\xspace}
\newcommand{\ph}{Persistent Homology\xspace}
\newcommand{\cubr}{{\tt Cubical Ripser}\xspace}
\newcommand{\HIPI}{HI4PI\xspace}

\newcommand{\dgr}{$N_{HI} / A_V$\xspace}
\newcommand{\hlnames}[1]{\parbox[t]{3.2cm}{\raggedright #1}}

\shorttitle{IVCs in the Solar Neighborhood}
\shortauthors{O'Neill et al.}

\defcitealias{RohserKerp2016}{R16}
\defcitealias{Wakker2001}{W01}
\defcitealias{KuntzDanly1996}{KD96}
\defcitealias{GladdersClarke1999}{G99}

\begin{document}

\title{The 3D Structure and Kinematics of the Local Disk-Halo Interface: \\ Intermediate-velocity Clouds are the Minority of High-altitude Clouds in the Solar Neighborhood}
 
\author[0000-0003-4852-6485]{Theo J. O'Neill}
\affiliation{Center for Astrophysics $|$ Harvard \& Smithsonian, 60 Garden St., Cambridge, MA 02138, USA}

\author[0000-0002-6561-9002]{Andrew K. Saydjari}
\altaffiliation{Hubble Fellow}
\affiliation{Department of Astrophysical Sciences, Princeton University, Princeton, NJ 08544, USA}

\author[0000-0002-2250-730X]{Catherine Zucker}
\affiliation{Center for Astrophysics $|$ Harvard \& Smithsonian, 60 Garden St., Cambridge, MA 02138, USA}

\author[0000-0001-9605-780X]{Eric W. Koch}
\affiliation{National Radio Astronomy Observatory, 800 Bradbury SE, Suite 235, Albuquerque, NM 87106, USA}

\author[0000-0002-8109-2642]{Robert A. Benjamin}
\altaffiliation{Professor Emeritus}
\affiliation{Dept of Physics, University of Wisconsin-Whitewater, 800 W. Main St, Whitewater, WI 53190 USA}
\affiliation{Dept of Astronomy, University of Wisconsin-Madison, 475 N. Charter St., Madison, WI 53706 USA}

\author[0000-0001-5610-3779]{Philipp Frank}
\affiliation{Max Planck Institute for Astrophysics, Karl-Schwarzschild-Straße 1, 85748 Garching, Germany}
\affiliation{Kavli Institute for Particle Astrophysics \& Cosmology (KIPAC), Stanford University, CA 94305, Stanford, USA}

\author[0000-0003-4015-9975]{Stephanie Yoshida}
\affiliation{Center for Astrophysics $|$ Harvard \& Smithsonian, 60 Garden St., Cambridge, MA 02138, USA}

\begin{abstract}
Studies of the Milky Way’s disk-halo interface have historically identified inflowing and outflowing gas incompatible with disk rotation on the basis of radial velocity, leading to the well-known categories of intermediate-velocity clouds (IVCs) and high-velocity clouds (HVCs).  In this work, we leverage recent progress in 3D dust mapping of the Solar Neighborhood to perform the first 3D spatial search for anomalous-velocity clouds at the local disk-halo interface.  We identify 1,695 dust clouds within 1.25 kpc of the Sun (with altitudes ranging between $z=-646$ pc to $z=+928$ pc) by applying a topological structure finding method to a parsec-resolution 3D dust map.  We then evaluate the morphological similarity between these clouds and \HI{} 21 cm emission to measure cloud kinematics, and construct a sample of 519 clouds with high-confidence distances, 3D morphologies, and radial velocities.  Among these are several IVCs (embedded within the well-known Intermediate Velocity Arch complex) now identified in 3D for the first time, enabling direct measurement of their distances, sizes, densities, masses, pressures, and dust-to-gas ratios.  We observe a pronounced asymmetry in the vertical distribution of all clouds in the Solar Neighborhood, with $(2.9 \pm 0.2) \times$ more clouds in the Northern Galactic hemisphere than the Southern above altitudes at which IVCs are present ($|z| \geq 480$ pc).  IVCs make up only 18\% of the total number of clouds located at these high-altitudes, with the remainder having low-velocities --- highlighting the importance of accounting for low-radial-velocity structures when evaluating the local disk-halo interface and modeling feedback-driven Galactic fountain flows. 
\end{abstract}

\section{Introduction}

The motion of gas in, around, and through galaxies represents one of the most significant mechanisms driving how galaxies form and evolve.  These ``flows'' of gas reshape the structure, chemistry, and dynamics of galaxies' circumgalactic and interstellar media (CGM and ISM, respectively), and determine the evolution of a galaxy over cosmic time.  In recent years, new simulations, observations, and theoretical insights have led to a major perspective shift in our view of the CGM as an intricate, complex ecosystem, populated by multiphase gas and highly sensitive to random and structured gas motion, with a strong connection to the ISM \citep[see][for reviews]{PutmanPeek2012,TumlinsonPeeples2017,FaucherGiguereOh2023}.

Traditionally, studies of the Galactic disk-halo interface have focused on identifying inflowing or outflowing gas that is clearly incompatible with the properties of gas in the disk, typically on the basis of velocity cuts.  Both inflowing and outflowing gas are expected to be kinematically distinct from gas corotating with the disk, and so candidate halo gas in the Milky Way is frequently observationally categorized into populations of high velocity clouds (HVCs) and intermediate velocity clouds (IVCs), whose kinematics are incompatible with standard Galactic rotation \citep[e.g.,][]{WakkervanWoerden1997}.  Note that this distinction is made based on \textit{radial} velocity being inconsistent with the disk gas population; a large population of low (radial) velocity halo clouds (LVHCs) is also predicted, but it is challenging to distinguish observationally from disk gas from radial velocity alone \citep[e.g.,][]{PeekHeiles2009, SaulPeek2012, SaulPeek2014, ZhengPutman2015, BishWerk2021, BishWerk2026}. 

HVCs are the most well-known and well-studied class of anomalous velocity halo gas in the Milky Way, with typical lower limits on Local Standard of Rest (LSR) frame velocity defined to be $|v_{\rm{LSR}}| \gtrsim 90$ km/s \citep[e.g.,][]{Wakker2001}.   HVCs are generally low-metallicity, dust-poor, and located at great distances \citep[e.g.,][]{Wakker2001,PutmanPeek2012}.  After their initial discovery in the 1960s via 21-cm \HI{} emission \citep{MullerOort1963}, a variety of explanations for the existence and origins of HVCs were immediately proposed, ranging from condensations in the hot halo gas to ejections from the disk to accretion from the IGM or satellite galaxies \citep{Oort1966}.  Sixty years later, these remain plausible origins for this class of halo clouds \citep[e.g,][]{LucchiniHan2024}.

IVCs possess lower radial velocities than HVCs (defined in the literature as anywhere from 20--40 km/s $\leq v_{\rm{LSR}} \leq 90$ km/s, e.g., \citealt{Wakker2001,RohserKerp2016}) that are still incompatible with disk rotation.  The majority of IVCs have metallicities similar to Solar \citep[e.g.,][]{Wakker2001} and many have been observed to contain molecular gas and dust \citep[e.g.,][]{HeilesReach1988, DesertBazell1990, MagnaniSmith2010, RohserKerp2016b}.  This suggests that IVC origins generally lie in Galactic fountain gas flows, where superbubbles vent hot interiors into the lower Galactic halo, which then condense into clouds and rain back down onto the Galactic plane \citep[e.g.,][]{ShapiroField1976, Bregman1980, Kahn1981, SpitoniRecchi2008, MarascoFraternali2022}.  

The question of if all IVCs originate from the same mechanism is still unresolved, as is the question of if IVCs and HVCs represent distinct populations \citep[e.g.,][]{MarascoFraternali2022}.  Recent work by \citet{HayakawaFukui2024} investigating the dust-to-\HI{} ratio in IVCs and HVCs suggests that a non-trivial fraction ($\sim20\%$) of IVCs are dust-poor and therefore likely low-metallicity, lending support to extragalactic accretion origins.  Similar conclusions were reached by \citet{ChoiWerk2024} who, as part of their analysis of the metallicity of ionized diffuse gas at the Galactic disk-halo interface, detected a wide range of cloud metallicities ($0.04-3.2 \ Z_\odot$), including a low-metallicity outflowing cloud likely incompatible with a purely fountain-based model.

The distribution of IVCs on-sky as traced by \HI{} has long been known to be non-uniform, with the vast majority of IV gas concentrated in a few negative-velocity complexes at high latitudes in the Northern Galactic Hemisphere \citep[e.g.,][]{WesseliusFejes1973,Heiles1984,Danly1989}.  \citet{KuntzDanly1996} described this Northern negative-velocity gas as being separated primarily into three distinct complexes: the Intermediate Velocity Arch (IV Arch), the Low-latitude Intermediate Velocity Arch (LLIV Arch), and the Intermediate Velocity Spur (IV Spur).  The dearth of positive-velocity Northern IVCs, and of Southern IVCs in general, is an open question, with proposed interpretations ranging from phase differences in inflowing vs. outflowing populations \citep[e.g.,][]{MarascoFraternali2012} to projection effects affecting a titled biconal outflow geometry \citep{MarascoFraternali2022} to asymmetric recent feedback events \citep[e.g.,][]{KuntzDanly1996}.

Whatever their origins, distances to IVCs have been challenging to measure.  The tightest constraints have historically come from bracketing detections vs. non-detections of intermediate-velocity absorption features in the spectra of background stars with known distances \citep[e.g.,][]{BenjaminVenn1996, GladdersClarke1998, Wakker2001}.  These studies have generally suggested individual IVC altitudes consistent with locations in the disk-halo interface.  More recent statistical analysis by \citet{LehnerHowk2022} of the IVC covering fraction in UV spectra towards 55 halo stars has similarly suggested that IVCs are primarily limited to $|z| \lesssim 1.5$ kpc.  However, no IVCs have yet been mapped in 3D; since estimates of the basic physical properties of these clouds (including masses, sizes, densities, and pressures) scale directly with distance, the direct measurement of IVC distances and 3D morphologies is a key next step in constraining models of Galactic fountain flows. 

The rapid improvement in the physical and angular resolution achieved by 3D maps of dust in the Solar Neighborhood over the last decade (enabled primarily by the \textit{Gaia} spacecraft) now allows for meaningful constraints to be placed on the properties of clouds (including IVCs) at the local disk-halo interface.  There is now a nearly decade-long history of combining distance-resolved maps of the ISM (like 3D dust maps) with velocity-resolved tracers of the ISM (including \HI{} and CO) to derive 4D fields (3D position and 1D line-of-sight velocity) of Galactic motion \citep[e.g.,][Frank \& Saydjari et al. in preparation]{TchernyshyovPeek2017,TchernyshyovPeek2018}, including through analysis of morphological similarity between tracers \citep[e.g.,][]{SolerZucker2023, LiChen2024, SolerMolinari2025}.  These studies have typically focused on low-Galactic-latitude regions, where the majority of molecular clouds reside.  

In this work, we perform a morphological matching analysis between clouds mapped with 3D dust extinction in the \citet{EdenhoferZucker2024} dust map of the Solar Neighborhood, and \HI{} 21 cm emission from clouds in the \HIPI all-sky survey \citep{HI4PICollaborationBenBekhti2016}.   We perform this matching for clouds at all latitudes, but focus the majority of our analysis on the kinematics of high-altitude, high-latitude, and intermediate-velocity clouds in the Solar neighborhood.  This work represents a departure from previous studies of IVCs and HVCs in that we first define clouds spatially in 3D before investigating anomalous kinematics, as opposed to most previous analysis which, in the absence of 3D spatial maps, has identified anomalous clouds in kinematic space before inferring distances.  This reversal is enabled by improvements in 3D dust resolution and sensitivity, as well as new methods for the spatial identification of clouds.

In \S\ref{S:data}, we describe the \citet{EdenhoferZucker2024} 3D dust map and the \citet{HI4PICollaborationBenBekhti2016} \HI survey.  In \S\ref{S:methods}, we discuss identifying clouds in the dust map using a topological structure finding method (``\perch'', O'Neill et al. in preparation), and the process of defining high-quality morphological matches between these dust clouds and \HI structures.  We then analyze the kinematics, positions, and properties of our sample of \HI-matched clouds in \S\ref{S:results}. We connect a subset of these matched clouds to well-known IVCs in \S\ref{S:discuss}, before concluding in \S\ref{S:conclusions}.

\section{Data}\label{S:data}

\subsection{Gaia XP 3D Dust Map}\label{S:dustmap}

We analyze clouds in the \citet{EdenhoferZucker2024} 3D dust extinction map, which probes out to distances of 1.25 kpc from the Sun.  This map was created through Bayesian inference based on a catalog of distances and extinctions for 220 million stars with \gaia XP BP/RP spectra created by \citet[][hereafter ZGR23]{ZhangGreen2023}.  We refer to the dust map constructed by \citet{EdenhoferZucker2024} as the GXP map.

The GXP dust extinction estimates are reported in differential extinction ($A'_{ZGR23} = d A_{ZGR23} / 1 \ \rm{pc}$), which can be converted to other bands using ZGR23's published extinction curve as $A_X = m_x \ A'_{ZGR23}$ (e.g., $A'_V = 2.8 \ A'_{ZGR23}$). If an extinction-to-\HI column density ratio is additionally assumed (or directly measured), the map can be converted to an estimate of total hydrogen nuclei volume density \citep{ONeillZucker2024}.

The GXP map was constructed in a \Hpx \citep{GorskiHivon2005} \nside{256} voxelization system with logarithmically spaced distance bins ranging between 0.4--7 pc in width, with the centers of nearest and furthest bins at 69 pc and 1244 pc. The GXP map provides 12 samples drawn from the inferred posterior distribution of 3D dust extinction on this \Hpx grid; we refer to the map samples as ``draws'' throughout this work.  These 12 draws are not independent, as they were obtained as pairs from 6 antithetically drawn samples spanning the posterior distribution.

We interpolate the 12 draws from \Hpx to a 3D cartesian grid with voxels of side length 2 pc (yielding a cube of size 1251$^3$ voxels) using the script {\tt interp2box.py} provided by \citet{EdenhoferZucker2024}.  We additionally make use of the interpolated (2 pc)$^3$ map of the standard deviation of extinction between draws released by \citet{EdenhoferZucker2024}.

We identify and analyze clouds in each of the 12 draws of the dust map independently (\S\ref{S:perch}), before matching clouds between draws (\S\ref{S:cluster}).  This allows us to constrain statistical uncertainties on the results of this work and minimize fingers-of-god effects that are introduced by averaging over realizations of clouds with uncertain distances.

\subsection{HI4PI 21 cm \HI Survey}\label{S:hi4pi}

We compare the GXP dust map to all-sky \HI 21 cm emission compiled by the \HIPI survey \citep{HI4PICollaborationBenBekhti2016}.  The \HIPI data products were created by combining data from the Effelsberg-Bonn \HI{} Survey \citep[EBHIS;][]{KerpWinkel2011} and the Galactic All-Sky Survey \citep[GASS;][]{McClureGriffithsPisano2009}.  The survey achieves an all-sky angular resolution of 16.2', and was released in various projections.  In this work, we make use of the \Hpx{} projections of the survey.  We stitch them together into a single \nside{1024} spectral cube, and then downsample each velocity channel to \nside{256} to match the resolution of the GXP dust map using the {\tt healpy} \citep{Zonca2019_healpy} function {\tt ud\_grade}.  

The \HIPI survey covers LSR velocities in the Northern portion of $|v_{\rm{LSR}}| \leq 600$ km/s and in the Southern portion of $|v_{\rm{LSR}}| \leq 470$ km/s.  We restrict our search for dust-\HI matches to $|v_{\rm{LSR}} | \leq 300$ km/s, providing a generous buffer around the velocities in which the vast majority of Galactic emission is contained.  Velocity channels in the map have widths of $\Delta v = 1.29$ km/s.  

The map is reported in terms of brightness temperature $T_B(v)$, and we convert this to an inferred $N_{HI}$ column density in the optically-thin limit as \citep{HI4PICollaborationBenBekhti2016},
\begin{equation}
    N_{HI} [\rm{cm}^{-2}] = 1.823 \times 10^{18} \int dv \ T_B(v) [\rm{K \ km \ s^{-1}}].
    \label{eqn:hi_col}
\end{equation}
The theoretical $5\sigma$ detection limit of the survey, derived by \citet{HI4PICollaborationBenBekhti2016} for a Gaussian profile with FWHM line width of 20 km/s, is $N_{HI} = 2.3 \times 10^{18}$ cm$^{-2}$.

\section{Methods}\label{S:methods}

\subsection{Structure Identification in 3D Dust with Persistent Homology}\label{S:perch}

To identify clouds in the GXP dust map, we apply the in-development topological structure identification python package \perch (O'Neill et al., in preparation).  In brief, \perch applies the technique of persistent homology to identify significant structures in 2D images and 3D volumes.

\subsubsection{A Persistent Homology Primer}

Persistent homology \citep{Edelsbrunner2002} is a mathematical tool designed to describe ``holes'' or ``voids'' present in a space.  Voids are classified into homology groups based on their dimensionality.  The zeroth homology group $H_0$ describes connected components in a space, which, in the context of astronomical images corresponds to distinct structures like molecular clouds and spiral arms.  The first homology group $H_1$ describes one-dimensional loops, while the second homology group $H_2$ describes two-dimensional voids.  In this work, we focus solely on the $H_0$ group; the properties of the $H_1$ and $H_2$ structures in the GXP map will be described in the future by O'Neill et al. (in preparation).

Persistent homology allows for the extraction of topological structures across all homology groups by performing a series of nested filtrations of a data space $D$ (here, a 3D data cube).  \perch implements superlevel set filtrations (i.e., for a threshold value $x$, creating a binary mask $M_t$ with the same shape as $D$ where $M_t = 1 \ \rm {if} \ D \geq t, \ 0 \ \rm{otherwise}$), with filtering proceeding from the highest value in the image, $t = \rm{max}(D)$, to the lowest value, $t = \rm{min}(D)$, and spanning all intermediate unique pixel values in the dataset.  Structures are tracked between filtrations, and the thresholding pixel values in the filtration at which a given structure appears (is ``born'') or disappears (``dies'') are recorded as that structure's birth and death values, respectively.  An $H_i$ structure (where $i$ is the dimensionality of the void) can die as the result of 1) merging into (becoming connected to) another $H_i$ structure born at an earlier filtration, or 2) the birth of an $H_{i+1}$ structure.  When performing superlevel set filtrations, a structure's birth will always be greater than its death.    

The interval between a structure's birth and death is referred to as its persistence, 
\begin{equation}
     \textrm{persistence} = \textrm{birth} - \textrm{death}.
\end{equation}
Structures with higher persistence, i.e., structures that exist across many filtrations, are conventionally interpreted as having higher significance \citep[e.g.,][]{Edelsbrunner2002,Carlsson2009,Edelsbrunner2010,Sousbie2011}.

In practice, \perch uses the \cubr algorithm \citep{KajiSudo2020} for \ph applied to images and cubes.  \cubr extends the Ripser algorithm \citep[developed for point clouds,][]{Bauer2021Ripser} for Vietoris-Rips complexes to the context of weighted cubical complexes (e.g., images).  \cubr tracks the pixels at which topological structures are born and die.  Since \cubr is sufficiently fast, we do not set quantized filtering values and \cubr uses all unique values in the dataset.
\perch makes use of this information to segment structures (in 2D or 3D), using the python package {\tt connected-components-3D} \citep{connectedcomponents3d}.  Segmented structures that share pixels are sorted into a hierarchical structure based on level in the filtration sequence at which structures appear; this process is similar to constructing a dendrogram.  Higher-significance structures in the $H_0$ group reflect structures that would be identified in a traditional dendrogram-based analysis (e.g., with the package {\tt astrodendro}, \citealt{rosolowsky_structural_2008}). 

\subsubsection{Significance Filtering}

We ran \perch on each of the 12 draws of the cartesian-interpolated GXP dust map.  This yielded between 28,778,158 and 28,793,800 $H_0$ structures of all significance levels in the individual draws, with a mean of 28,787,419 structures per draw.

We then need to define ``significant'' topological structures in each draw.  For each structure, within the voxel at which the topological structure dies (with indices $d_x, d_y, d_z$, marking the lowest-valued pixel at which the structure exists in the filtration sequence), we access the standard deviation of the values of that voxel across the $N_{\rm{draw}} = 12$ draws of the GXP dust map, $\sigma_{\rm{death}}$.  We then compute the ratio of the structure's persistence to this value, 
\begin{equation}
    p_n = \frac{\rm{persistence}}{\sigma_{\rm{death}}}
\end{equation}
in order to define a measure of significance that is sensitive to spatial variation in the level of uncertainty in the GXP dust map.  Based on experimentation, we define a threshold for significant $H_0$ structures of $p_n > 10$.  This yields between 1249 and 1307 significant $H_0$ structures in the individual draws, with a median of 1276.5 structures per draw.

Note that, since \perch performs superlevel set filtrations and proceeds from the highest-valued pixels to the lowest-valued, the $H_0$ structure that is born at the highest-valued pixel (birth = max(data)) in the image is not guaranteed to have a valid death value (since all subsequently-born $H_0$ structures will eventually merge into the hierarchy defined by the first-born structure) and so have infinite persistence.  In order to address this edge case, we run \perch twice: once on the original data cube, and once on a padded version of the cube.  Specifically, we pad the original data cube (which consists of a finite-valued sphere of radius $1250$ pc centered on the Sun, surrounded by undefined voxels) by performing a binary dilation of the finite-valued sphere and filling the new edge pixels with an arbitrary large value ($10 \times$ max($D$)).  This ensures that, in the padded \ph computation, the original highest-valued pixel will be assigned a finite death and persistence.  We then modify the \ph resulting from the original image so that the death of the maximum-valued-birth structure is replaced by the death of the same structure in the padded run.   

We segment our subset of significant structures in each draw using \perch, which yields 3D maps the same shape as the original cube segmented by each structure's ID number. A variety of descriptive properties of each structure are additionally calculated by \perch (described in \S\ref{S:catalog}).

\subsection{Clustering Structures between Draws of the 3D Dust Map
}\label{S:cluster}

Our next goal is to match equivalent topological structures between draws of the dust map.  To do so, we make use of our expectation that 3D dust maps like the GXP map are most consistent in their plane-of-the-sky (POS) constraints of dust cloud  morphology, and that more significant uncertainty arises in determining where a cloud is along the line-of-sight (LOS, i.e., its distance from the Sun).  This expectation leads us to conclude that matching the morphology of clouds on the POS between draws will be the most productive approach in capturing the variations in reconstructed cloud properties between draws, rather than clustering clouds directly in 3D cartesian space.

We must first project the \perch-segmented 3D cartesian map to a grid of spherical \Hpx shells as a function of distance.  Specifically, we project the map to the same grid on which the GXP dust map was constructed: \Hpx \nside{256} with logarithmically spaced distance bins.  We perform nearest-neighbors interpolation to map the segmented maps onto radial distances matching those of the centers of the native distance bins.  We refer to this \Hpx by distance map of structure IDs as $\mathcal{S}(p,r)$, where $p$ indexes the \Hpx pixels and $r$ indexes the distance bins.  If multiple structures share pixels at a given index in $\mathcal{S}(p,r)$, the single structure ID reported is the highest-level hierarchical structure (i.e., the ``child'' or ``leaf''). 

For each unique structure $\mathcal{S}_i$ with ID $i$, we first create a 3D \Hpx map of its extinction on the POS as a function of distance,
\begin{equation}
\mathcal{A}'_i(p,r) = 
\begin{cases}
  A'_{ZGR23}(p,r) & \text{if } \mathcal{S}(p,r) = (i \text{\ or descendants of\  } i)\\
  0 & \text{otherwise}
\end{cases}
\end{equation}
\perch performs hierarchical segmentation, which is why the masks of $A'_i(p,r)$ must include the contribution of pixels identified as descendants (children and descendants of children) of a given structure.

We then create an integrated \Hpx map of the structure's 2D extinction on the POS, 
\begin{equation}
    \mathcal{A}_i(p) = \sum_{k=1}^{N_{dr}(\mathcal{S}_i)} A'_i(p, r_k) dr_k
    \label{eqn:integrated_Ai}
\end{equation}
with units of magnitudes, where $p$ is the \Hpx pixel, $dr_k$ is the distance between distance bin $r_k$ and $r_{k+1}$ in parsecs, and $N_{dr}$ is the number of radial bins in which $\mathcal{S}_i$ is defined.  

We additionally create a radially-integrated map of its extinction along the LOS, 
\begin{equation}
    \mathcal{R}_i(r) = \sum_{k=1}^{N_{\rm{pix}}(\mathcal{S}_i)} A'_i(p_k, r)
    \label{eqn:integrated_Ri}
\end{equation}
with units of mag/pc, where $N_{\rm{pix}}$ is the number of \Hpx pixels on the POS occupied by $\mathcal{S}_i$.

\begin{figure*}
    \centering
\includegraphics[width=\textwidth]{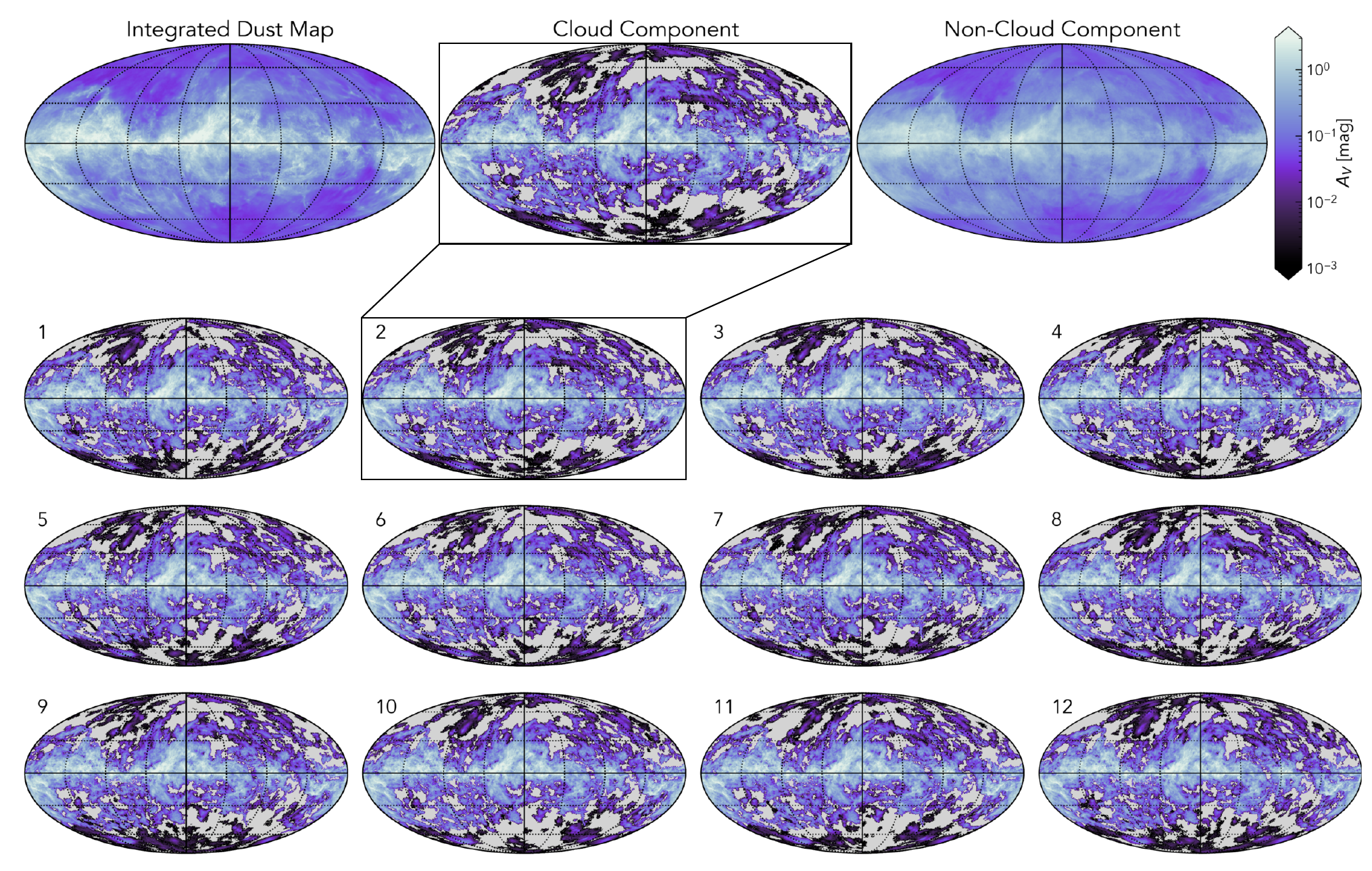}
    \caption{\textit{Top row:} Integrated extinction, $A_V$, for one draw (Draw 2) of the GXP dust map.  The left column shows the full dust map, the center column shows cloud-like structures segmented by \perch, and the right column shows the non-cloud (residual) extinction not segmented by \perch.  Figures are in Mollweide projection, centered towards the Galactic center at $\ell=0^\circ$.  \textit{Lower rows:} As center column of top row, but for all twelve draws of the dust map.  \underline{\textit{Interactive Component:}} An interactive figure is available at \url{https://theo-oneill.github.io/HACs_and_IVCs/segmentation/}, showing the integrated dust map, cloud component, and non-cloud component for each draw of the dust map in Mollweide, polar, and Cartesian projections. }
    \label{fig:perch_segs}
\end{figure*}

Figure \ref{fig:perch_segs} shows the total segmented $\mathcal{A}(p)$ (converted to V-band extinction $A_V$) for all cloud-like structures in each draw.  We additionally show the integrated extinction of the full map for each draw, as well as the residual non-segmented (non-cloud-like) extinction.  By comparing the segmented component vs. non-segmented component, one can observe that \perch identifies nearly all dense, ``cloud''-like structures, with the non-segmented component made up primarily of diffuse dust.  

From inspecting Figure \ref{fig:perch_segs}, it is also apparent that many realizations of the ``same'' clouds are identified between draws.  We describe our fiducial method for creating clusters of clouds between draws in Appendix \ref{ap:cluster}.  Our final catalog contains 1,695 clouds, each appearing in at least 3 draws of the GXP dust map ($N_{\rm{draw}} \geq 3$, i.e., one-quarter of draws).

\subsection{Morphological Matching with HI Emission}\label{S:morph_match}

We now describe the procedure via which we identify significant morphological matches between structures identified in 3D dust and HI emission.  We provide a high-level summary of this process in this section, with more technical details given in Appendix \ref{ap:morphological}.

\subsubsection{Morphological Similarity Metric}

We evaluate the morphological similarity between dust and \HI{} emission as a function of \HI{} velocity using the Structural Similarity Index Measure \citep[SSIM, ][]{WangBovik2004}.  The SSIM was designed to quantify similarity between images while accounting for dependencies between pixels, and has achieved widespread use across many fields of the sciences, especially in computer vision, since its introduction \citep[e.g.,][]{NilssonAkenineMoller2020,VenkataramananWu2021,BakurovBuzzelli2022} 

The SSIM is calculated over local neighborhoods of pixels in two images (e.g., integrated dust extinction $A_V$ and \HI{} emission at a given velocity H$\mathrm{I}$(v)) and then pooled into a single summary statistic.  We provide the full derivation of the SSIM in Appendix \ref{ap:ssim}.  The metric simplifies to,
\begin{equation}
    \rm{SSIM}(x,y) = \frac{(2 \mu_x \mu_y + C_1)(2 \sigma_{xy} + C_2)}{(\mu_x^2 + \mu_y^2 + C_1)(\sigma_x^2 + \sigma_y^2 + C_2)}
\end{equation}
where $\mu_x$ ($\mu_y$) is the mean value in Image $x$ (Image $y$) within the local neighborhood under consideration, $\sigma_x$ ($\sigma_y$) the local standard deviation in $x$ ($y$), and $\sigma_{xy}$ the local covariance between Images $x$ and $y$.  $C_1$ and $C_2$ are small stabilizing constants.  Lower values of the SSIM indicate lower morphological similarity between the local regions of the images, while higher values indicate greater morphological similarity. 

As described in Ap. \ref{ap:ssim}, we pool the local values of the SSIM between each pair of images under consideration by computing the dust-weighted average SSIM,
\begin{equation}
    \rm{SSIM}(v) = \frac{\Sigma_k^{N_{\rm{pix}}} \ SSIM(H\mathrm{I}(v, p_k),\mathcal{A}_i(p_k)) \cdot \mathcal{A}_{i}(p_k)}{\Sigma_k^{N_{\rm{pix}}}  \mathcal{A}_i(p_k)}
\end{equation}
where $A_i$ is POS dust extinction as defined in Equation \ref{eqn:integrated_Ai} and  H$\mathrm{I}$(v) is \HI{} emission at a given velocity.  Dust-weighting is critical in avoiding confusion with co-moving (but not co-spatial) \HI{} at the diffuse edges of our dust clouds that are unrelated to the local structures in consideration.

\subsubsection{Velocity Scanning and Similarity Peak Detection}\label{S:scan}

\begin{figure*}
    \centering
\includegraphics[width=\textwidth]{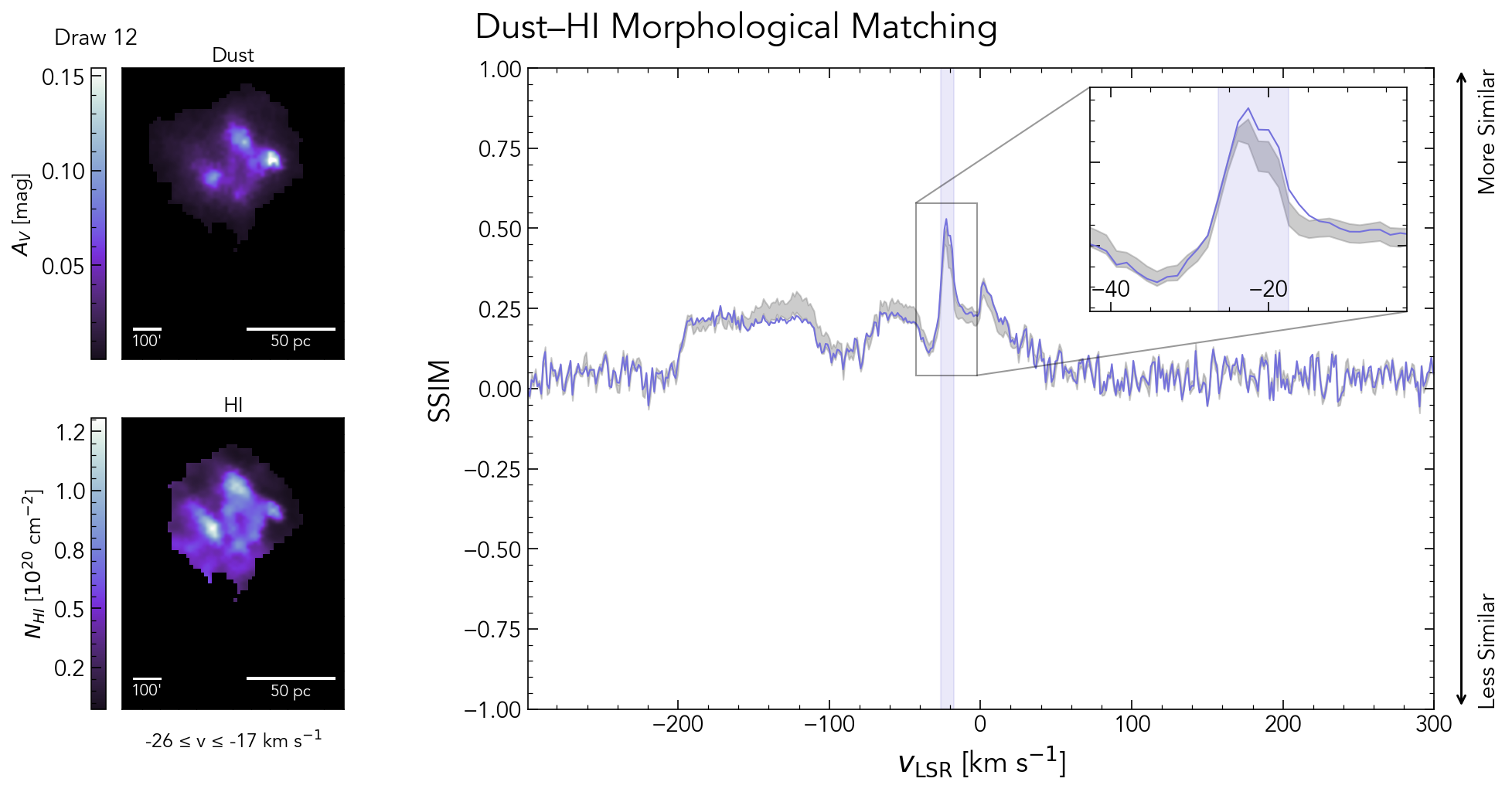}
    \caption{An example dust-\HI{} match for Cloud 352 (identified as the Draco cloud in \S\ref{S:discuss}).  \textit{Upper left:} integrated POS extinction $A_V$ for Draw 12 of the cloud.  \textit{Right:} Morphological quality metric SSIM between dust extinction (upper left) and \HI{}, as a function of \HI{} velocity.  The purple curve shows SSIM(v) for Draw 12, while the gray shaded region shows the $1\sigma$ variation in SSIM(v) across all draws that the cloud appears in.  The purple shaded region shows the half-prominence width surrounding the SSIM peak, defining the minimum and maximum velocity of the match.  The inset in the upper right displays a closer look at the SSIM(v) peak region.  \textit{Lower left:} Integrated $N_{HI}$ within the width of the SSIM peak (shown by the shaded purple region in the SSIM(v) panel).  \underline{\textit{Interactive Component}}: An interactive figure at \url{https://theo-oneill.github.io/HACs_and_IVCs/ssim/index.html} displays the same plot for the other 11 draws of the cloud, along with versions of this plot for many of the other named clouds highlighted later in \S\ref{S:discuss}.}
    \label{fig:example_ssim}
\end{figure*}

We apply our morphological matching procedure to all clusters identified in \S\ref{S:cluster}.  For each individual cluster, we perform the following set of operations.  We first preprocess the $N_{\rm{draw}}$ maps of $A_i(p)$.  The value of the SSIM is sensitive to the mean value of the pixels in each image, so in order to ensure consistency between draws, we mask those \Hpx pixels that are not populated by at least half of the $N_{\rm{draw}}$ $A_i(p)$ maps.  For draws where pixels that pass this criterion are not populated within that specific draw, we fill those pixels with extinctions of 0 mag.  We define the subset of \Hpx pixels that are populated in $\geq \frac{1}{2} N_{\rm{draw}}$ as the ``median-extent'' cloud.  Later in this work, we will define ``maximum-extent'' clouds as the set of all pixels in which the cloud appears in any draw. 

We then compute the effective angular resolution of the GXP dust map, FWHM$_0$, within the 3D volume occupied by each cloud; this procedure is described in Appendix \ref{ap:fwhm}.  If FWHM$_0$ for the cloud is greater than the native resolution of the data, we smooth the \HIPI data to match the resolution of the dust cloud.  Specifically, we smooth each channel of the \HIPI \Hpx{} cube, HI(v), with a Gaussian beam with FWHM = $\sqrt{(\rm{FWHM}_0)^2 - (\rm{FWHM}_{HI4PI})^2}$; if FWHM$_0$ is less than the native resolution of the \HIPI data, we do not smooth the data cube and instead match as is.  We then mask the pixels in the \HI{} cube that are not populated by the masked dust cloud.  Finally, we scale each $A_i(p)$ and each \HI{} channel to a constant [0,1] range to ensure consistency of the scale of the SSIM between velocity channels and draws.

For each velocity channel in the \HIPI cube (which have widths of 1.29 km/s) between $|v_{\rm{LSR}}| \leq 300$ km/s, we then calculate SSIM(v) as described in Appendix \ref{ap:ssim}.  Figure \ref{fig:example_ssim} shows an example of the results of this procedure for one draw of one cloud (which we later identify in \S\ref{S:lit} as the Draco cloud); an interactive figure is available showing the same plot for other draws of this cloud, as well as similar plots for other clouds of interest.  

A clear peak in the SSIM curve is present, indicating the maximum-similarity velocity between the \HI{} and dust extinction.  For each draw, we identify the velocity at which the maximum SSIM peak occurs.  We then evaluate the width of the peak by finding the half-prominence boundary of the peak (as calculated in \citet{ONeillZucker2024}, with prominence here evaluated relative to a $\pm 50$ km/s window around the peak).  This enables us to identify a minimum and maximum velocity within which the dust and \HI are morphologically similar.  We then derive various properties from the \HI{} (described in \S\ref{S:hiprops}), including column density $N_{HI}$ between this minimum and maximum velocity.  Figure \ref{fig:example_ssim} additionally compares the integrated $N_{HI}$ within the SSIM peak to the integrated dust emission; by eye, the two images appear to show the ``same'' cloud observed with different tracers.  Figure \ref{fig:example_ssim} also shows the $1\sigma$ range of the curves of SSIM(v) resulting from all draws of the cloud considered; consistency between draws is generally high.

We note that defining the boundaries of a peak is generally a poorly-defined problem, and that the widths of some of the more complex (e.g., asymmetric, or multiply-peaked) SSIM peaks shown in the interactive component of Figure \ref{fig:example_ssim} may be better served by more complex boundary definitions. This appears to primarily affect high-latitude peaks with low amounts of confused low-velocity emission; we note a potential systematic effect underestimating the velocity widths (and, therefore properties such as \HI{} column densities) of these high-latitude clouds, and recommend that future work generalize this peak-width definition to a larger variety of morphological peak shapes.

\subsubsection{Statistical Significance of Dust-HI Match}\label{S:match_sig}

\begin{figure*}
    \centering
\includegraphics[width=\textwidth]{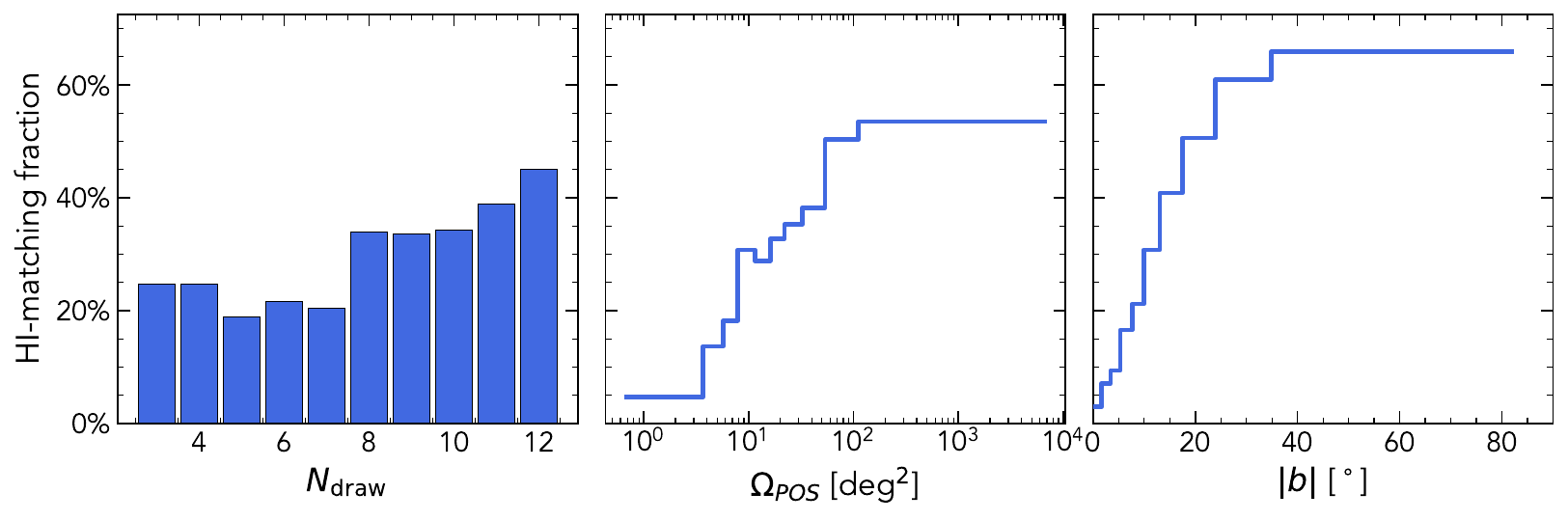}
    \caption{\textit{Left:} The percentage of dust clouds successfully matched to \HI{}, as a function of the number of draws the cloud appears in, $N_{\rm{draw}}$.  \textit{Center:} As left, but as a function of area on the POS, $\Omega_{\rm{POS}}$.  Points were divided into 10 equal percentiles, with percentile boundaries shown by the horizontal width of each step.  \textit{Right:} As center, but for absolute Galactic latitude $|b|$.}
    \label{fig:match_fraction}
\end{figure*}

\begin{figure*}
    \centering
\includegraphics[width=\textwidth]{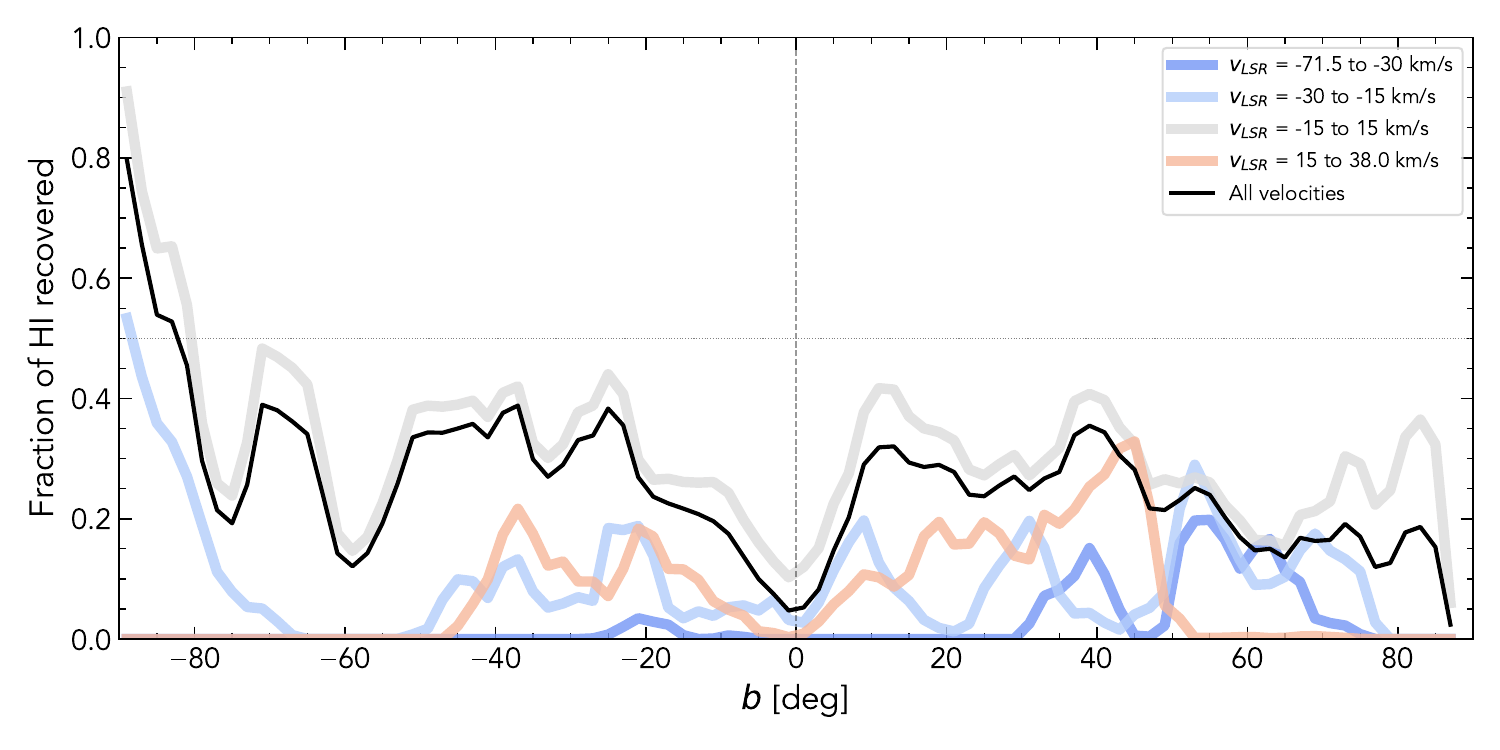}
    \caption{Fraction of total \HI{} emission in the \HIPI map recovered by our dust-matched cloud sample as a function of Galactic latitude $b$ (calculated in $2^\circ$ bins), separated by velocity ranges from -71.5 to -30 km/s (dark blue), -30 to -15 km/s (light blue), -15 to 15 km/s (gray), and 15 to 38.0 km/s (light red). The solid black curve shows the fraction integrated over all velocities between -71.5 to 38.0 km/s (representing the minimum $v_{min}$ to maximum $v_{max}$ range in our matched cloud sample). The horizontal dotted line marks 50\% recovery, and the vertical dashed line marks $b=0^\circ$.}
    \label{fig:frac_hi_recovered}
\end{figure*}

We must now evaluate what makes a ``good'' match between dust and \HI{}.  In order to do so, we require an understanding of the distribution of statistics derived from SSIM(v) for erroneous dust-\HI matches (essentially, their distribution under the null hypothesis of no good match between dust and \HI).  To approximate these distributions for each structure, we translate $\mathcal{A}_i$ to other locations on the sky and repeat the process described in \S\ref{S:scan}.  This provides a basis to help us distinguish between SSIM peaks that occur by chance vs. from genuine matches.  We describe this process in \S\ref{ap:sig_matches}.  In total, 519 dust clouds (representing 30.6\% of all clouds) pass our imposed quality cuts and are successfully matched to \HI{} emission.  These clouds range in altitude between $z=-646 \pm 10.8$ pc to $+928 \pm 13.2$ pc.

Figure \ref{fig:match_fraction} shows the fraction of clouds passing cuts on the quality of our morphological match between dust extinction and \HI, as a function of number of draws $N_{\rm{draw}}$, area on the POS $\Omega_{\rm{POS}}$, and absolute Galactic latitude $|b|$.  Match success rate is lower for low-$N_{\rm{draw}}$ structures, suggesting that clouds that appear in fewer draws of the GXP dust map may be less physically meaningful.  Matching rate is also lowest for clouds that are smaller on the POS, which we interpret as indicating that having a smaller number of information-carrying pixels makes it more challenging to distinguish high-quality matches; using current and future high-angular-resolution dust maps \citep[e.g., ][]{ZuckerSaydjari2025} may prove more useful in matching these smaller and/or more distant structures.  Finally, match rate is lowest at low $|b|$, suggesting that confusion in the Galactic plane at low latitudes significantly interferes with our ability to match dust to \HI.  Therefore, our analysis in the remainder of this work is mainly restricted to higher latitude targets.  

We present the total fraction of \HI{} emission recovered within our local clouds as a function of Galactic latitude and velocity in Figure \ref{fig:frac_hi_recovered}.  We observe that, above $|b| > 20^\circ$ or so, we recover on average 20--40\% of all \HI{} emission within the low-velocity range (-15 to 15 km/s).  We expect that our recovered \HI{} fractions are lower limits on the true amount of \HI{} emission coming from distances within 1.25 kpc, owing to the low-matching fractions between 3D dust clouds and \HI{}.  Future work should experiment with more sophisticated methods of associating dust with \HI{}, in order to obtain a more precise estimate of local neutral gas.

\subsection{Cloud Catalog}\label{S:catalog}

Here we describe the properties of clouds released in catalogs with this work.  All properties are computed for all draws, and we provide their values both in all individual draws and the median value across draws.  Systematic uncertainties are derived from computing the standard deviation of each property across draws.  A subset of these properties are presented for selected clouds later in this work in Table \ref{tab:highlights_hac} and Table \ref{tab:highlights_nonhac}; the full catalogs are available online at \url{https://doi.org/10.5281/zenodo.20349203}.

\subsubsection{3D Dust Properties}

For each cloud, we provide the following properties.

\paragraph{IDs} Unique identification number in full \perch generator list, and a filtered identification number solely within structures passing our topological significance cuts

\paragraph{Topological Properties} Birth and death threshold values, pixel indices at which birth and death occur, and persistence.  We additionally report the local noise level at the death pixel $\sigma_{death}$ and the noise-normalized persistence $p_n$.

\paragraph{Clustering Properties} The number of draws the cloud appears in, $N_{\rm{draw}}$.

\paragraph{Hierarchical Properties} Embedded flag (for full set and for HI-matched set), parent structure ID (if embedded), number of children (if a parent).

\paragraph{3D and 2D Maps} 3D cartesian pixel indices of cloud's segmentation in each draw, as well as 2D \Hpx maps of $A_V$, derived from $\mathcal{A}_i$ as,
\begin{equation}
    A_V(p) = m_V \mathcal{A}_i(p)
\end{equation}
where $m_V = 2.8$ \citep{ZhangGreen2023}.  We additionally provide 1D radial maps of the integrated extinction along the LOS, $\mathcal{R}_i$, converted to V-band with the same constant.

\paragraph{Size} Number of pixels $N_{\rm{pix}}$ in 3D space, as well as the equivalent radius (in parsecs) of a sphere occupying the same volume as the structure,
\begin{equation}
    R_{eff} = \left(\frac{3 N_{\rm{pix}}}{4\pi}\right)^{1/3} \Delta x
\end{equation}
where $\Delta x = 2$ pc.  We additionally report the projected area on the POS occupied by $\mathcal{A}_i$,
\begin{equation}
    \Omega_{\rm{POS}} = N_{\rm{pix, Hpx}} \Delta \rm{Hpx}
\end{equation}
where $N_{\rm{pix, Hpx}}$ is the number of \Hpx pixels containing the cloud and $\Delta \rm{Hpx}$ is each pixel's area (in sq. degrees).

\paragraph{Centroid and Spatial Extent} The extinction-weighted centroid of each cloud in 3D coordinates, $(x_c,y_c,z_c)$, calculated using {\tt skimage.measure.centroid}.  We also report the centroid converted to Galactic coordinates ($\ell_c, b_c, d_c$).  We additionally calculate the bounding box of each cloud (minimum and maximum $x, y, z$) and the minimum and maximum distances of the cloud (from $\mathcal{R}_i$, $d_{min}$, $d_{max}$).  

\paragraph{Extinction} The minimum, maximum, median, and integrated values of $A'_V$ in 3D space, as well as the minimum, maximum, and median values of $\mathcal{A}_i$ on the POS.

\paragraph{Angular Resolution} The effective angular resolution of the cloud, $\rm{FWHM}_0$, in the GXP dust map.

\paragraph{Elongation along the line-of-sight} We calculate the relative elongation of each cloud along the LOS relative to its area on the POS,
\begin{equation}
    \mathbb{L}_{\rm{LOS}} = \frac{dr}{2 \sqrt{\Omega_{\rm{POS}} / \pi} \ d_0}
\end{equation}
where $dr$ and $d_0$ are defined in Eqns. \ref{eqn:dr} and \ref{eqn:d_0}, respectively.  $\mathbb{L}_{\rm{LOS}} < 1$ corresponds to a ``pancake''-like cloud mostly tangent to the POS, $\mathbb{L}_{\rm{LOS}}=1$ corresponds to a perfect sphere, and $\mathbb{L}_{\rm{LOS}} > 1$ corresponds to a ``cigar''-like cloud elongated along the LOS.

\subsubsection{HI Properties}\label{S:hiprops}

For clouds with significant HI matches, we additionally provide the following properties.

\paragraph{Morphological Significance} The metrics of morphological significance described in \S\ref{S:match_sig} and \S\ref{ap:sig_matches} ($P_1$, $P_1/P_2$, $\delta_1$), as well as their values normalized by the maximum in the rotation set

\paragraph{Velocity} The LSR velocity at which the maximum SSIM occurs, $v_{HI}$, as well as the minimum and maximum velocities of the SSIM peak, $v_{min}$ and $v_{max}$.  The width of the peak is $\Delta v = v_{max}-v_{min}$.

\paragraph{HI Column Density} Within the range of the SSIM peak ($v_{min}$ to $v_{max}$), we calculate a \Hpx map of the HI column density using Eqn. \ref{eqn:hi_col}.

\paragraph{Intensity-Weighted Velocity} A \Hpx map of the intensity-weighted mean velocity (first moment), $\langle v \rangle$%
\begin{equation}
    \langle v \rangle(p) = \frac{\int_{v_{min}}^{v_{max}} dv \ v \ T_B(v)}{\int_{v_{min}}^{v_{max}} dv \ T_B(v)}
\end{equation}

\paragraph{Line width} A \Hpx map of the \HI line width (square root of the second moment), $\sigma_{HI}$.
\begin{equation}
    \sigma_{HI}(p) = \sqrt{\frac{\int_{v_{min}}^{v_{max}} dv \ (v - \langle v \rangle)^2 \ T_B(v)}{\int_{v_{min}}^{v_{max}} dv \ T_B(v)}}
\end{equation}

\paragraph{Gas-to-Dust Ratio} A \Hpx map of the dust extinction to \HI column density ratio, $ N_{HI} / A_V$

\paragraph{Dust-weighted Mean Properties} The $A_V$-weighted means ($\sum_P X(p) A_V(p) / \sum_p A_V(p)$) of the maps of $N_{HI}$, $\langle v \rangle$, $\sigma_{HI}$, $N_{HI} /A_V$.  Dust-weighting ensures these single summary values reflect regions with significant dust content rather than diffuse edges of clouds.

\paragraph{Volume Density and Mass} The minimum, median, and maximum neutral hydrogen volume densities within each cloud, estimated from the measured $A_V/N_{HI}$, and calculated from the GXP dust map following \citet{ONeillZucker2024} as,
\begin{equation}
    n_{HI} = \frac{m_V}{(A_V / N_{HI})} \frac{A'_{ZGR23, i}}{(3.086 \times 10^{18})}
\end{equation}
for $n_i$ in units of cm$^{-3}$, where the factor $3.086 \times 10^{18}$ is introduced to convert between parsecs and centimeters.  The estimated neutral ISM mass of each cloud is then
\begin{equation}
      M = 1.37 \ m_p \sum_i \ n_{HI,i}  \ dv_i 
\end{equation}
where  $m_p$ is the mass of a proton, 1.37 is a factor derived from cosmic abundances to convert from hydrogen mass to total mass including helium, and $dv_i = (2 \rm{pc})^3$.

\paragraph{Surface Density} The mean neutral atomic ISM surface density,
\begin{equation}
    \Sigma = \frac{M}{\pi R_{eff}^2}
\end{equation}
 
\section{Results}\label{S:results}

\subsection{Kinematics of 3D Dust Clouds: Defining Intermediate Velocity Clouds}\label{S:kinematic}

\begin{figure*}
    \centering
\includegraphics[width=.8\textwidth]{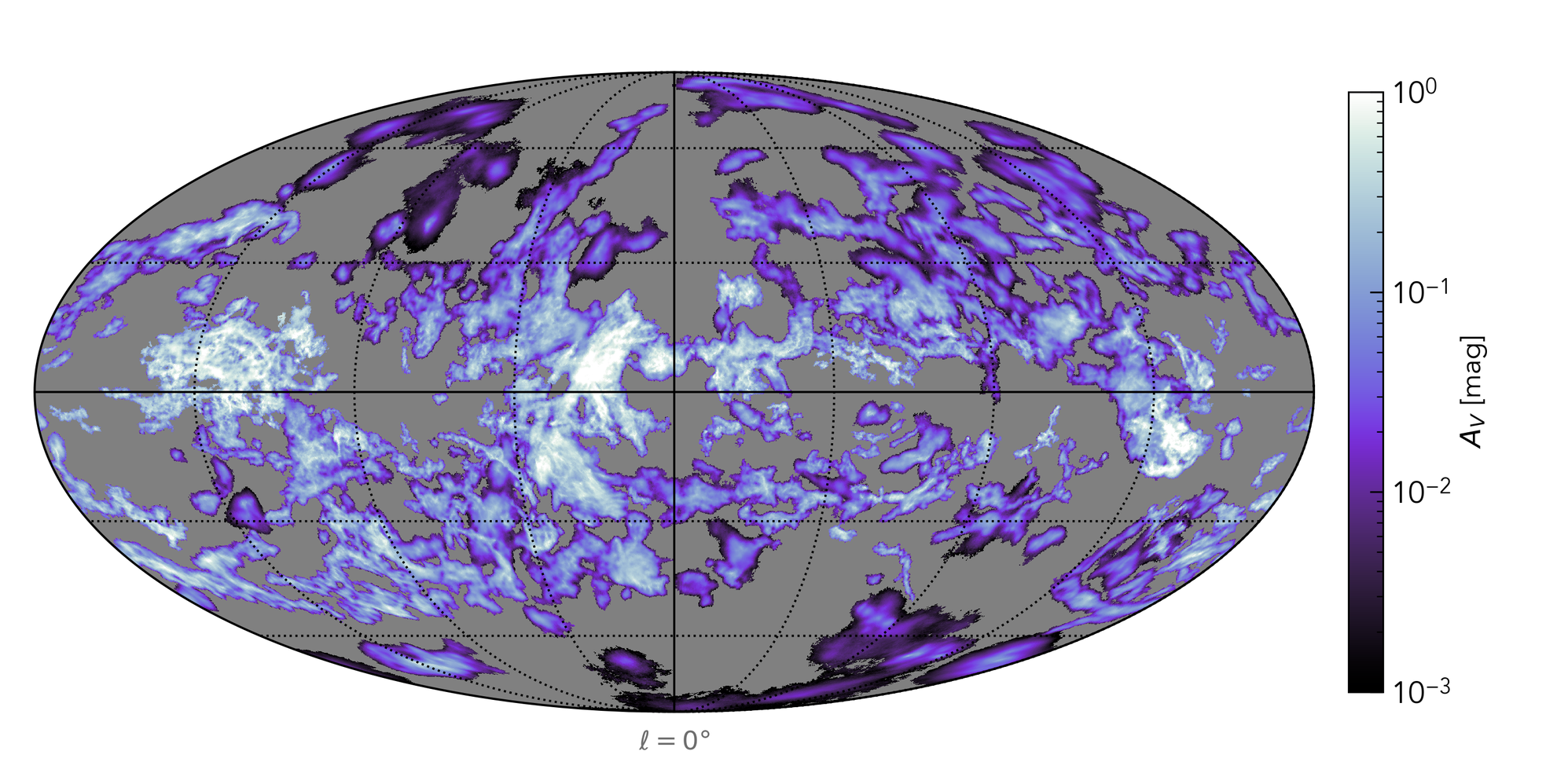}
\includegraphics[width=.8\textwidth]{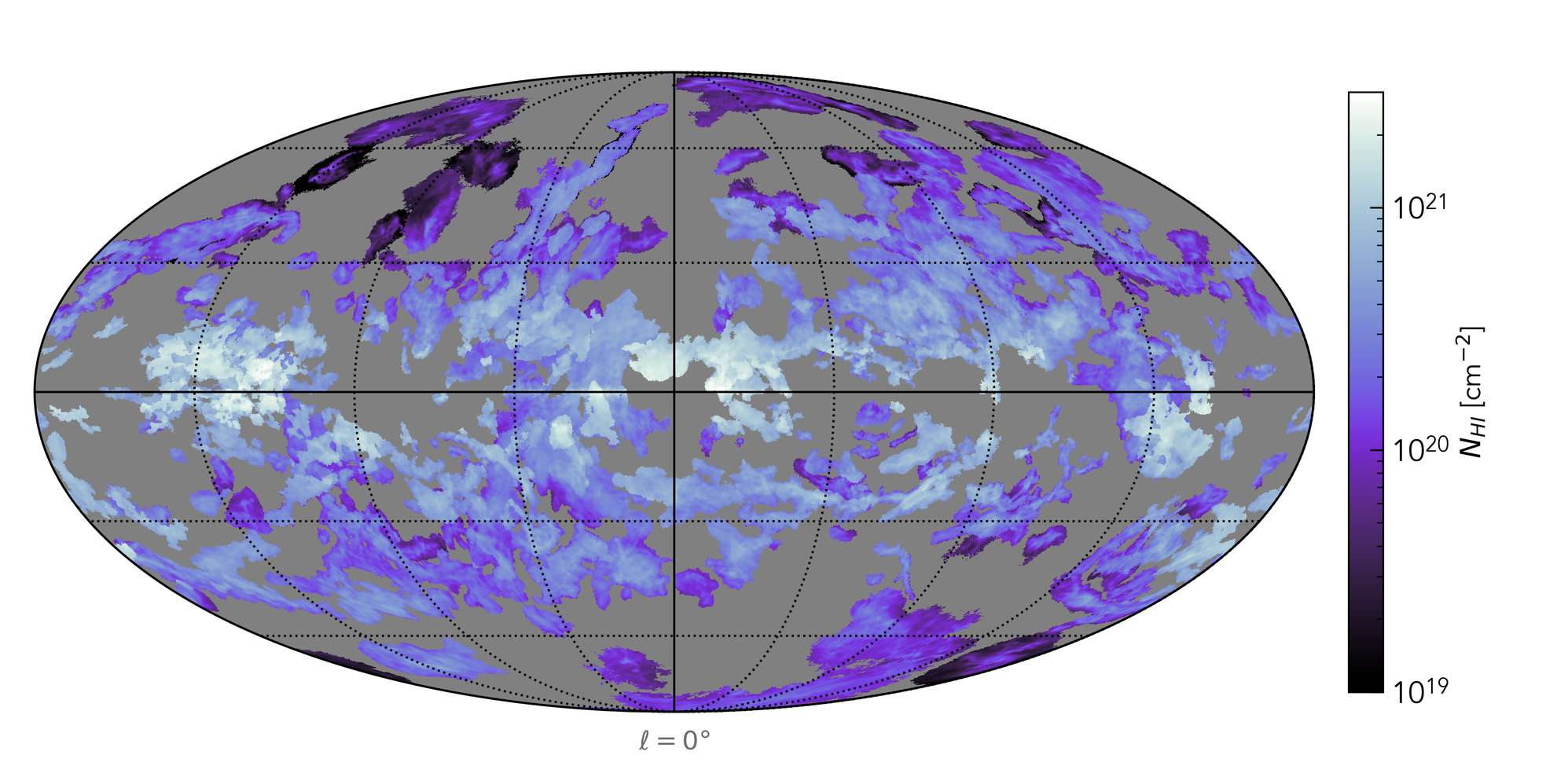}
\includegraphics[width=.8\textwidth]{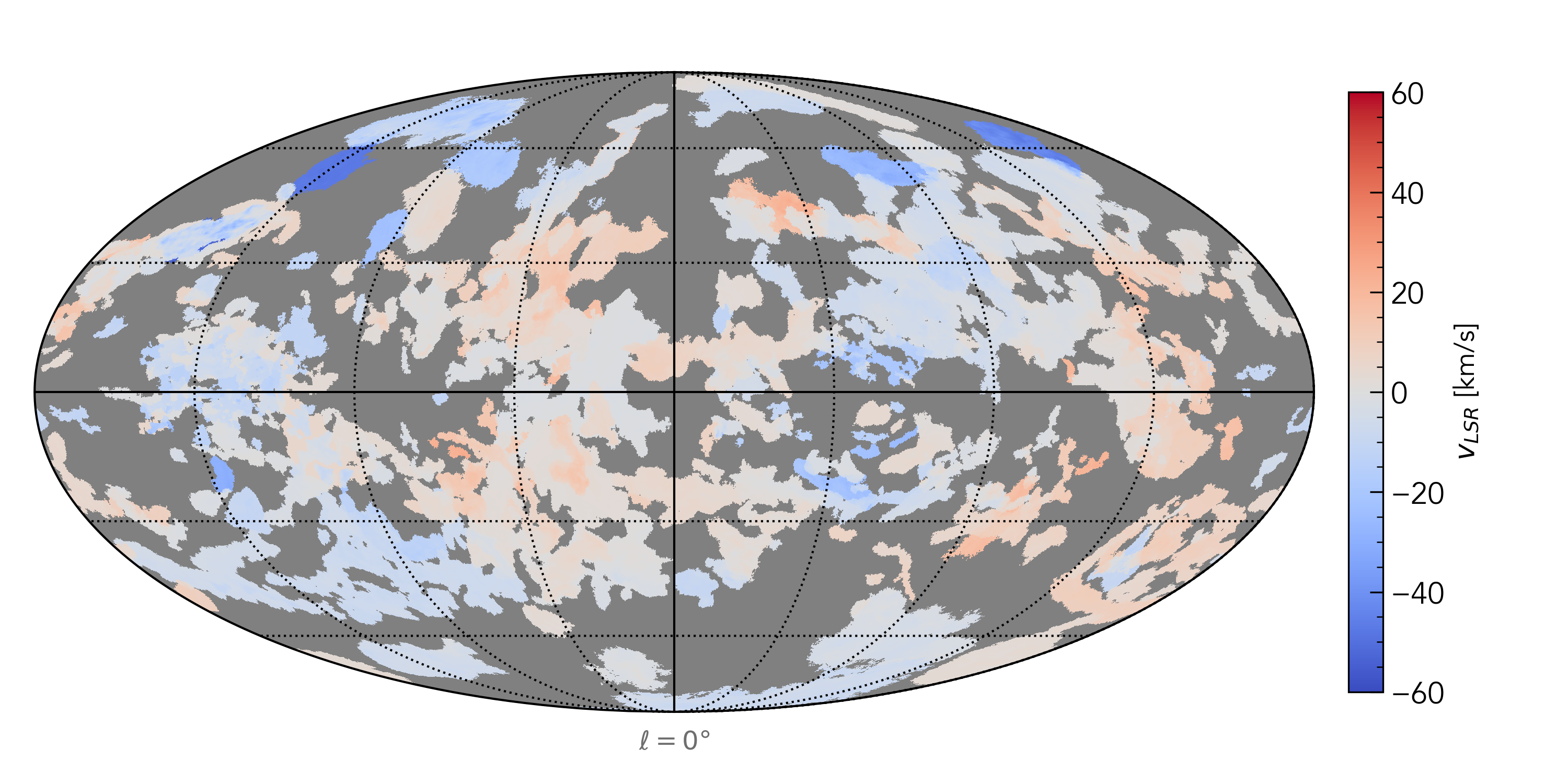}
    \caption{\textit{Top:} Integrated extinction $A_V$ for all dust clouds matched to \HI{}.  Cloud areas are limited to pixels that appear in at least half of each cloud's draws.  \textit{Center:} As top, but showing \HI{} column density $N_{HI}$.  \HI{} is shown at the native angular resolution of the \HIPI dataset; note that matching to dust was performed after smoothing the \HI{} to the effective angular resolution of each cloud.  \textit{Bottom:} Moment 1 map of all matched \HI{} velocity in the LSR frame, weighted by \HI{} intensity. \underline{Interactive Component:} An interactive viewer of various cloud properties, subsets, and projections is available at \url{https://theo-oneill.github.io/HACs_and_IVCs/moments/}.}
    \label{fig:moll_matches}
\end{figure*}

\begin{figure*}
    \centering
\includegraphics[width=.9\textwidth]{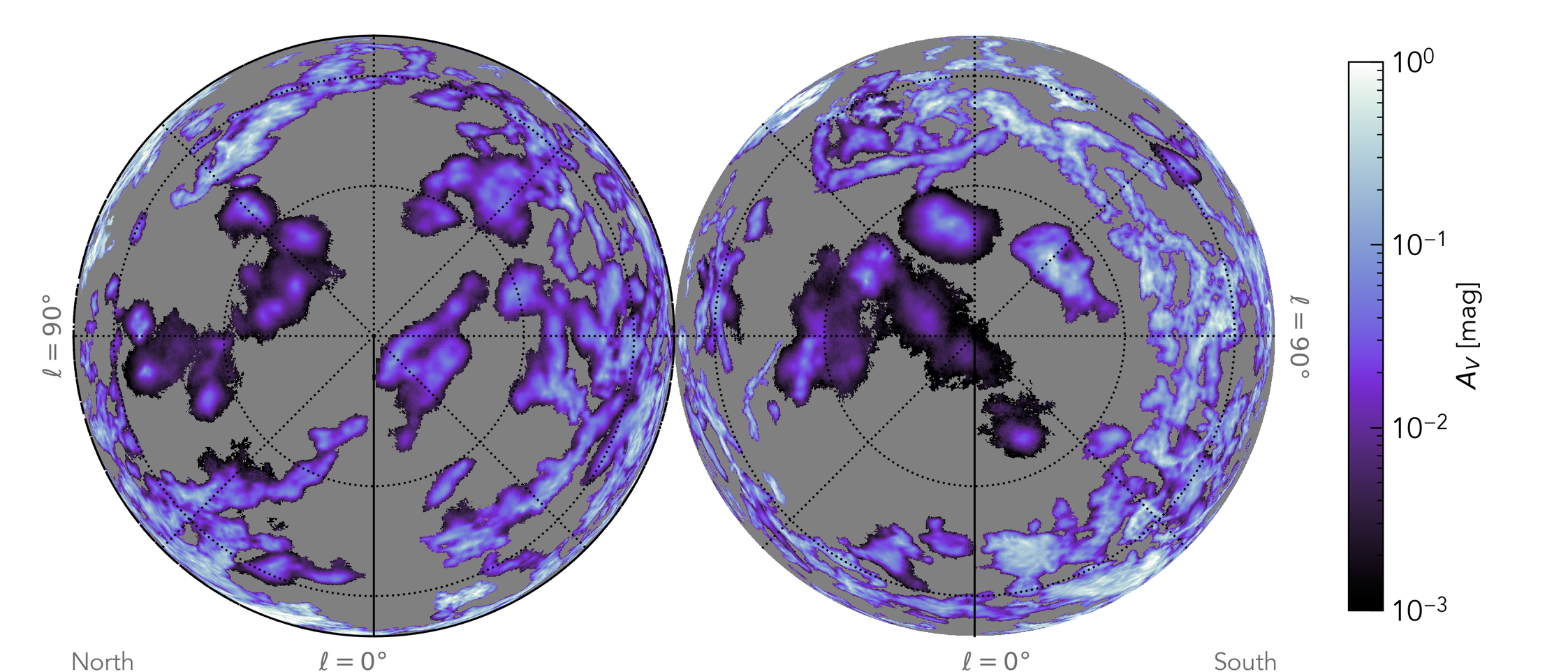}
\includegraphics[width=.9\textwidth]{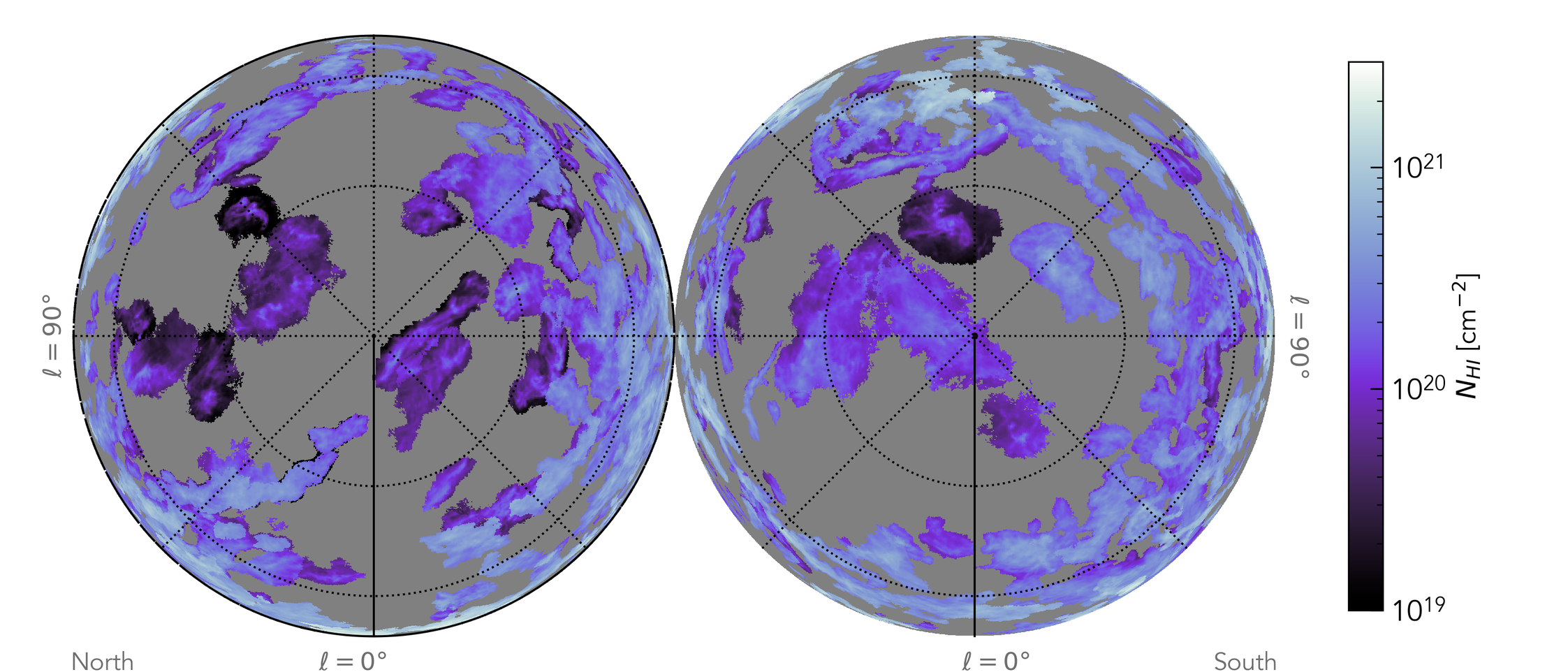}
\includegraphics[width=.9\textwidth]{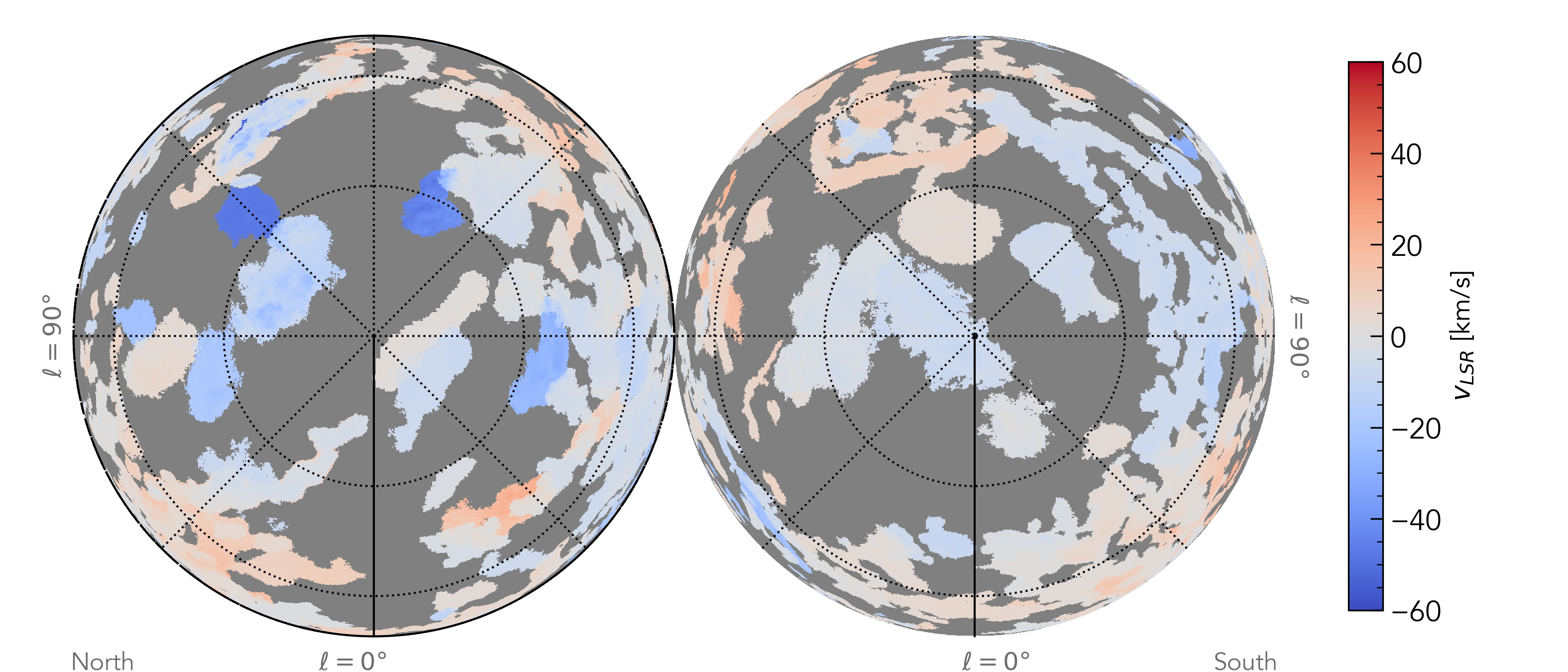}
    \caption{As Figure \ref{fig:moll_matches}, but in polar projection.  The northern hemisphere is shown on the left of each row and the southern on the right, with $\ell=0^\circ$ oriented towards the bottom.}
    \label{fig:polar_matches}
\end{figure*}

\begin{figure*}
    \centering
\includegraphics[width=\textwidth]{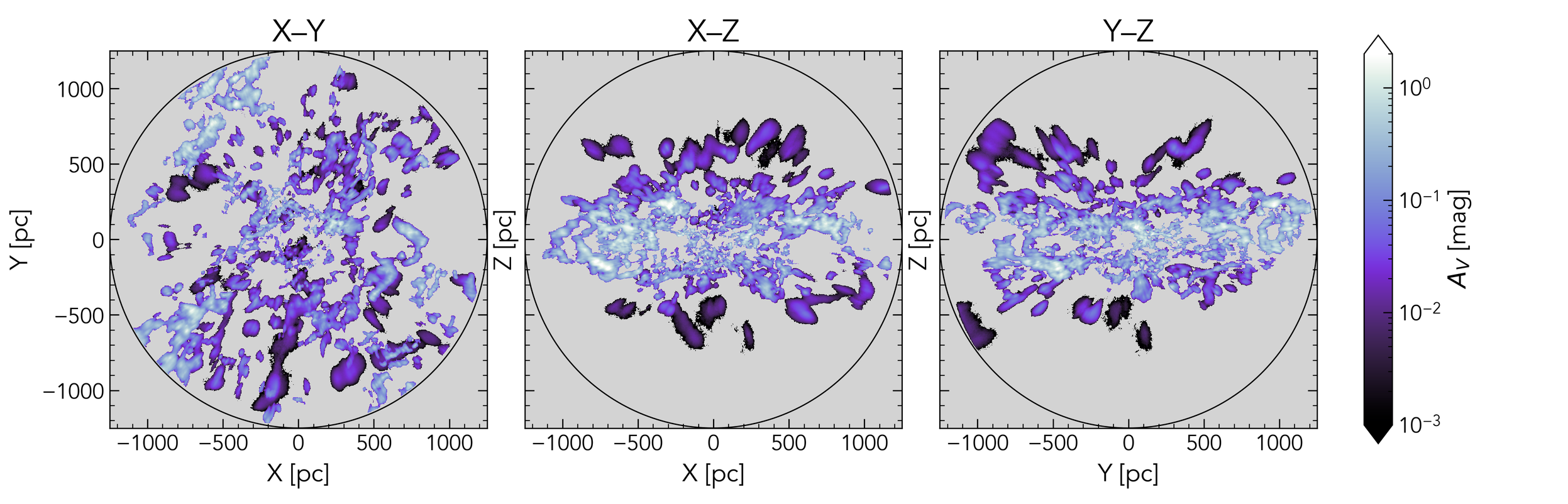}
\includegraphics[width=\textwidth]{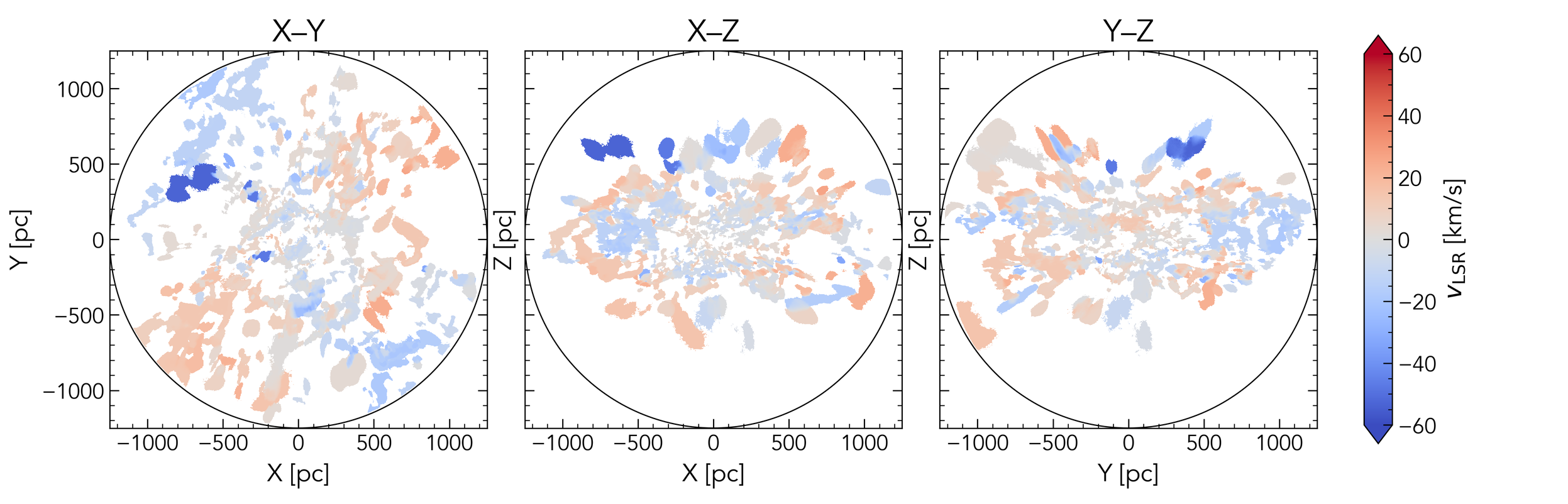}
\includegraphics[width=\textwidth]{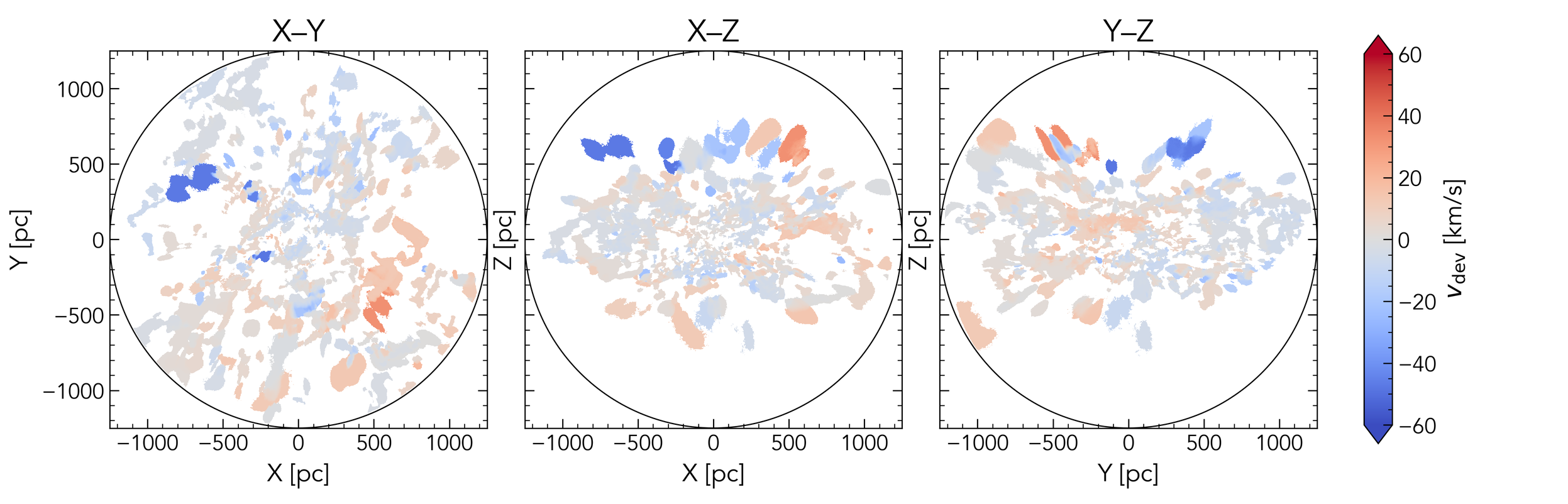}
    \caption{\textit{Top:} Integrated extinction $A_V$ for all dust clouds matched to \HI{}, in edge-on Cartesian 3D projections.  The left column shows the X-Y plane, the center column X-Z, and the right column Y-Z.  The black circle shows the limit of the GXP dust map at $d=1250$ pc.  Cloud areas are restricted to pixels that appear in at least half of each cloud's draws.  \textit{Center:} Moment 1 map of LSR \HI{} velocity, weighted by $A_V$.  \textit{Bottom:} As center, but for deviation velocity $v_{\rm{dev}}$.}
    \label{fig:cart_kinematics}
\end{figure*}

\begin{figure*}
    \centering
\includegraphics[width=\textwidth]{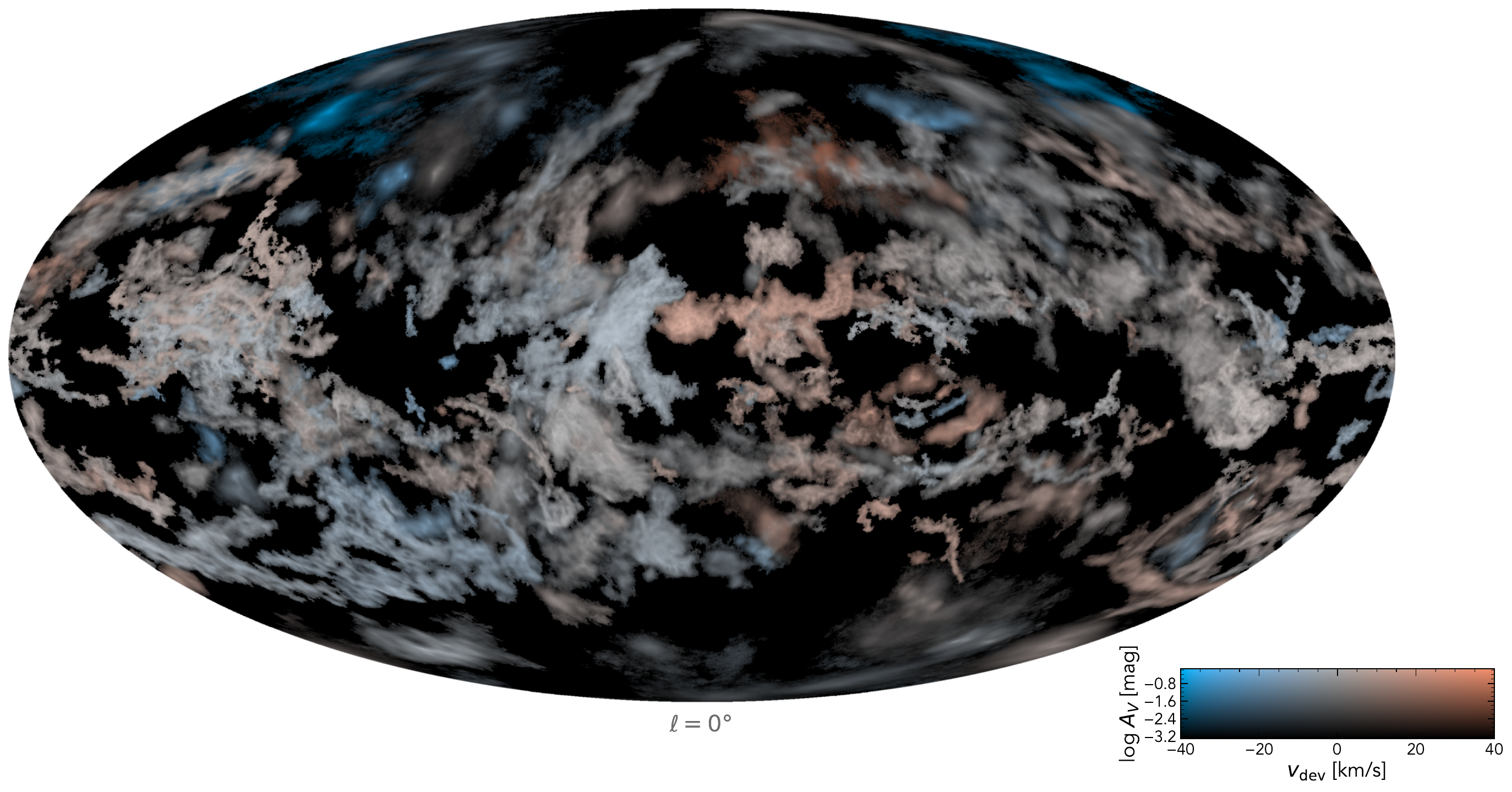}
\includegraphics[width=\textwidth]{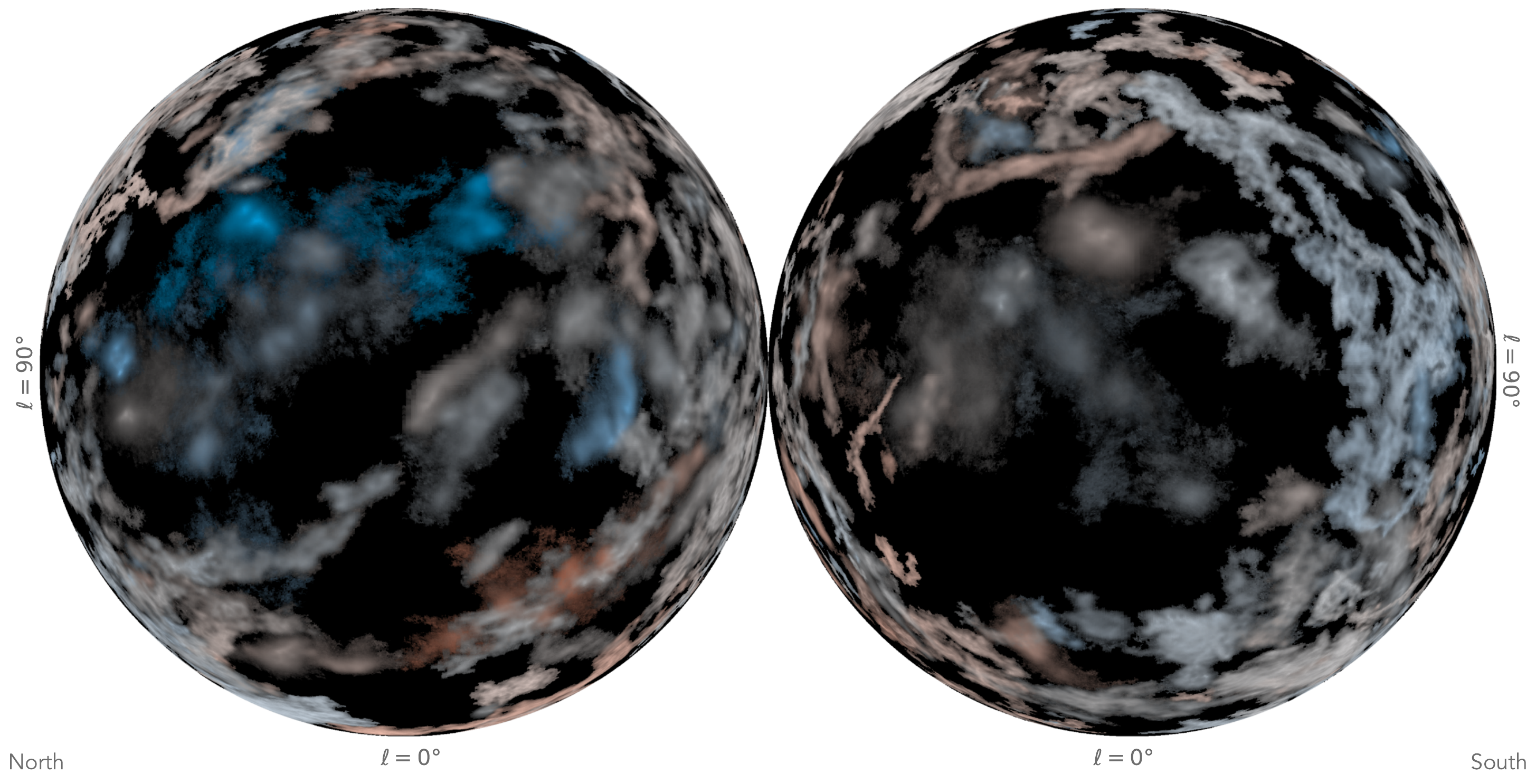}
    \caption{\textit{Top:} Composite RGB figure of maximum-extent \HI{}-matched dust clouds in Mollweide projection.  The Moment 0 map of total cloud extinction $A_V$ is mapped to brightness, and the Moment 1 map (extinction-weighted deviation velocities) of $v_{\rm{dev}}$ is mapped to hue.  \textit{Bottom:} As top, but in polar projection.}
    \label{fig:rgb_mollweide}
\end{figure*}

\begin{figure*}
    \centering
\includegraphics[width=.9\textwidth]{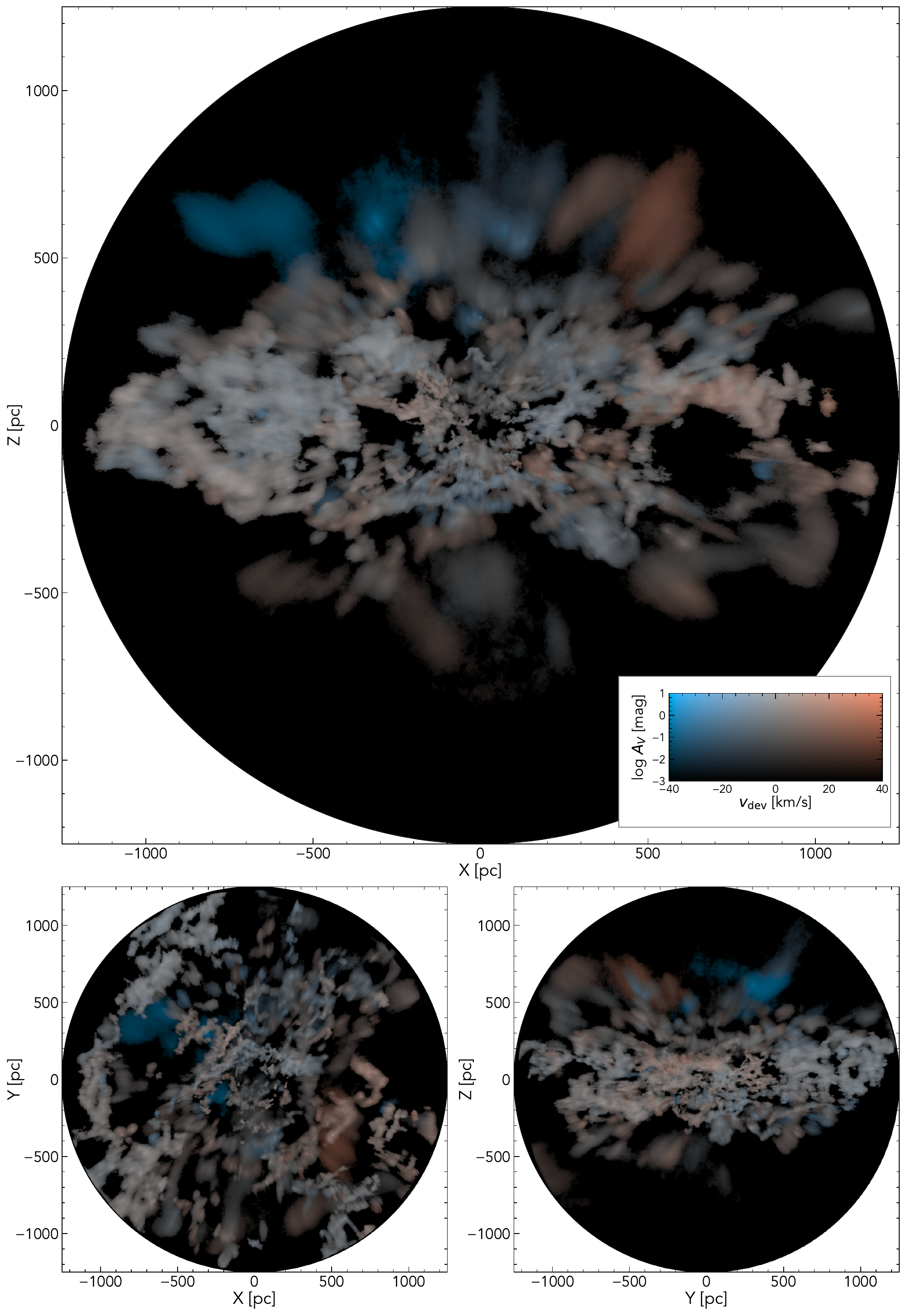}
    \caption{As Figure \ref{fig:rgb_mollweide}, but showing 3D cartesian edge-on projections of maximum-extent clouds in the X-Z plane (top), X-Y plane (lower left), and Y-Z plane (lower right). }
    \label{fig:rgb_cart}
\end{figure*}

\begin{figure*}
    \centering
\includegraphics[width=\textwidth]{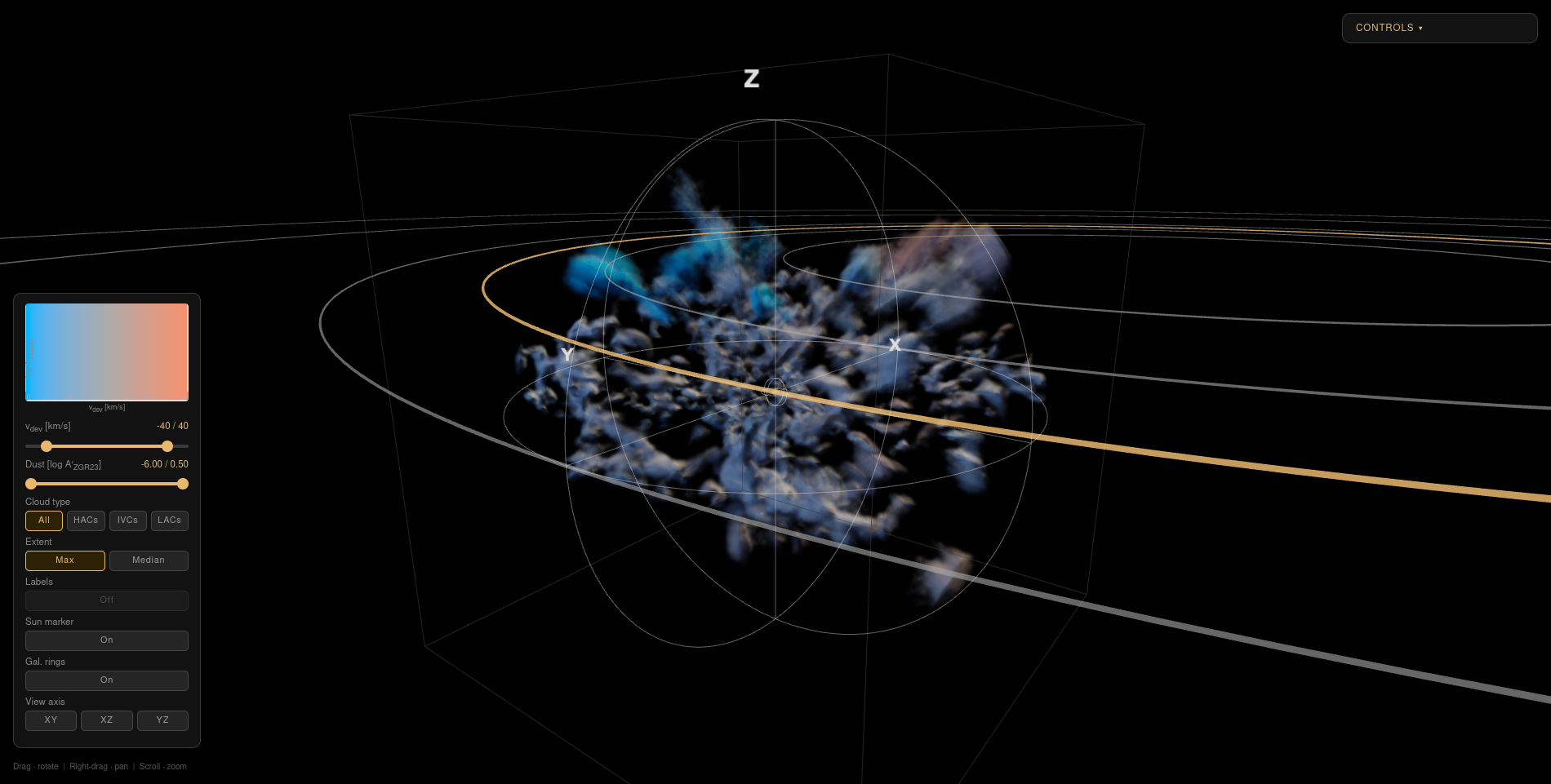}
    \caption{A screenshot of an interactive 3D figure, showing RGB renderings of cloud extinction and velocity similar to Figure \ref{fig:rgb_cart}.  The edges of the GXP dust map are shown by the white spheres.  Curves of constant galactocentric radius $R_{gal}$ are plotted, assuming the Galactic center is positioned at (x,y,z)=(8.5 kpc, 0 kpc, 0 kpc); the Solar circle at $R_{gal}=8.5$ kpc is marked in yellow.  \underline{\textit{Interactive Component:}} A 3D interactive figure is available at \url{https://theo-oneill.github.io/HACs_and_IVCs/all_clouds_3d/index.html}.  A pre-rendered video of the interactive figure is also available at \url{https://theo-oneill.github.io/HACs_and_IVCs/all_clouds_3d_video/}. }
    \label{fig:rgb_3d}
\end{figure*}

We first inspect the kinematics of our sample of \HI{}-matched 3D dust clouds.  Figures \ref{fig:moll_matches} and \ref{fig:polar_matches} present Mollweide and Polar projections, respectively, of the mean extinctions $A_V$, \HI{} column densities $N_{HI}$, and velocities $v_{\rm{LSR}}$ of our sample of matched clouds on-sky.  Figure \ref{fig:cart_kinematics} presents x-y, x-z, and y-z projections of the 3D cartesian view of our cloud sample.  

To identify a class of anomalous-velocity clouds at the disk-halo interface, we must understand which clouds are moving at velocities incompatible with Galactic rotation.  To do so, we calculate the deviation velocity of each cloud, a term introduced to the IVC/HVC literature by \citet{Wakker1991} to quantify the departure between a cloud's predicted and actual velocity\footnote{We note that our definition of $v_{\rm{dev}}$ departs from \citet{Wakker1991}, owing to the known distances to our sample of clouds provided by 3D dust.  The \citet{Wakker1991} definition computed the difference between clouds' LSR velocities and the \textit{range} of velocities allowed by galactic rotation at the $(\ell,b)$ position of each cloud.}, as:
\begin{equation}
    v_{\rm{dev}} = v_{\rm{LSR}} - v_{\rm{rot}}(\ell_c, b_c, d_c)
\end{equation}
where $v_{\rm{LSR}}$ is the \HI{}-matched LSR velocity, $(\ell_c, b_c, d_c)$ are the dust-weighted centroid coordinates of the cloud, and $v_{\rm{rot}}$ is the cloud's expected velocity under Galactic rotation.  We calculate $v_{\rm{rot}}(\ell_0, b_0, d_0)$ under the \citet{ReidMenten2019} Galactic rotation curve, using the python package {\tt KDUtils} \citep{KDUtils2017}.  We calculate $v_{\rm{dev}}$ for each individual draw of a cloud, with the median and standard deviation of $v_{\rm{dev}}$ calculated across draws used to summarize this information.  Figures \ref{fig:rgb_mollweide} and \ref{fig:rgb_cart} present composite RGB figures of cloud $A_V$ colored by $v_{\rm{dev}}$ in various 2D and 3D projections; Figure \ref{fig:rgb_3d} highlights an accompanying interactive 3D figure showing the 3D spatial + 1D velocity structure of our sample. 

\begin{figure*}
    \centering
\includegraphics[width=\textwidth]{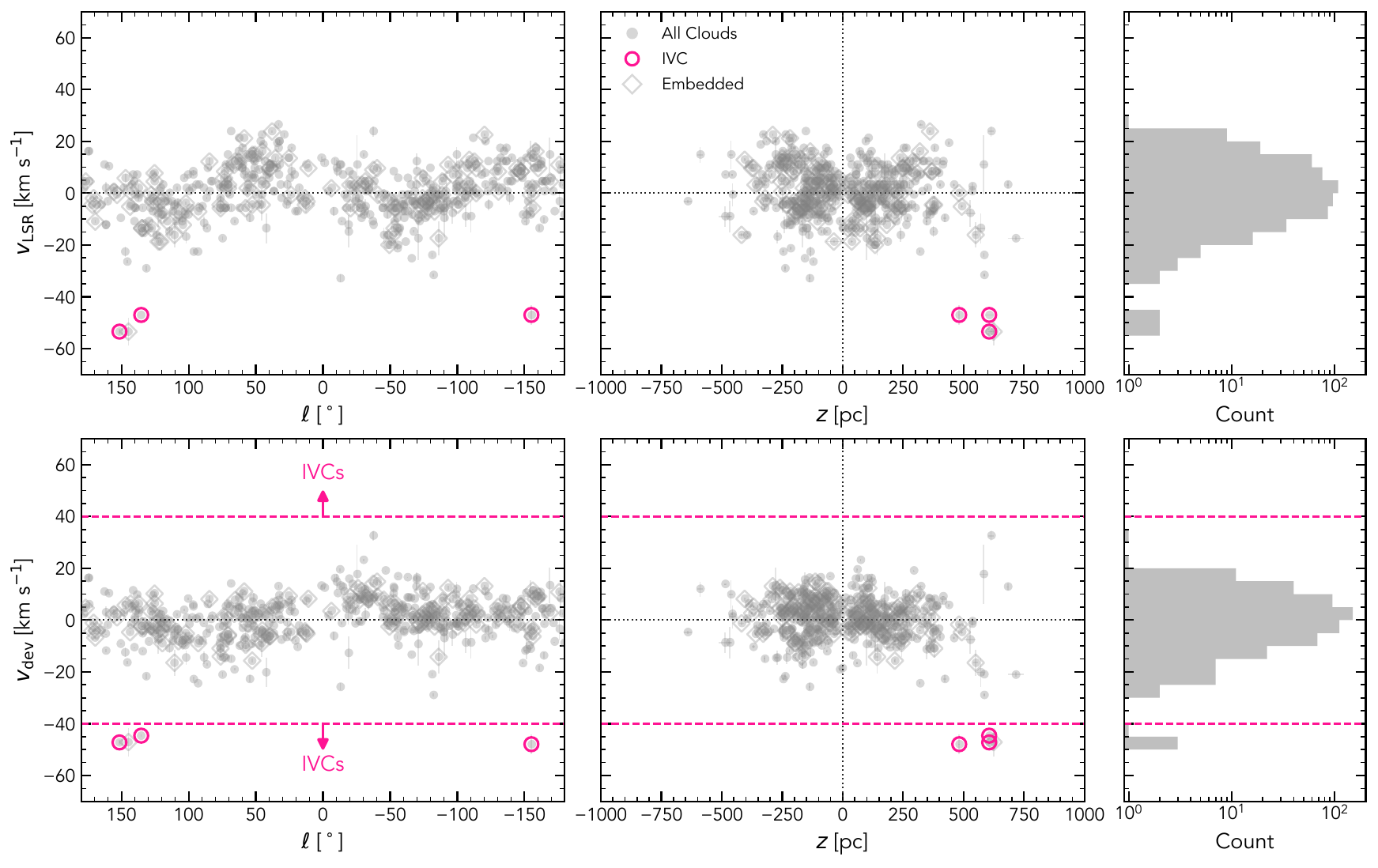}
    \caption{\textit{Top:} LSR velocity $v_{\rm{LSR}}$ as a function of (left) Galactic longitude $\ell$ and (center) altitude $z$, with the rightmost panel showing its global distribution.  Clouds are shown by the gray circles, with error bars representing the standard deviations across draws.  Embedded clouds are marked with a gray diamond, and IVCs (defined from the bottom row) are marked with pink circles.  \textit{Bottom:} As top, but with deviation velocity from rotation $v_{\rm{dev}}$.  The horizontal pink dashed lines show our fiducial threshold to define IVCs, $|v_{\rm{dev}}|>40$ km/s.}
    \label{fig:l_v_z}
\end{figure*}

\begin{figure*}
    \centering
\includegraphics[width=\textwidth]{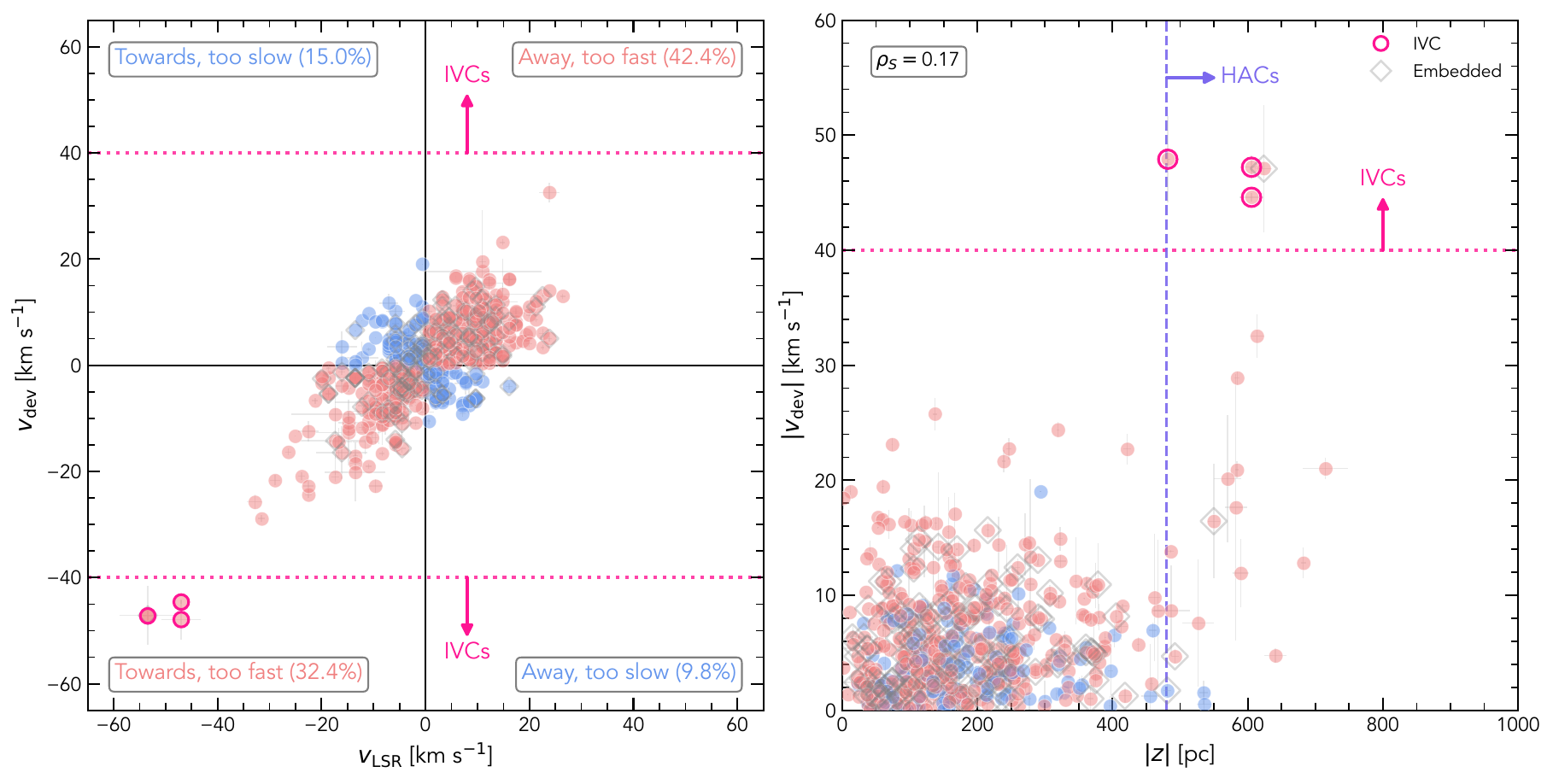}
    \caption{\textit{Left:} $v_{\rm{dev}}$ as a function of $v_{\rm{LSR}}$, with physical interpretation of $v_{\rm{dev}}$ by-quadrant labeled in each panel.  The fraction of all matched clouds in each quadrant is also reported.  Clouds that are moving ``too fast'' for rotation are shown in pink, while clouds moving ``too slow'' are shown in blue.  The pink IVC dotted thresholds are as in Fig. \ref{fig:l_v_z}. \textit{Right:} $|v_{\rm{dev}}|$ as a function of $|z|$, with colors for too fast or too slow as on the left.  The purple vertical dashed line marks our fiducial threshold for high-altitude clouds (HACs) defined in \S\ref{S:vertical}.  Embedded clouds and IVCs are shown by the gray diamonds and pink circles, respectively.}
    \label{fig:vdev_vlsr_z}
\end{figure*}

Figure \ref{fig:l_v_z} presents $\ell-v$, $\ell-v_{\rm{dev}}$, $z-v$, and $z-v_{\rm{dev}}$ diagrams of our cloud sample.  The traditional sinusoidal pattern in $\ell-v$ space imparted by Galactic rotation is present \citep[e.g.,][]{Oort1927,DameHartmann2001}, and is greatly reduced (although not entirely eliminated) in $\ell-v_{\rm{dev}}$ space.  The structured residuals visible in $\ell-v_{\rm{dev}}$ may imply that refinements to the local rotation curve are needed.  The majority of clouds have $v_{\rm{dev}}$ compatible with disk rotation, with a median of 1.2 km/s and standard deviation of 8.8 km/s, and one-$\sigma$ percentiles (16th and 84th percentiles) of -6.7 km/s and 8.3 km/s.  A small number of clouds have significantly larger $v_{\rm{dev}}$, which we judge as likely incompatible with disk rotation and corresponding to a class of IVCs.

As shown in Figure \ref{fig:vdev_vlsr_z}, interpretation of the value of $v_{\rm{dev}}$ depends on the sign of both $v_{\rm{LSR}}$ and $v_{\rm{dev}}$.  For negative-velocity clouds ($v_{\rm{LSR}}<0$), $v_{\rm{dev}} > 0$ implies a cloud moving towards us slower than predicted by rotation; $v_{\rm{dev}} < 0$ implies moving towards us faster than under rotation.  For positive-velocity clouds ($v_{\rm{LSR}}>0$), $v_{\rm{dev}} > 0$ implies a cloud moving away from us faster than predicted by rotation; $v_{\rm{dev}} < 0$ implies moving away from us slower than under rotation.  As noted by \citet{Wakker2001}, turbulence within a cloud can easily reach levels of tens of km/s, meaning that small values of $|v_{\rm{dev}}|$ should not be interpreted to mean that a cloud is inconsistent with rotation.

We observe that $(74.9 \pm 0.5)\%$ of our cloud sample is moving faster than expected under the \citet{ReidMenten2019} rotation curve (where uncertainties on this fraction are derived from 5,000 iterations of bootstrap resampling over individual cloud draws).  We perform simple numerical experiments injecting various amounts of isotropic 3D turbulence (described in Appendix \ref{ap:turbulence}, with turbulent velocities drawn from Gaussian distributions as $\sim \mathcal{N}(0,\sigma^2)$) into our cloud sample's predicted rotation velocities and computing the resulting fraction of clouds that are too fast vs. too slow as assessed via the combination of $v_{\rm{LSR}}$ and $v_{\rm{dev}}$.  Figure \ref{fig:turbulence_constraint} in Appendix \ref{ap:turbulence} presents the results of this test.  We find that a preference for clouds that move faster than rotation under this metric emerges naturally from the injection of isotropic turbulence, and that our observed fraction of $(74.9 \pm 0.5)\%$ moving too quickly is consistent with levels of 1D turbulence $\sigma \sim 5.5-8.0$ km/s (as assessed by computing two-sided $p$-values from comparing the observed vs simulated distributions, and defining inconsistency as $p < 0.05$).  Despite the greatly simplified nature of this comparison, this is entirely consistent with expectations for turbulence on 10---100 parsec scales in molecular clouds \citep[e.g.,][]{larson1981,solomon_mass_1987,HeyerDame2015}.

Figure \ref{fig:vdev_vlsr_z} additionally shows the correlation between $|z|-|v_{\rm{dev}}|$.  There is a weak but significant trend between these quantities, with a Spearman $\rho_s$ correlation coefficient of $\rho_s = 0.17$ ($p \ll 0.001$).  It is possible that some or all of this trend may be explained by the proposed vertical lag in the Galactic rotation curve as altitude above the midplane increases; previous models have estimated lags ranging between $-22 \pm 6$ km s$^{-1}$ kpc$^{-1}$ \citep{LevineHeiles2008} to $-15 \pm 4$ km s$^{-1}$ kpc$^{-1}$ \citep{MarascoFraternali2011}.  In future work, we will perform detailed kinematic and spatial modeling of our sample of \HI{}-matched dust clouds in order to investigate potential signatures of lag and other kinematic effects.

No universally adopted definitions of ``intermediate-velocity'' or ``high-velocity'' gas exist; for the purposes of this work we follow \citet{Wakker2001}, though we use a threshold in deviation velocity rather than in the LSR frame. We define low-velocity clouds (LVCs) as clouds with $|v_{\rm{dev}}| \leq 40$ km s$^{-1}$, IVCs as clouds with $|v_{\rm dev}| > 40$  km s$^{-1}$, and HVCs as $|v_{\rm dev}| \geq 90$ km s$^{-1}$.  We exclude embedded children and middle structures (which are morphological duplicates of their parents, as described in Appendix \ref{S:embedded}) from our analysis, because they are not independent ``clouds.''  Under this observational definition, we identify 395 LVCs and 3 IVCs.  We note that there is a clear break in the distribution of $v_{\rm{dev}}$ around our selected threshold, and that any $|v_{\rm{dev}}|$ threshold between $33-44$ km/s would yield the same number of IVCs.    

All 3 IVCs are ``inflowing'' (negative LSR velocities and negative deviation velocities, i.e., moving towards us in the radial direction more quickly than expected under Galactic rotation) clouds located in the Northern Galactic Hemisphere.  Their $z$-heights (median across draws) range between 482 to 606 parsecs; this is consistent with observations that generally suggest that most IVCs should be confined to $|z| <1500$ pc \citep{LehnerHowk2022}.  Our sample of entirely Northern and inflowing IVCs is consistent with the well-known global asymmetries in the distribution of Northern vs. Southern IVCs and inflowing vs. outflowing IVCs \citep[e.g.,][]{KuntzDanly1996,RohserKerp2016,PanopoulouLenz2020}; these asymmetries are at least partially driven by the Northern Intermediate Velocity Arch feature (of which our 3 IVCs appear to be a part, discussed further in \S\ref{S:lit}).

Our observed IVCs represent the largest deviation velocities in our sample (with $v_{\rm{dev}}$ ranging between $-44.6$ km/s to $-47.9$ km/s), so we therefore detect no HVCs in the Solar Neighborhood.  This is unsurprising, given the expectation that HVCs are generally much more distant ($d \gtrsim 5$ kpc, \citealt{Wakker2001}) than the limits of the GXP dust map ($d \leq 1.25$ kpc); HVCs are also generally dust-poor \citep{Wakker2001,PutmanPeek2012,HayakawaFukui2024}, and so even a nearby dust-poor HVC (typically $\sim3-10$ mmag $A(V)$) might not be detectable with current dust map sensitivity limits ($\sim30$ mmag $A(V)$, see \S\ref{S:cv_cluster}).

\subsection{Vertical Distribution of Clouds: High-altitude Asymmetries in the Northern vs. Southern Galactic Hemispheres}\label{S:vertical}

Although we find only 3 IVCs in our sample of 398 non-embedded clouds, a number of other clouds are located at equally high-altitudes.  The aim of this paper is to study local clouds of all velocities that exist at the border between the disk and lower halo.  To this end, we consider the vertical distribution of clouds identified by \perch in the Solar neighborhood to search for clouds located at altitudes that are meaningfully distinct from disk gas.  

\begin{figure*}
    \centering
\includegraphics[width=\textwidth]{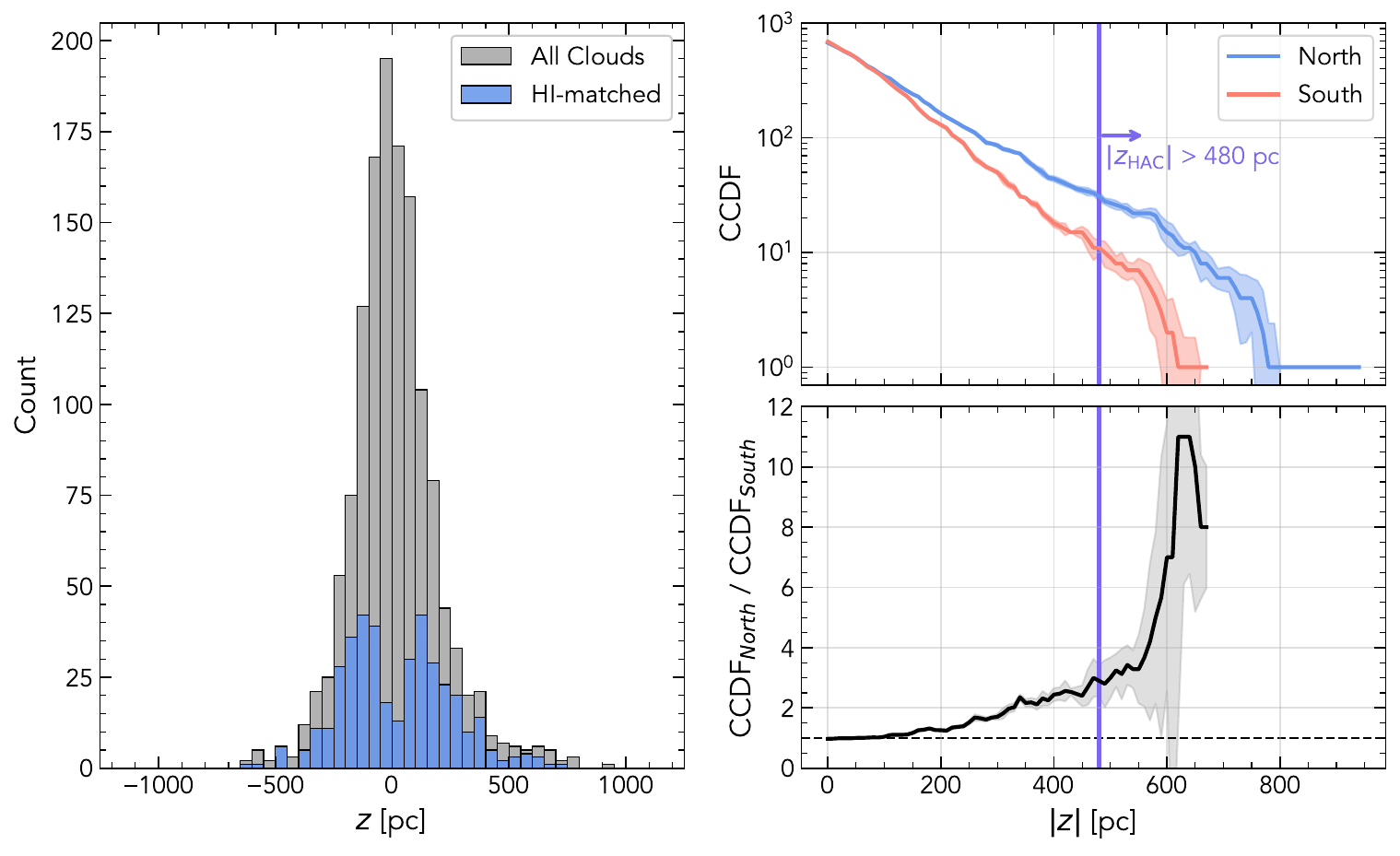}
    \caption{\textit{Left:} Histogram of cloud altitudes $z$, for our full sample of clouds (in grey) and our \HI{}-matched subsample of clouds (in blue).  \textit{Top right:} Complementary cumulative distribution functions (CCDFs) of absolute cloud altitude $|z|$, for clouds in the North ($z>0$, in blue) and in the South ($z<0$, in pink).  CCDFs were derived from bootstrap resampling of cloud altitudes across draws, with the shaded regions around each draw showing the $3\sigma$ uncertainties.  \textit{Bottom right:} The ratio of the Northern CCDF to the Southern CCDF as a function of $|z|$ is shown by the black curve.  The gray shaded region shows the $3\sigma$ uncertainties.  The purple vertical line shows the position of the minimum IVC altitude, $|z|\geq480$ pc, which we use to define a minimum $z$-height for high-altitude clouds (HACs).}
    \label{fig:zdist_ccdf}
\end{figure*}

For each cloud in our full sample of clusters (pre-\HI{} matching, to avoid issues with our selection effects at low-latitudes), we calculate the median $z$-height of the dust-weighted center across each draw it appears in.  Figure \ref{fig:zdist_ccdf} shows the distribution of the median $z_{center}$ for all clouds across draws, for our full sample of clouds.  The distribution is centered at a median $z = 0$ pc, with a standard deviation of $\sigma = 183$ pc.  Since the distribution is consistent at being centered around $z=0$, we do not recenter around the estimated position of the Galactic midplane between $z \simeq -25$ pc to $z \simeq -5$ pc \citep[e.g., as measured with stellar vs. gaseous tracers, ][]{MaizApellaniz2001_spatial,JuricIvezic2008,AndersonWenger2019}.  The distribution of vertical cloud heights is significantly positively skewed (Fisher-Pearson skew coefficient of 0.58, with a skew test $p \ll 0.001$), i.e., with a more prominent tail extending towards Galactic north than Galactic south. 

We evaluate where the departure between Northern and Southern clouds becomes significant in Figure \ref{fig:zdist_ccdf}.  We generate 5,000 bootstrapped samples of the distributions of $z_{center}$ for each of our clusters, by randomly selecting with replacement the measurements of $z_{center}$ for the individual draws in which the cluster appears.  We then calculate the complementary cumulative distribution functions (CCDFs) of $|z_{center}|$ for Northern and Southern clouds (i.e., the number of Northern or Southern clouds with $|z_{center}| > |z|$ as a function of minimum $|z|$) in bins of 10 pc in $|z|$.   We additionally plot the mean bootstrapped CCDFs in Figure \ref{fig:zdist_ccdf}, along with the range of $3 \sigma$ uncertainties on each mean CCDF.  We show the ratio of the Northern vs. Southern CCDFs, CCDF(North)/CCDF(South), with corresponding $3\sigma$ uncertainties.   

We observe that the imbalance between Northern and Southern clouds becomes significant (defined as the lower 5$\sigma$ boundary exceeding a ratio of 1, for more than one $z$-bin in a row) at $z \geq 100$ pc.  The imbalance increases continuously at higher altitudes, until a maximum Northern-to-Southern ratio of $11.0 \pm 3.1$ (where the reported uncertainty is 1$\sigma$) is reached for $z \geq 620$ pc (note, though, that at these altitudes in the map, there is only one southern cloud and small number statistics cause the uncertainties on the ratio of CCDFs to grow large).  

In a galactic disk, pressure-regulated feedback-modulated models imply vertical symmetry in the distribution of gas around the midplane when averaged over kiloparsec-scales over several vertical dynamical times \citep{OstrikerKim2022}.  Although our sample represents only an instantaneous sample of clouds within a kiloparsec-radius sphere, we find that clouds at lower altitudes in the GXP map are consistent with this expectation of vertical symmetry.  Our observation of increasing asymmetry as $|z|$ increases suggests a transition in the properties or origins of the cloud sample being observed at high vs. low altitudes.  

We observe that the north-south imbalance begins to dramatically increase shortly after the height of lowest-altitude IVC (at $z=482$ pc) is reached; the imbalance at the nearest 10-pc bin threshold, $|z|\geq480$ pc, is $2.9 \pm 0.2$ times more clouds in the North than South.  We therefore come to a working definition of ``high-altitude'' clouds \citep[HACs, a term dating back in the literature to at least][]{Malhotra1994} as those with $|z|$-heights greater than or equal to our minimum observed IVC altitude, $|z|\geq480$ pc.   This observationally-driven threshold is well above the expected scale heights (in the Solar Neighborhood) of a few hundred parsecs for \HI{} \citep[e.g.,][]{DickeyLockman1990} and less than a hundred parsecs for molecular gas \citep[e.g.,][]{HeyerDame2015}.

In total, this threshold yields a sample of 17 (non-embedded) \HI{}-matched HACs, of which only 3 ($\sim$18\%) are IVCs.  The interactive component of Figure \ref{fig:moll_matches} visualizes the HAC sample in various projections.  We observe that 13 of our 17 HACs (76.5\%) are located in the Galactic Northern hemisphere.  In total, the HAC sample has a mass of $(3.38 \pm 1.04) \times 10^5 \ M_\odot$, with 71\% of the mass located in the North ($(2.40 \pm 0.88) \times 10^5 \ M_\odot$ in the North vs. $(0.98 \pm 0.57) \times 10^5 \ M_\odot$ in the South).  In the following subsection, we investigate various cloud properties for any revealing differences between cloud groupings and environments.

\subsection{Physical Properties of Cloud Populations}\label{S:corr_props}

\subsubsection{Disentangling Trends with Cloud Altitude and Distance}\label{S:trends}

\begin{deluxetable*}{llccc}
\tablecaption{Summary statistics of cloud properties \label{tab:summary}}
\tablehead{
\colhead{Property} & \colhead{Units} & \colhead{Min} & \colhead{Median} & \colhead{Max}
}
\startdata
\multicolumn{5}{c}{\textit{3D Dust} ($N = 1695$)} \\
\hline
$R_{\rm eff}$ & pc & 1.24 & $18.74 \pm 0.57$ & $223.5 \pm 4.6$ \\
$\Omega_{\rm POS}$ & deg$^2$ & 0.682 & $16.1 \pm 7.4$ & $6625 \pm 960$ \\
$\max(A'_{\rm 3D})$ & mag pc$^{-1}$ & $(2.3 \pm 1.1) \times 10^{-5}$ & $(2.77 \pm 0.19) \times 10^{-2}$ & $0.834 \pm 0.094$ \\
$\max(A_{\rm POS})$ & mag & $(5.0 \pm 2.1) \times 10^{-4}$ & $0.414 \pm 0.021$ & $5.57 \pm 0.31$ \\
$\Sigma A'_{\rm 3D}$ & mag pc$^{-1}$ & $(1.82 \pm 0.91) \times 10^{-2}$ & $12 \pm 28$ & $(2.007 \pm 0.040) \times 10^{4}$ \\
$\mathbb{L}_{\rm LOS}$ & \ldots & $0.284 \pm 0.022$ & $1.360 \pm 0.014$ & $3.91 \pm 0.11$ \\
\hline
\multicolumn{5}{c}{\textit{3D Dust + \HI{}} ($N = 519$)} \\
\hline
$V_{\rm LSR}$ & km s$^{-1}$ & $-53.43 \pm 0.92$ & $0.64 \pm 0.59$ & $26.39 \pm 0.48$ \\
$V_{\rm dev}$ & km s$^{-1}$ & $-47.9 \pm 3.7$ & $1.18 \pm 0.49$ & $32.5 \pm 1.9$ \\
$\max(N_{\rm HI})$ & cm$^{-2}$ & $(1.049 \pm 0.065) \times 10^{20}$ & $(4.67 \pm 0.20) \times 10^{20}$ & $(3.56 \pm 0.22) \times 10^{21}$ \\
$\sigma_{\rm HI}$ & km s$^{-1}$ & $0.82 \pm 0.18$ & $2.450 \pm 0.065$ & $8.30 \pm 0.76$ \\
$N_{\rm HI}/A_V$ & cm$^{-2}$ mag$^{-1}$ & $(4.71 \pm 0.18) \times 10^{20}$ & $(3.31 \pm 0.19) \times 10^{21}$ & $(3.9 \pm 1.7) \times 10^{22}$ \\
$X_{n_H}$ & pc cm$^{-3}$ mag$^{-1}$ & $427 \pm 17$ & $3002 \pm 175$ & $(3.5 \pm 1.6) \times 10^{4}$ \\
$\max(n_{\rm HI})$ & cm$^{-3}$ & $0.76 \pm 0.22$ & $17.20 \pm 0.95$ & $123 \pm 21$ \\
$M$ & $M_{\odot}$ & $23.0 \pm 3.8$ & $3553 \pm 1912$ & $(3.62 \pm 0.64) \times 10^{5}$ \\
$\Sigma$ & $M_{\odot}$ pc$^{-2}$ & $0.273 \pm 0.051$ & $2.44 \pm 0.13$ & $15.7 \pm 2.6$ \\
$\alpha_{\rm vir}$ & \ldots & $1.040 \pm 0.048$ & $26.9 \pm 3.4$ & $601 \pm 37$ \\
$P_e / k_B$ & K cm$^{-3}$ & $65 \pm 10$ & $2458 \pm 335$ & $(5.7 \pm 1.6) \times 10^{4}$ \\
$T_k$ & K & $85 \pm 35$ & $727 \pm 58$ & $8356 \pm 1481$ \\
$P_k / k_B$ & K  cm$^{-3}$ & $48.6 \pm 7.6$ & $1946 \pm 251$ & $(4.4 \pm 1.2) \times 10^{4}$ \\
\enddata
\tablecomments{Uncertainties on the minimum/maximum values are the standard deviations of that cloud's properties across draws.  Uncertainties on the median are the asymptotic standard error of the sample median for a normal distribution, $\frac{1.2533 \sigma}{\sqrt{n}}$.  Extinctions ($\max(A'_{\rm 3D})$, $\max(A_{\rm POS})$, $\Sigma A'_{\rm 3D}$) are reported in V-band using our assumed conversion factor from the native units of the GXP dust map ($A_V = 2.8\,A_{\rm ZGR23}$, derived from \citealt{ZhangGreen2023}).}
\end{deluxetable*}

\begin{deluxetable*}{l|cc|ccc|cc}
\centering
\tablecaption{Cloud Property Correlations \label{tab:corr}}
\tablehead{
\colhead{Property} & \colhead{3D Dust} & \colhead{\HI{}} & \colhead{$\rho_S$ with $|z|$} & \colhead{$\rho_S$ with $d$} & \colhead{$\rho_S$ with $d$ ($|z| < 100$ pc)} & \colhead{Robust?} & \colhead{$|z|$ Trend Strength}
}
\startdata
\multicolumn{8}{c}{\textit{Dependent on 3D Dust} ($N = 1695$, $N_{|z|<100\,\mathrm{pc}} = 883$)} \\
\hline
$R_\mathrm{eff}$                         & \checkmark &  & $+0.32$ ($<0.01$) & $+0.58$ ($<0.01$) & $+0.67$ ($<0.01$) & $\times$ &  \\
$\mathbb{L}_\mathrm{LOS}$                & \checkmark &  & $+0.16$ ($<0.01$) & $+0.13$ ($<0.01$) & $+0.20$ ($<0.01$) & $\times$ &  \\
$\max(A'_{\mathrm{3D}})$                 & \checkmark &  & $-0.60$ ($<0.01$) & $-0.33$ ($<0.01$) & $-0.31$ ($<0.01$) & $\times$ &  \\
$\Sigma A'_{\mathrm{3D}}$                & \checkmark &  & $-0.11$ ($<0.01$) & $+0.43$ ($<0.01$) & $+0.56$ ($<0.01$) & $\times$ &  \\
$\max(A_{\mathrm{POS}})$                 & \checkmark &  & $-0.55$ ($<0.01$) & $-0.03$ ($0.27$) & $+0.15$ ($<0.01$) & $\times$ &  \\
\hline
\multicolumn{8}{c}{\textit{Dependent on 3D Dust and \HI{}} ($N = 519$, $N_{|z|<100\,\mathrm{pc}} = 138$)} \\
\hline
$\max(N_\mathrm{HI})$                    &  & \checkmark & $-0.37$ ($<0.01$) & $+0.20$ ($<0.01$) & $+0.45$ ($<0.01$) & $\times$ &  \\
$N_\mathrm{HI}/A_V$                      & \checkmark & \checkmark & $+0.20$ ($<0.01$) & $+0.16$ ($<0.01$) & $-0.01$ ($0.87$) & \checkmark & Weak \\
$\max(n_\mathrm{HI})$                    & \checkmark & \checkmark & $-0.67$ ($<0.01$) & $-0.37$ ($<0.01$) & $-0.17$ ($0.05$) & \checkmark & Strong \\
$M$                                      & \checkmark & \checkmark & $+0.22$ ($<0.01$) & $+0.60$ ($<0.01$) & $+0.60$ ($<0.01$) & $\times$ &  \\
$\alpha_\mathrm{vir}$                    & \checkmark & \checkmark & $+0.27$ ($<0.01$) & $-0.25$ ($<0.01$) & $-0.43$ ($<0.01$) & $\times$ &  \\
$T_{k}$                                  & \checkmark & \checkmark & $+0.31$ ($<0.01$) & $+0.24$ ($<0.01$) & $+0.29$ ($<0.01$) & $\times$ &  \\
$P_e / k_B$                              & \checkmark & \checkmark & $-0.29$ ($<0.01$) & $-0.19$ ($<0.01$) & $-0.07$ ($0.40$) & \checkmark & Weak \\
$P / k_B$                                & \checkmark & \checkmark & $-0.33$ ($<0.01$) & $-0.15$ ($<0.01$) & $+0.03$ ($0.77$) & \checkmark & Moderate \\
\enddata
\tablecomments{Robust = $|\rho_S| \geq 0.1$ and $p < 0.01$ with $|z|$, and $|\rho_S| < 0.1$ or $p > 0.01$ with $d$ at $|z| < 100$ pc. Trend strengths are defined as: weak $0.1 \leq |\rho_S| < 0.3$, moderate $0.3 \leq |\rho_S| < 0.6$, strong $|\rho_S| \geq 0.6$.}
\end{deluxetable*}

\begin{figure*}
    \centering
\includegraphics[width=\textwidth]{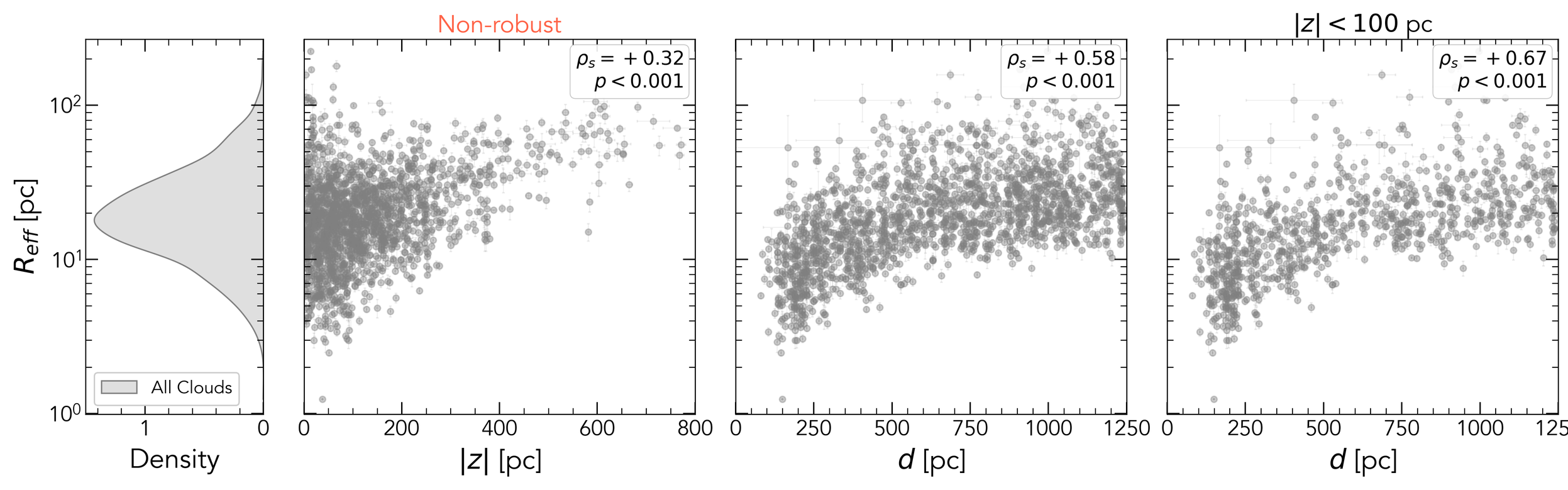}
\includegraphics[width=\textwidth]{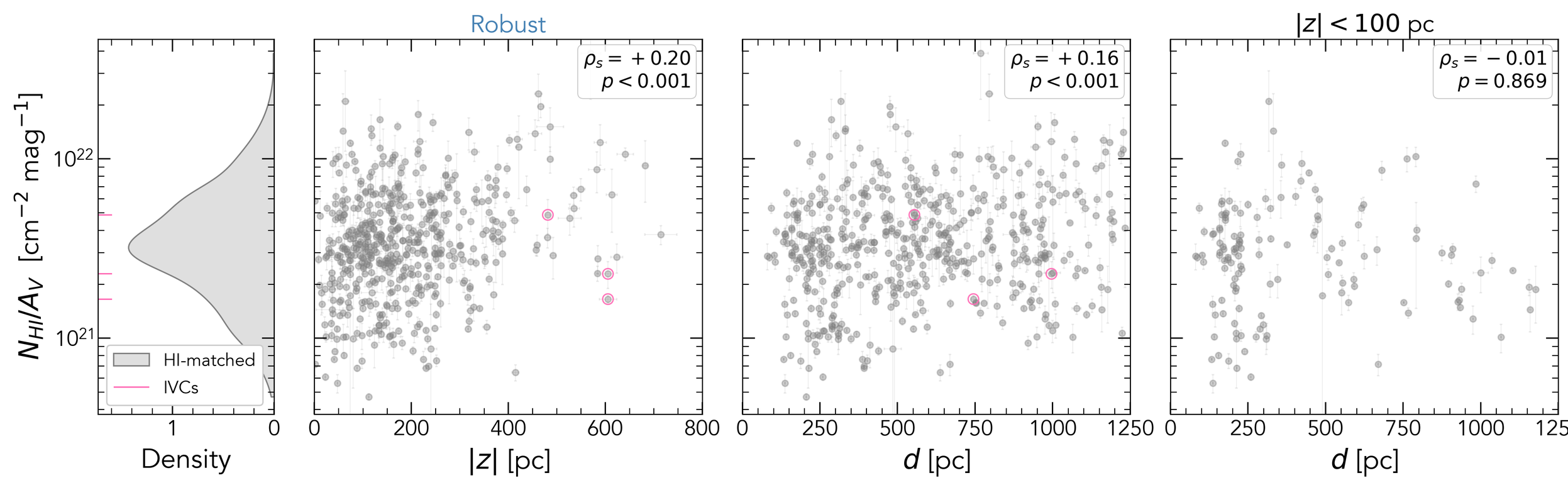}
    \caption{\textit{Top:} An example of a non-robust trend between cloud properties (here, equivalent radius $R_{eff}$) and altitude $|z|$.  From left: KDE of $R_{eff}$, relationship between $R_{eff}$ and $|z|$, relationship with distance $d$, and relationship with $d$ for low-altitude ($|z|<100$ pc) clouds.  Each scatter plot reports Spearman correlation coefficients and associated p-values.  Points are drawn from our full sample of identified dust clouds.  \textit{Bottom:} As top, but for a robust-trend between gas-to-dust ratio $N_{HI}/A_V$ and $|z|$.  IVCs are marked by pink scatter points.  Points are drawn from our subsample of \HI{}-matched dust clouds.  }
    \label{fig:ztrends_ex}
\end{figure*}

The physical conditions affecting cloud properties and evolution change significantly with altitude above or below the Galactic midplane.  Higher-altitude clouds on average likely experience lower confining pressures \citep{BoularesCox1990,Cox2005}, decreased interstellar radiation field (ISRF) strengths \citep{StrongMoskalenko2000, PopescuYang2017}, and weaker magnetic fields \citep{Cox2005,JanssonFarrar2012}, among other effects.  Our catalog of 3D (+1D velocity) clouds offers an opportunity to investigate various trends between cloud properties and altitude $|z|$ that may result from this change in environment.  However, in order to test these trends, we must additionally consider potential biases introduced by the construction of the 3D dust map and/or our \HI{}-matching procedure.  

We summarize the minimum, median, and maximum values of various properties derived for our sample of clouds in Table \ref{tab:summary}.  To identify trends with $|z|$ that are potentially physical, we assume that our sample of low-altitude disk clouds (defined as $|z| < 100$ pc, a threshold selected to ensure a significant sample size and robust azimuthal sampling in distance) are mostly drawn from the same population and should have properties that are constant with distance $d$ from the Sun.  Under this assumption, any trends with $d$ in this low-altitude group would indicate the property is biased by systematics in the 3D dust map or \HI{}-matching procedure.  For each property considered (listed in Table \ref{tab:corr}), we calculate the Spearman correlation coefficient $\rho_S$ between that property and 1) $|z|$, 2) $d$, and 3) $d$ for clouds with $|z| < 50$ pc.  We identify properties that have ``robust'' trends with $|z|$ as those whose trend with $|z|$ has $|\rho_S| \geq 0.1$ and $p < 0.01$ \textit{and} whose low-altitude trend with $d$ has $|\rho_S| < 0.1$ or $p \geq 0.01$.

We summarize the results of these tests in Table \ref{tab:corr}.  We calculate correlations for properties solely based on 3D dust using our full cloud sample of 1,695 clouds, while correlations for properties dependent on \HI{} are calculated using our matched cloud sample of 519 clouds.  Trends that we find are robust with $|z|$ are: the gas-to-dust ratio $N_{HI}/A_V$, the maximum 3D volume density $\rm{max}(n_{HI})$, and measures of external and internal pressure $P_e/k_B$ and $P/k_B$ (discussed further in \S\ref{S:phases}).  All other considered trends $\rm{max}(A_{POS})$, ($R_{eff}$, $\mathbb{L}_{\rm{LOS}}$, $\rm{max}(A'_{3D})$, $\Sigma(A'_{3D})$, $\rm{max}(N_{HI})$, $M$, and $\alpha_{\rm{vir}}$ [discussed in \S\ref{S:phases}]) have systematic trends with $d$ within the low-altitude group.  Figure \ref{fig:ztrends_ex} displays these correlations for one non-robust trend ($R_{eff}$) and one robust trend ($N_{HI}/A_V$); Appendix \ref{ap:cloud_corr} presents figures for the other considered trends. 

The non-robust trends in properties depending solely on 3D dust suggest that the GXP dust map contains subtle biases, possibly attributable to the construction of the map on a \Hpx grid with a distance-dependent prior.  For the case of cloud sizes and elongation along the LOS (known as fingers-of-god effects) in particular, we expect that this may additionally be the result of uncertainties on individual stars' distances being greater than the uncertainties on their POS position, leading to a ``smearing'' effect along the LOS.  The median $\mathbb{L}_{\rm{LOS}}$ of all clouds is $1.36$, suggesting that on average fingers-of-god effects affect most clouds in the map at all altitudes and distances.    

The non-robust trends with $M$ and $\alpha_{\rm{vir}}$ are likely direct consequences of the non-robust trends with $R_{eff}$ and $\Sigma(A'_{3D})$; it is also likely the result of the selection function of which clouds are successfully matched to \HI{}, as we are more successful at matching clouds with larger angular sizes, potentially prioritizing larger and more massive clouds in the more distant regime of the map.  We discuss the robust trend with $N_{HI}/A_V$ in \S\ref{S:dgr}.  We interpret the robust trend of decreasing $\rm{max}(n_{HI})$ with $|z|$ as revealing vertical stratification in the average ISM density as distance from the midplane increases.  

The dominance of non-robust trends suggests that significant attention should be paid to developing priors in the construction of future 3D dust maps to minimize such systematics in future reconstructions.

\subsubsection{Gas-to-Dust Ratios}\label{S:dgr}

\begin{figure*}
    \centering
\includegraphics[width=\textwidth]{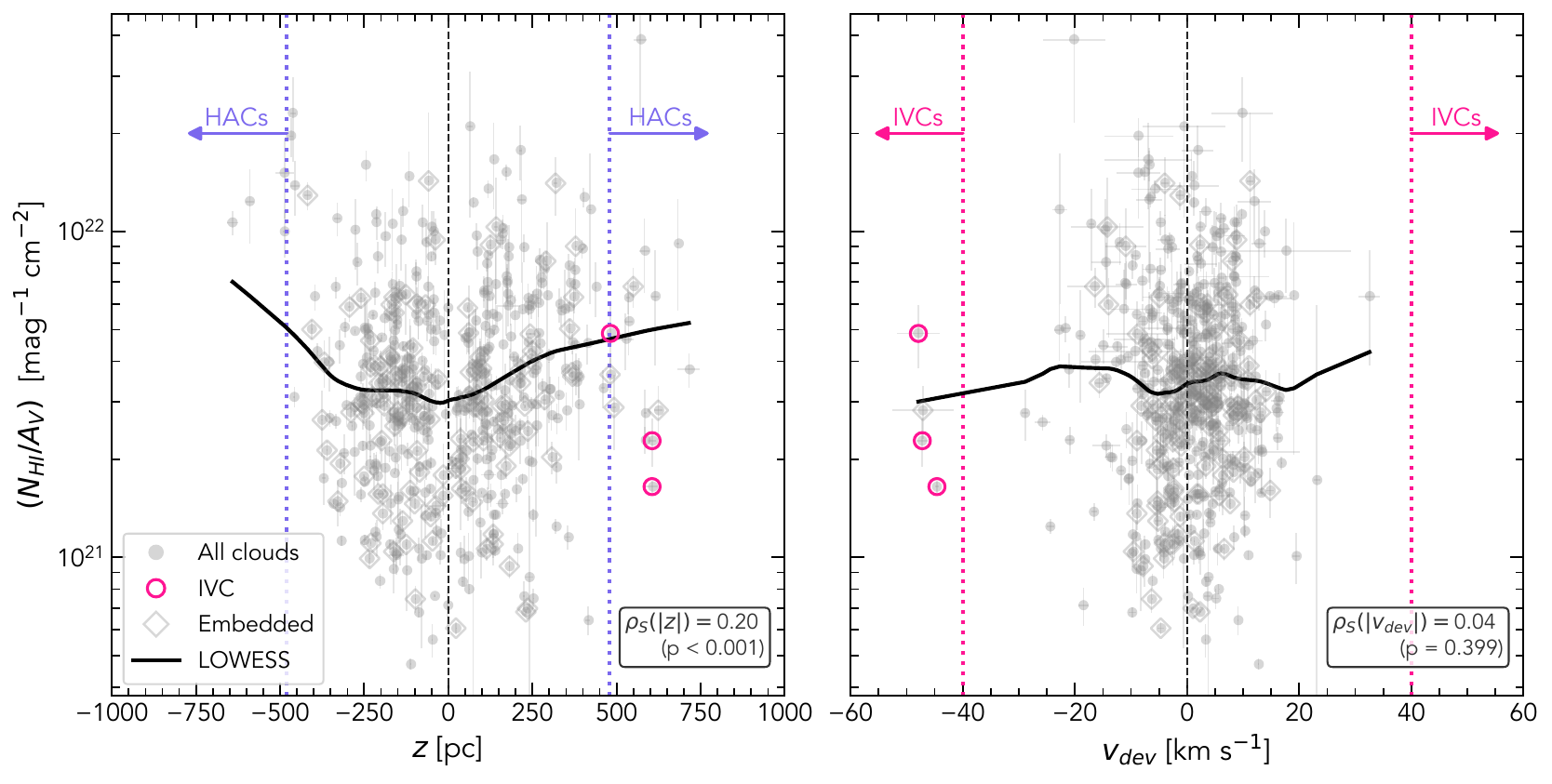}
    \caption{\textit{Left:} Neutral gas to dust ratio $N_{HI}/A_V$ as a function of cloud altitude $z$.  Embedded clouds are marked by diamonds and IVCs by pink circles.  The black curve shows a LOWESS regression including 50\% of the local data.  The purple dotted lines show our fiducial HAC threshold at $z=\pm480$ pc.  The black dashed line marks $z=0$ pc.  The Spearman correlation coefficient between $N_{HI}/A_V$ and $|z|$ is shown in the lower right.  \textit{Right:} As left, but as a function of cloud deviation velocity $v_{\rm{dev}}$.  The pink dotted lines show our fiducial IVC threshold at $v_{\rm{dev}}=\pm40$ km/s, and the black dashed line marks $v_{\rm{dev}}=0$ km/s.     }
    \label{fig:dgr_z_vdev}
\end{figure*}

We pay particular attention to the ratio of \HI{} column density to $V$-band dust extinction, \dgr, within our matched cloud sample; this quantity is imperative in accurately estimating quantities such as masses and volume densities, and may also reveal key insights into the mechanisms driving the physical origins of different classes of clouds (e.g., disk clouds vs. those participating in Galactic fountain flows).  Note that our measured ratios only account for the contribution of atomic neutral hydrogen, and does not reflect any contribution from molecular hydrogen.  In this work, we refer to the quantity \dgr as the atomic-gas-to-dust ratio, to differentiate it from the typical quest in the literature to measure the gas-to-dust ratio $N_H/E(B-V)$ (where $N_H = N_{HI} + 2 N_{H_2}$).   Although our ratio only includes the contribution of $N_{HI}$, our clouds are generally low extinction (and therefore likely have a low molecular fraction; related work by \citet{ShullPanopoulou2024} found that including H$_2$ in their sample of high-latitude sightlines increased the total $N_H$ by a factor of only 2-3\% on average), so we expect that this should have a minimal effect on our comparisons.

A high degree of variation in \dgr is present in our matched cloud sample, with values ranging between a minimum of $(4.7 \pm 0.2) \times 10^{20} \ \rm{mag}^{-1} \ \rm{cm}^{-2}$ to a maximum of $(3.9 \pm 1.7) \times 10^{22} \ \rm{mag}^{-1} \ \rm{cm}^{-2}$.  The median measured \dgr of the full sample is $3.3 \times 10^{21} \ \rm{mag}^{-1} \ \rm{cm}^{-2}$  with a standard deviation of $3.5 \times 10^{21} \ \rm{mag}^{-1} \ \rm{cm}^{-2}$.  \citet{ShullPanopoulou2024} reported an average value from previous literature for the Galactic disk of $N_H/E(B-V)$ of $(6.0 \pm 0.2) \times 10^{21} \ \rm{mag}^{-1} \ \rm{cm}^{-2}$ \citep{BohlinSavage1978, GudennavarBubbly2012, ShullDanforth2021, LisztGerin2023}.  In their sample of high latitude quasars with low-velocities (excluding HVCs) and $0.01 \leq E(B-V) \leq 0.1$,  \citet{ShullPanopoulou2024} derived an average $N_H/E(B-V)$ of $(9.2 \pm 0.3) \times 10^{21} \ \rm{mag}^{-1} \ \rm{cm}^{-2}$.   Assuming a standard $R_V=3.1$ \citep{CardelliClayton1989}, these values corresponds to a typical literature $N_H/A_V$ of $(1.9 \pm 0.1) \times 10^{21} \ \rm{mag}^{-1} \ \rm{cm}^{-2}$ in the disk and $(3.0 \pm 0.1) \times 10^{21} \ \rm{mag}^{-1} \ \rm{cm}^{-2}$ at high latitudes.  This latter value agrees extremely well with our sample of clouds, which, as previously discussed, is primarily located at high latitudes with low extinctions.  Studies considering solely \HI{} have found similar ratios at high-latitudes and low extinctions; e.g., \citet{Liszt2014a} derived a typical $N_{HI}/E(B-V)=8.3 \times 10^{21}$ cm$^{-2}$ mag$^{-1}$ (for $|b|\gtrsim20^\circ$ and $E(B-V)\lesssim 0.1$ mag) while \citet{LenzHensley2017} derived an average $N_{HI}/E(B-V)=8.8 \times 10^{21}$ cm$^{-2}$ mag$^{-1}$ (for gas with $N_{HI}<4 \times 10^{20}$ cm$^{-2}$ and  $|v_{\rm{LSR}}|<90$ km/s).  

It stands to reason that \dgr may vary with environment or cloud origins.  Figure \ref{fig:dgr_z_vdev} compares the distributions of \dgr among our cloud sample as a function of $z$ and $v_{\rm{dev}}$.  We observe that \dgr is generally lowest at low altitudes and increases with increasing $|z|$, albeit with significant scatter; as established in \S\ref{S:trends}, a weak but significant positive correlation exists between $|z|$ and \dgr ($\rho_s = 0.20$, $p \ll 0.001$).  We observe that, within the set of low-velocity high-altitude clouds, there is an asymmetry in the typical \dgr in the Northern vs. Southern hemispheres; the 4 southern low-velocity HACs have a substantially higher median \dgr (median $1.1 \times 10^{22} \ \rm{mag}^{-1} \ \rm{cm}^{-2}$) than the 10 northern low-velocity HACs (median $5.8 \times 10^{21} \ \rm{mag}^{-1} \ \rm{cm}^{-2}$).  This may suggest distinct origins or histories for the Northern vs. Southern HAC population.

We observe no significant correlation between $|v_{\rm{dev}}|$ and \dgr ($\rho_s = 0.04$, $p = 0.40$).  The three IVCs on-average have \dgr broadly consistent with trends in $|z|$, with a median \dgr of $2.3 \times 10^{21} \ \rm{mag}^{-1} \ \rm{cm}^{-2}$.  We note that the two higher-altitude IVCs (later identified as IVC 135 and the LLIV Arch in \S\ref{S:lit}) sit on the lower-tail of values for clouds at their altitudes, with values consistent with lower-altitude disk clouds.  This may suggest a past origin in the disk for these clouds.  

The weak dependency of \dgr on cloud altitude, but not on deviation velocity, suggests that the gas-to-dust ratio is more sensitive to cloud environment than kinematics.  We speculate this trend of decreasing dust content relative to gas at higher altitudes could be the result of some combination of decreasing metallicity, dust destruction, and dust settling at high altitudes \citep[e.g., as discussed in ][]{ShullPanopoulou2024}.  The lack of dependence on deviation velocity, and the consistency of IVCs with overall trends, suggests that either the IVCs are drawn from the larger population of disk clouds, or that the response timescale of dust in IVCs to environmental effects is sufficiently short for IVCs observed at present to convincingly masquerade as disk clouds. 

\subsubsection{Recommendations for Converting the Gaia XP Dust Map to Estimated Volume Density}

\begin{figure}
    \centering
\includegraphics[width=.47\textwidth]{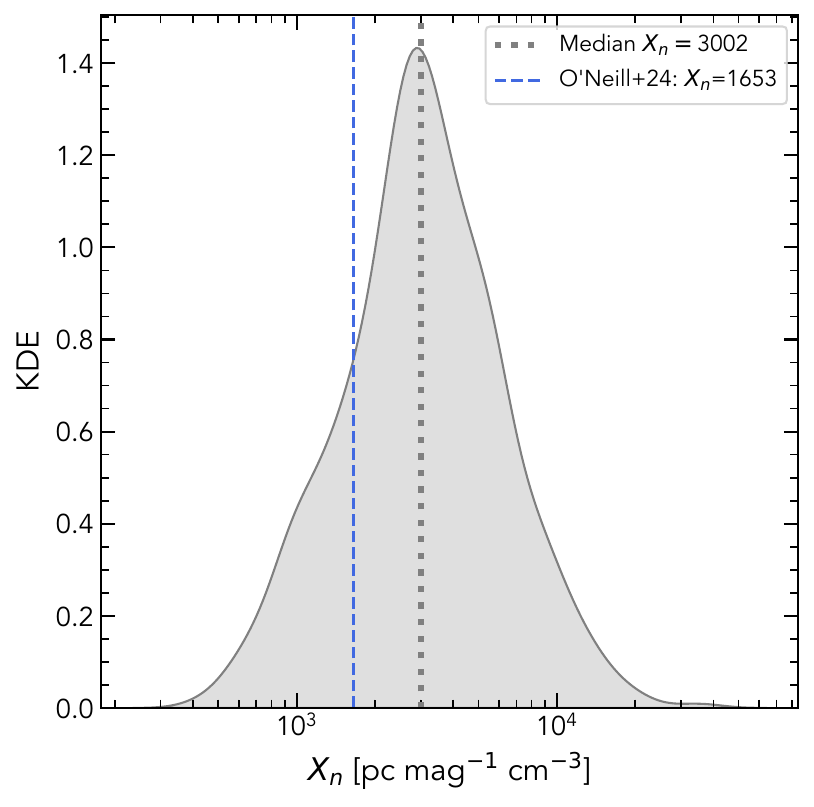}
    \caption{KDE of derived conversion factors $X_n$ between ZGR23 differential extinction $A'_{ZGR23}$ and neutral hydrogen volume density $n_{H}$.  The median $X_n=3023$ pc mag$^{-1}$ cm$^{-3}$ is marked by the dotted gray line.  The conversion factor (to total hydrogen volume density) adopted by \citet{ONeillZucker2024} of 1653 pc mag$^{-1}$ cm$^{-3}$ is marked by the blue dashed line.}
    \label{fig:convert_ZGR23}
\end{figure}

As described in \S\ref{S:hiprops}, \citet{ONeillZucker2024} assumed a constant conversion factor between $A'_{ZGR23}$ and \textit{total} neutral hydrogen volume density ($A_G / N_{H} = 4 \times 10^{-22}$ mag cm$^2$ derived from \citet{Draine2003} and \citet{Draine2009} by an assumed $R_V = 3.1$ \citep{CardelliClayton1989}, and assuming ZGR23's extinction curve such that $m_G = 2.04$), to derive an appropriate conversion factor from extinction in the GXP dust map to approximate total neutral hydrogen volume density, such that 
\begin{equation}
n_{H} =  X_n \ A'_{ZGR23}, 
\end{equation}
where $n_{H}$ is in cm$^{-3}$, $X_n$ is in pc mag$^{-1}$ cm$^{-3}$, and $A'_{ZGR23}$ is in mag pc$^{-1}$.   \citet{ONeillZucker2024} derived $X_n = 1653$ pc mag$^{-1}$ cm$^{-3}$ under the assumptions described above.  The direct measurement of \dgr in this work, revealing a factor of nearly-two-dex variation in this quantity in our sample of 519 matched clouds, suggests that a constant conversion factor is not necessarily a well-founded assumption.  

As shown in Figure \ref{fig:convert_ZGR23}, if converted to these units, the conversion factors $X_n$ derived from this work's cloud sample would range from $X_n = 427$ pc mag$^{-1}$ cm$^{-3}$  to 35156 pc mag$^{-1}$ cm$^{-3}$ , with a median and standard deviation of the full distribution of $3002 \pm 175$ pc mag$^{-1}$ cm$^{-3}$.  This median value is $1.8\times$ larger than the conversion factor derived by \citet{ONeillZucker2024}.  Given that the \citet{ONeillZucker2024} factor is meant to estimate total neutral hydrogen, and our derived values only account for atomic hydrogen, this suggests that the value assumed by \citet{ONeillZucker2024} is an underestimate of the total amount of neutral hydrogen.  We recommend that the median conversion factor of $3002$ derived in this work be used if a single number is necessary, but caution that the large amount of variation even within this sample of clouds in the Solar Neighborhood ensures that uncertainties on subsequent results will be large.

\subsubsection{Cloud Phases}\label{S:phases}

\begin{figure*}
    \centering
\includegraphics[width=\textwidth]{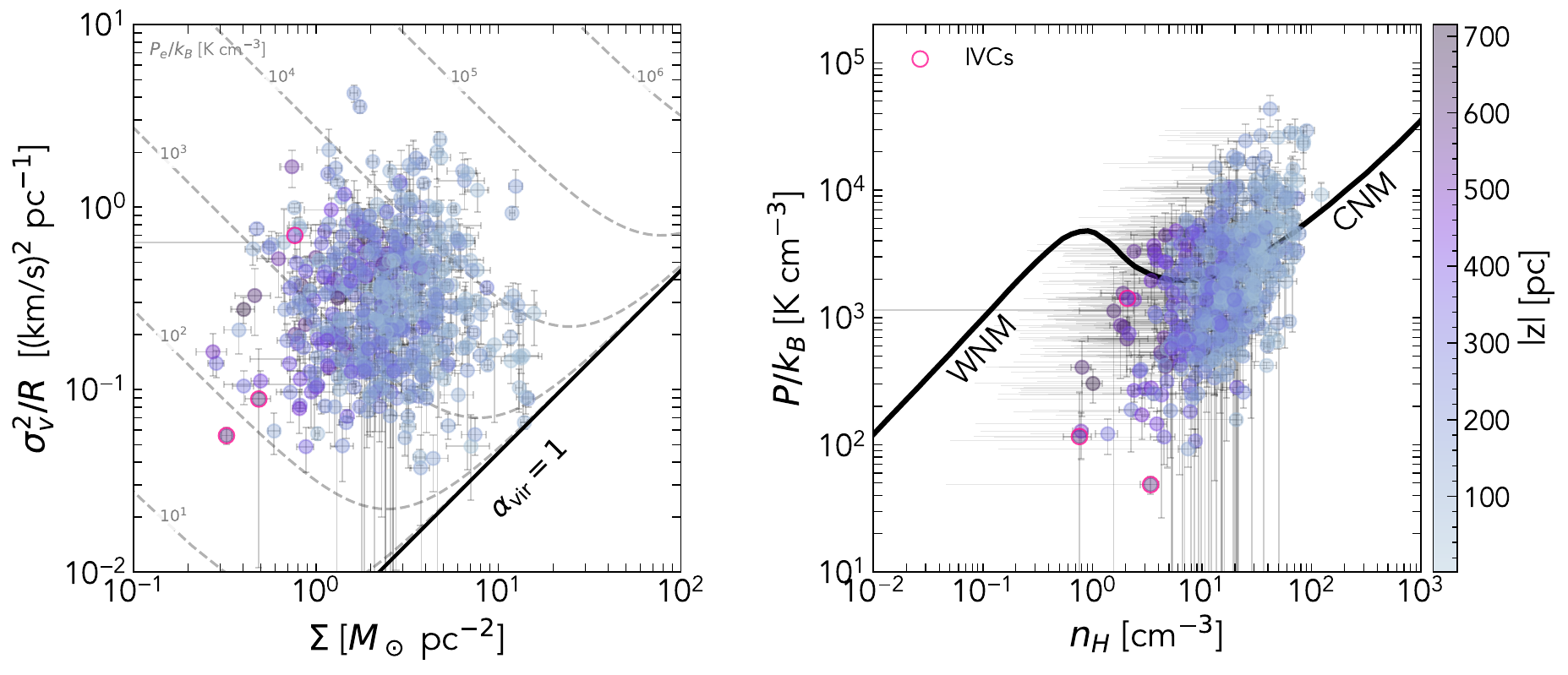}
    \caption{\textit{Left:} Size-linewidth parameter $\sigma_v^2/R$ vs. surface density $\Sigma$ for the \HI{}-matched cloud sample.  Points are colored by altitude $|z|$, with error bars showing uncertainties across draws.  IVCs are marked with pink circles.  The black line shows the position for virial equilibrium ($\alpha_{vir}=1$), with $\alpha_{vir}$ increasing up and to the left from this line.  The gray dashed V-shaped curves show lines of constant external pressure $P_e/k_B$ (in units of K cm$^{-3}$) required to achieve virial equilibrium.  \textit{Right:} Thermal pressure $P/k_B$ vs. neutral hydrogen density $n_H$.  Points are colored as on the left.  The gray horizontal lines extending from each point show the range of internal densities within each cloud, with points plotted at the maximum density within the cloud.  The black curve shows a solution for the 2-phase ISM for $R_{gal}=8.5$ kpc derived by \citet{WolfireMcKee2003}, with the WNM and CNM branches labeled.  Note that the maximum volume densities are likely lower limits, and that the thermal pressure estimates are uncertain in that they depend on the product of volume density (lower limits) and kinetic temperature (upper limits).}
    \label{fig:sizelinesigma_phases}
\end{figure*}

Although several of the measured dust cloud properties display trends with distance at low-altitudes, indicating potential subtle biases in the dust-map-based measurements, we now investigate what can be inferred about the physical states of our cloud sample even under these uncertainties.  Comparisons of cloud sizes $R$, line widths $\sigma_v$, surface densities $\Sigma = M / (\pi R^2)$, and related properties as probes of cloud physical states and phases dates back to the pioneering relationships identified by \citet{larson1981}.  Here, we place our sample of \HI-matched dust clouds in this context; we emphasize that, while we believe the overall conclusions of this section are reasonable, uncertainties on individual clouds are large and should not be interpreted as definitive individual property or phase measurements. 

For a virialized spherical cloud following a power-law internal density profile ($\rho \propto r^{-k}$), a relationship 
\begin{equation}
    \frac{\sigma_v^2}{R} = \frac{(3-k)}{3(5-2k)} \pi G \Sigma,
\end{equation}
is expected to hold \citep[as derived in e.g.,][]{ONeill2022_codark}.  Figure \ref{fig:sizelinesigma_phases} populates our cloud sample in the space of $\sigma_v^2 / R$ vs $\Sigma$.  We assume an internal density profile for clouds on average of $k=2$, based on the work of \citet{zucker2021} who analyzed the radial density profiles of molecular clouds in the Solar Neighborhood mapped with 3D dust and found, at large cloud radii, average power-law density slopes between $k= 1.5 - 2.5$.  

We observe that nearly \textit{all} of our cloud measurements fall well ``above'' virial equilibrium in the space of $\sigma_v^2 / R$ vs $\Sigma$.  We derive virial parameters for our clouds \citep{BertoldiMcKee1992}
\begin{equation}
    \alpha_{vir} = \frac{M_{vir}}{M} = \frac{3(5-2k) R \sigma_v^2}{(3-k) G} \frac{1}{M},
\end{equation}
where $\alpha_{vir} = 2$ corresponds to virial equilibrium and $\alpha_{vir} > 2$ suggests being gravitationally unbound.  We find $\alpha_{vir}$ of 1 to 601, with a median of 27, i.e., we would infer our clouds are $\textit{severely}$ unbound.

High derived values of $\alpha_{\rm{vir}}$ have been frequently observed among clouds of all scales, with a well known anti-correlation between $\alpha_{\rm{vir}}$ and cloud mass \citep[e.g.,][]{BertoldiMcKee1992,MivilleDeschenesMurray2017,TraficanteDuarteCabral2018}.  \citet{field_does_2011} examined trends in observations of apparently unbound Galactic giant molecular clouds by \citet{heyer_re-examining_2009}, and proposed that external confining pressures realistic for the ISM should be sufficient to prevent clouds from dispersing rapidly.  When confining pressure is present, the size-linewidth parameter and surface density should follow the relationship,
\begin{equation}
    \frac{\sigma_v^2}{R} = \frac{1}{3} \left(\pi \gamma G \Sigma + \frac{4 P_e}{\Sigma} \right),
\end{equation}
which is shown in Figure \ref{fig:sizelinesigma_phases} (for $\gamma = 1$, corresponding to $k=2$) by the V-shaped curves for values of external pressure ranging between $P_e/k_B = 10^1$ K cm$^{-3}$ to $P_e/k_B = 10^6$ K cm$^{-3}$.  We can use this relationship to derive an estimate of the confining pressure that our clouds may be subject to if they are in virial equilibrium.  We emphasize that the assumption of virialization is not necessarily well-founded for clouds participating in a Galactic fountain flow.  This method yields estimates of $P_e/k_B$ ranging between $6.5 \times 10^1$ K cm$^{-3}$ to $5.7 \times 10^{4}$ K cm$^{-3}$, with a median of $2.5 \times 10^{3}$ K cm$^{-3}$.  The total midplane pressure in the Solar Neighborhood is estimated to be roughly $P/k \simeq 2.8 \times 10^4$ K cm$^{-3}$ \citep{BoularesCox1990}, and is expected to decrease with altitude \citep{Cox2005}; most of our clouds are located well above the midplane, and, in agreement with theoretical expectations, we observe a significant albeit weak negative correlation between $P_e/k_B$ and $|z|$.  However, we note that the reliability of this correlation is perhaps questionable, given that the component properties used to estimate $P_e/k_B$ individually display biases with distance.

We additionally estimate the internal pressure within each cloud using the kinetic temperature for \HI{},
\begin{equation}
T_{\rm{kin}} \leq \frac{m_H \ \sigma^2}{k_B}
\end{equation}
which provides an upper limit on temperature, so that an upper limit on pressure follows as,
\begin{equation}
\frac{P}{k_B} = n T_{\rm{kin}}
\end{equation}
Note that if the mean volume density within the cloud $\bar{n}=M/(\mu m_p N_{\rm{pix}} dV)$ is used to estimate $P$, the ratio between the confining external pressure and the internal pressure simplifies to $P_e / P = \mu (1 - \alpha_{vir}^{-1})$.  Through this estimate, cloud kinetic temperatures range between $T_{\rm{kin}} = (85 \pm 35)$ K to $(8.4 \pm 0.2) \times10^3$ K, with a median of 727 K.  Pressures range between $P/k_B = (4.9 \pm 0.8) \times 10^1$ K cm$^{-3}$ to $(4.4 \pm 1.2) \times 10^4$ K cm$^{-3}$, with a median of $1.9 \times 10^3$ K cm$^{-3}$.  This is consistent with the typical range of pressures derived for local CNM clouds by \citet{JenkinsTripp2011}, who found a mean of $P/k=3.8 \times 10^3 $ K cm$^{-3}$.  As with $P_e/k_B$, we observe a significant negative correlation between $P/k_B$ and $|z|$.

This estimate of pressure provides constraints on the clouds' phases.  The right panel of Figure \ref{fig:sizelinesigma_phases} compares $P / k_B$ to mean cloud volume density $\bar{n}$, with the range of minimum to maximum volume density inside each cloud shown by the horizontal lines.  We also plot one solution for the 2-phase ISM derived by \citet{WolfireMcKee2003} at $R_{gal} = 8.5$ kpc.  We observe that our cloud population measurements reasonably occupies this parameter space where expected, although we emphasize that our maximum volume densities are likely lower limits on true maximum density in a cloud, as 3D dust maps are known to underestimate density peaks \citep[e.g.,][]{zucker2021}.  Based on the position of our cloud sample in this space, it would appear that while some clouds are composed of the cold neutral medium (CNM), many others fall in the regime of the unstable neutral medium (UNM). A recent estimate of the vertical scale height of the CNM in the Solar Neighborhood, based on inferring distances to \HI{} absorption measurements, ranged from $\sigma_z \sim 50 - 90$ pc \citep{RybarczykWenger2024}, well below the altitudes of many of our HACs (which appear to tend to fall in the region of the UNM).

While biases with distance on the input cloud parameters to this analysis limit the scope of our conclusions, we have demonstrated (1) the powerful constraints on cloud phases accessible through joint analysis of HI and 3D dust maps ($\alpha_{\rm{vir}}$, $P_e / k_B$, $T_{\rm{kin}}$, $P/k_B$), (2) that even with potential biases, our inferred cloud properties reasonably agree with physical intuition, and (3) the pressing need for 3D dust mapping regularization to consider uniformity in cloud properties with distance, possibly going beyond imposing isotropic cartesian Gaussian process priors.

\section{Discussion}\label{S:discuss}

\subsection{Comparison to Literature IVCs}\label{S:lit}

Catalogs of IVCs exist in abundance in the literature, with tracers ranging from neutral \HI{} to cold gas absorption to ionized gas absorption.  A subset of intermediate velocity gas detections are part of structures sufficiently well-studied to have been assigned names, with a plethora of follow-up studies in the literature attempting to place constraints on their distances and related properties.  Distances to these well-known complexes have typically been derived through brackets from detection/non-detection of absorption in background stellar spectra, but uncertainties are large due to uncertainties on stellar distances, absorption non-detection limits, and the clumpiness of varying tracers.  Since basic physical parameters (including volume density, size, mass, and pressure) scale with cloud distance, directly constraining the 3D morphologies of IVCs is a key step forward in understanding the physics driving their origins.

In this work, we cross-match our cloud catalog against five catalogs of IVCs and high-latitude clouds in the literature.  Each catalog has its own selection parameters to define its target clouds:

\paragraph{\citet{RohserKerp2016}} \citetalias{RohserKerp2016} targeted high-latitude ($|b|>20^\circ$) molecular IVCs (MIVCs), identified via {\textit{Planck}} far-IR excess relative to the EBHIS and GASS \HI surveys; to define their sample of MIVCs, they required that selected components from their Gaussian decomposition of the \HI spectra fulfill 1) $20 \leq |v_{\rm{LSR}}| \leq 100$ km s$^{-1}$, 2) $N_{HI} \geq 10^{20}$ cm$^{-2}$, 3) HI amplitudes $>10$ K and FWHM linewidths $< 5$ km/s, along with additional cuts related to distinguishing between intermediate-velocity IR excess vs. confused low-velocity IR excess.  We use both their primary and secondary catalogs, and record secondary velocity components as separate candidate detections.  This yields a catalog of 253 detections.

\paragraph{\citet{Wakker2001}}  \citetalias{Wakker2001} compiled a catalog of IVC and HVC absorption line detections (including Na I, Ca II, and more ionized ions) in the literature, with IVCs defined as $\sim40 \leq |v_{\rm{LSR}}| \leq 90$ km s$^{-1}$, and derived updated distance brackets using measured or inferred stellar distances, along with updated \HI{} column densities.  We restrict their catalog to $|v_{\rm{LSR}}|<100 $ km/s for our matching consideration.  This yields a catalog of 531 detections.  

\paragraph{\citet{KuntzDanly1996}} \citetalias{KuntzDanly1996} presented a catalog of high-to-intermediate-velocity \HI{} clouds and clumps within the Northern Galactic hemisphere (primarily concentrated within three structures known as the IV Arch, IV Spur, and the LLIV Arch), using \HI{} observations from the Bell Laboratories 21 cm survey, with IVCs defined as $40 \leq |v_{\rm{LSR}}| \leq 100$ km s$^{-1}$.    \citetalias{KuntzDanly1996} provided ranges of most-negative to least-negative velocities for their clumps, and for many clumps also provided a peak velocity.  They estimated that uncertainties on the most-negative side were of order 5 km/s, while those for the least-negative side were approximately 15 km/s due to confusion with disk gas.  Their catalog includes 34 clumps.\footnote{Several of their clouds have least-negative velocities simply reported as ``Disk''; we infill these rows with -20 km/s, slightly lower than the lowest non-disk velocities reported for other clouds.  We additionally correct a presumed typo in their entry for their clump S1, which reports a non-physical least-negative velocity of -272.5 km/s (more-negative than their reported most-negative velocity for the same cloud); we update this to -22.5 km/s under the suspicion that a ``7'' from the previous row was incorrectly inserted.  Additionally, for clouds with no peak velocity reported, we infill a peak velocity at the average of the most and least negative velocities for the purpose of our tie-breaking matching system, described later in this section.}

\paragraph{\citet{GladdersClarke1999}} \citetalias{GladdersClarke1999} presented preliminary results from a dedicated program at the David Dunlap Observatory to constrain distance brackets to high-latitude \HI{} clouds (including IVCs) using detections/non-detections of the Na I doublet in stellar spectra; targets for this program were sourced from the \citet{HeilesReach1988} catalog of isolated degree-scale clouds with strong IRAS 100$\mu$m emission (many of which were found to contain CO emission by \citealt{ReachKoo1994}).  Distance constraints towards two of the fourteen sources targeted by \citetalias{GladdersClarke1999} were formally published by \citet[][for the Draco cloud]{GladdersClarke1998} and \citet[][for the G139.6+47.6 cloud]{BurnsTycner2003}. 

\paragraph{\citet{MagnaniHartmann1996}} performed a survey for high-latitude CO detections.  We obtain an updated catalog of high-latitude  ($|b|>25^\circ$) CO detections obtained as of the year 2014 from L. A. Magnani (private communication), which we match to our catalog with a primary goal of identifying known LVCs in our sample. This catalog includes 334 detections.

\vspace{2em}

For each detection in these catalogs with velocity $v_{cat}$, we retrieve the set of clouds in our catalog with pixels on the POS at the detection's coordinates.  If only one cloud in our catalog is found at that coordinate, we record it as a candidate match.  If multiple clouds are found, we admit the cloud whose velocity range distance,
\begin{equation}
    \Delta v = \begin{cases}
        |v_{cat}-v_{min}| & \rm{if } \ v_{cat} < v_{min} \\
        0 & \rm{if } \  v_{min} \leq v_{cat} \leq v_{max} \\
        |v_{cat} - v_{max}| & \rm{if} \ v_{cat} > v_{max}
    \end{cases}
\end{equation}
is smallest (or, in the case of \citetalias{KuntzDanly1996} which provides their own min-max velocities, the cloud whose velocity range is closer to the \citetalias{KuntzDanly1996} range).  In the case of ties in $\Delta v$ between multiple overlapping clouds, we then select the cloud whose peak distance ($|v_{cat} - \langle v \rangle|$, where $\langle v \rangle$ is the intensity-weighted \HI{} velocity at that coordinate within the clouds) is smallest.  For our set of winning candidate clouds, we then store $\Delta v$ for each match.  We define a high-quality literature match as $\Delta v \leq 2.6$ km/s (two velocity channels of the \HIPI map) in order to obtain a catalog of high-confidence detections.  We emphasize that, since most of our IVCs are relatively broad features extended over tens of km/s in velocity and uncertainties on $v_{cat}$ vary by-catalog, this may overlook several relevant matches.  

Using this criteria, we find 78 detections within \citetalias{RohserKerp2016}, 20 detections within \citetalias{Wakker2001}, 8 detections within \citetalias{KuntzDanly1996}, 11 detections within \citetalias{GladdersClarke1999}, and 251 detections within the Magnani catalog.  These are localized within 71 clouds in our catalog (including embedded sub-clouds), of which 45 have names in the matched catalogs.  

\begin{figure*}
    \centering
\includegraphics[width=.85\textwidth]{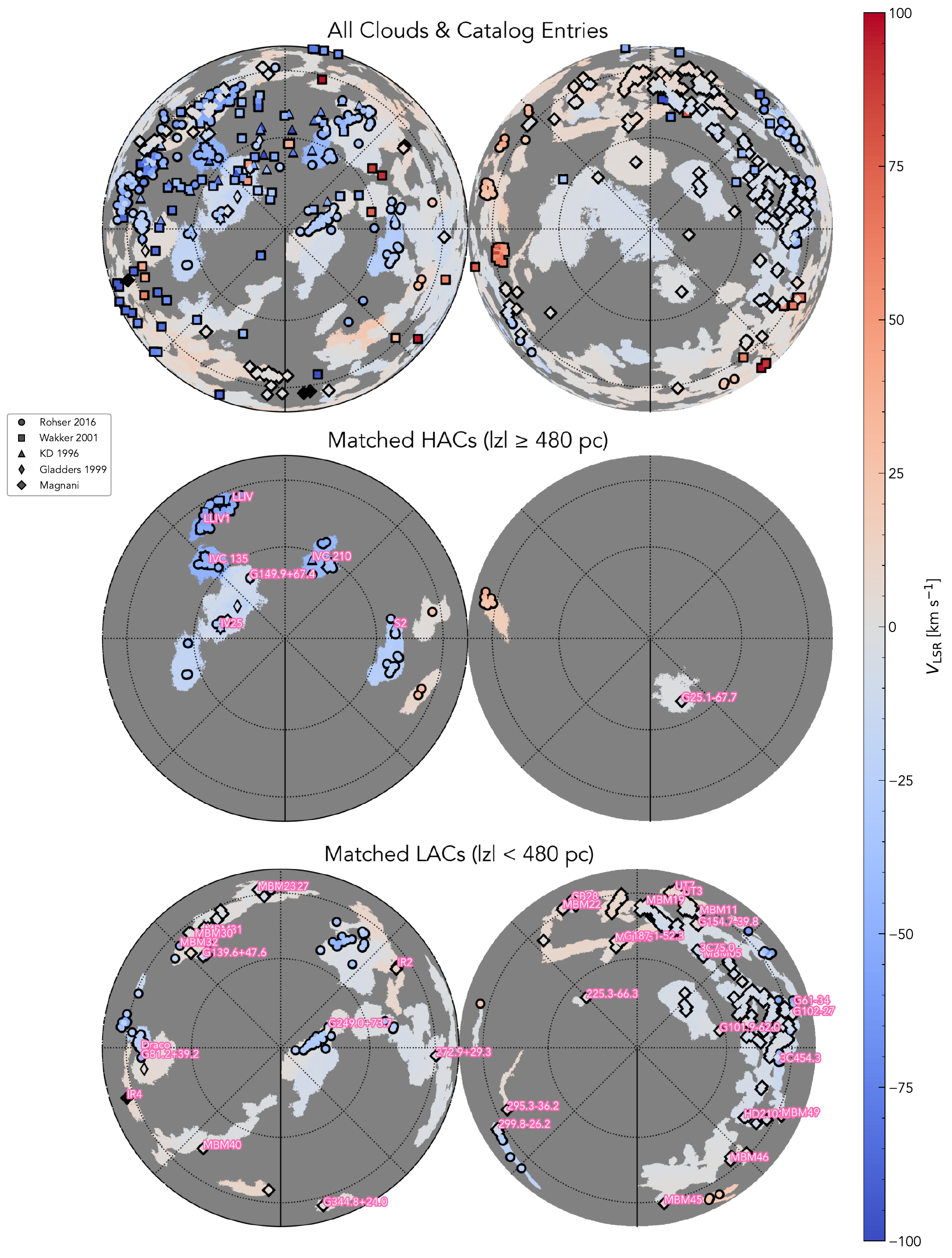}
 \caption{\textit{Top:} Polar projections of moment-1 map of \HI{} velocity for all matched dust clouds, overlaid with all catalog detections of IVCs and high-latitude clouds against which we compare our dust cloud sample.  Catalog detections are drawn from \citet[][circles]{RohserKerp2016}, \citet[][squares]{Wakker2001}, \citet[][triangles]{KuntzDanly1996}, \citet[][thin diamonds]{GladdersClarke1999}, and \citet[][thick diamonds]{MagnaniHartmann1996}  Catalog LSR velocities are plotted on the same velocity colorscale as our moment-1 map.  \textit{Center:} As top, but for the subset of HACs and catalog entries between which we find matching detections.  If names for the matched cloud exist in the catalogs, we plot them in white text outlined in pink; if multiple names exist, we plot only our preferred name (bolded in Table \ref{tab:highlights_hac} and Table \ref{tab:highlights_nonhac}).  \textit{Bottom:} As center, but for matched LACs. }
    \label{fig:ivcs_pos}
\end{figure*}

Figure \ref{fig:ivcs_pos} presents these clouds on the POS, with catalog detections overlaid on the same velocity color scale.  We additionally plot HACs without catalog detections.  In Table \ref{tab:highlights_hac}, we provide the properties of HACs with catalog detections along with non-matched HACs; we present a similar table of matched LACs in Table \ref{tab:highlights_nonhac}.  We include any names attached to these clouds in the catalogs, with our preferred names that we use to refer to each cloud in the remainder of this work marked by bold text.  

We note that, since this work represents the first treatment of rotation-corrected definitions of IVCs informed by actual 3D positions, we must re-classify some clouds previously referred to as IVCs in the literature as ``literature IVCs'' separate from the three $|v_{\rm{dev}}| > 40$ km/s IVCs we defined in \S\ref{S:kinematic}.  This is a purely definitional choice, not intended to communicate any assumptions about these lower-deviation-velocity clouds' origins.  

All named IVCs and literature IVCs found in this work are located in the Northern Galactic Hemisphere, with all named Southern clouds coming from the Magnani catalog.  In the following paragraphs we discuss a subset of individual named clouds, in order of lowest to highest centroid z-height $z_c$.  These include {\textit{all}} of the known high-latitude molecular (literature) IVCs with high inferred $H_2$ column densities – Draco, IVC 135, IVC 210, and a complex observed by \citet{MagnaniSmith2010} containing CO detections G283.8+54.9, G288.4+53.2, and G295.0+57.1. 

\begin{figure*}
    \centering
\includegraphics[width=\textwidth]{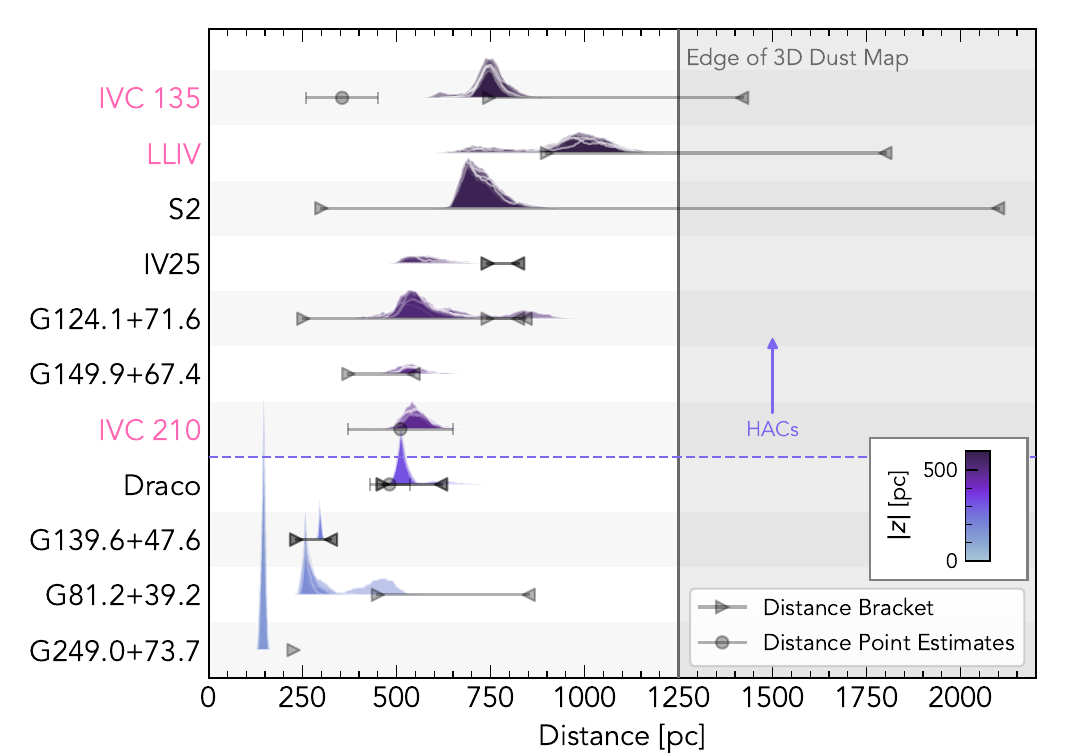}
    \caption{Radial dust extinction profiles $\mathcal{R}$ for our sample of highlighted clouds (including all HACs, IVCs, and clouds with literature detections).  Each row (labeled by cloud name or ID) shows draws of differential dust extinction as a function of distance from the Sun.  All radial dust profiles are shown on the same scale, so that larger peaks correspond to higher differential dust extinction.  Draws are colored by altitude $|z|$.  If constraints on the distance of this cloud were given in the literature, they are marked by black lines or markers; constraints that represent upper/lower brackets on distance are marked by triangles, while central distance estimates with uncertainties are marked by circles.  Clouds that are IVCs have their names written in pink.  The edge of the dust map at 1250 pc is marked by the gray shaded region.  Clouds are sorted by z-height, from lowest to highest $z$.  The $|z|$-threshold defining HACs is shown by the purple dashed lines at $\pm 480$ pc. }
    \label{fig:ivc_lit_dist}
\end{figure*}

\begin{figure*}
    \centering
\includegraphics[width=\textwidth]{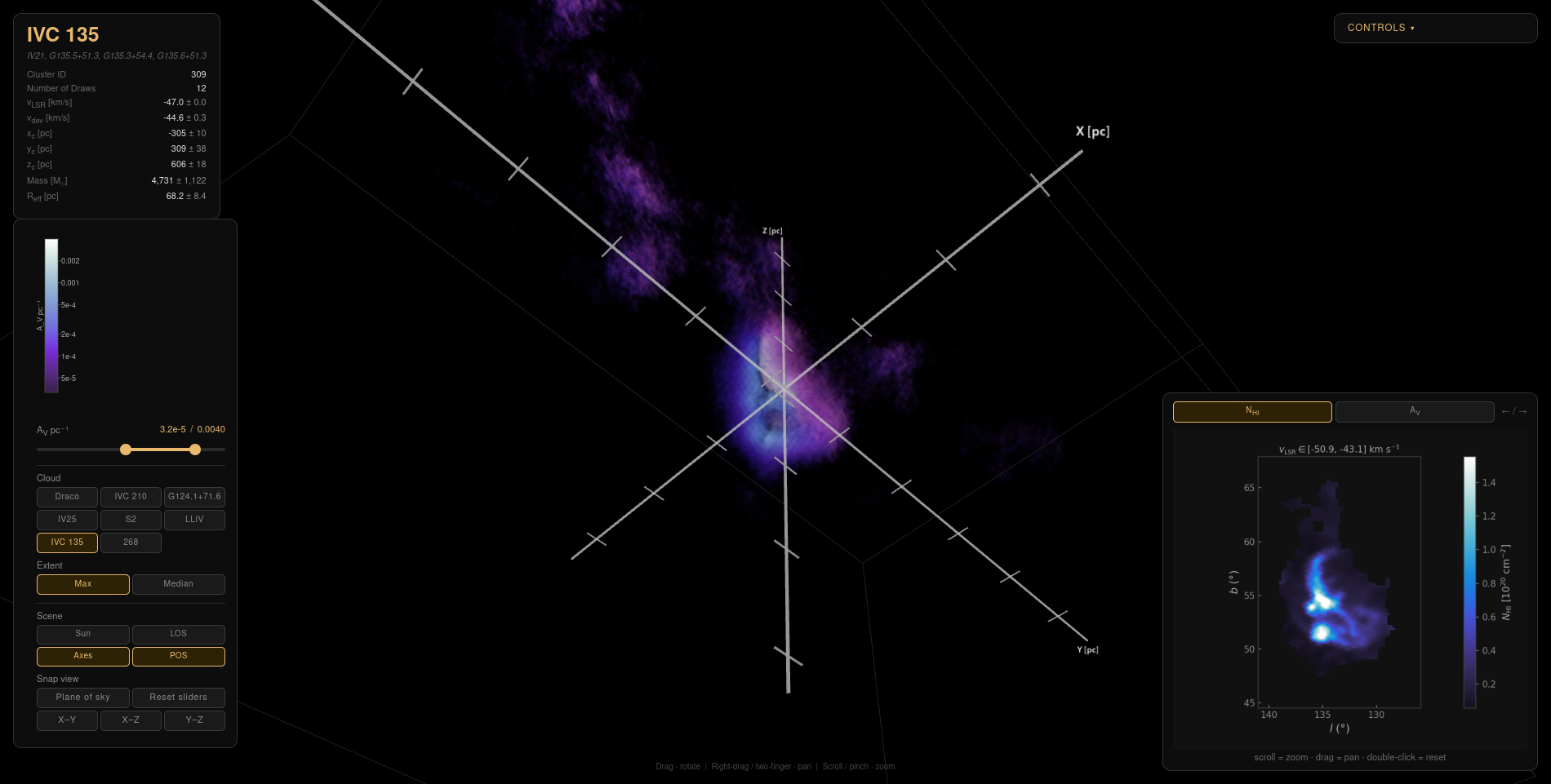}
 \caption{A screenshot of an \underline{interactive figure}, available at \url{https://theo-oneill.github.io/HACs_and_IVCs/ivc_clouds_3d/index.html}, presenting the 3D density structure of IVCs and other HACs of interest.  In this screenshot, the cloud IVC 135 is shown.  Users may click, pan, and zoom to manipulate the clouds' 3D structures, and compare their appearance to projected maps of $N_{HI}$ and $A_V$ on the POS.  Coordinate axes show heliocentric cartesian x, y, z centered at the cloud's dust-weighted centroid position.}
    \label{fig:ivcs_3d}
\end{figure*}

Figure \ref{fig:ivc_lit_dist} compares the radial density profiles of our literature-matched clouds as measured with 3D dust to the previous constraints on their distances from the literature (for the 11 of 71 clouds where we find previous constraints in our catalogs or the literature).  Cloud distances generally agree within uncertainties.  Figure \ref{fig:ivcs_3d} presents an interactive figure allowing for interaction with the 3D structures of the literature-matched clouds described below.

\begin{deluxetable*}{lccccccccccc l cc}
\rotate
\tabletypesize{\tiny}
\tablecaption{Properties of highlighted high-altitude clouds (HACs) \label{tab:highlights_hac}}
\tablehead{
\colhead{ID} & \colhead{$N_{\rm{draws}}$} & \colhead{\shortstack{Parent\\ID}} &\colhead{$\ell_c$} & \colhead{$b_c$} & \colhead{$d_c$} & \colhead{$z_c$} & \colhead{$v_{\rm LSR}$} & \colhead{$v_{\rm dev}$} & \colhead{$R_{\rm eff}$} & \colhead{$M$} & \colhead{$N_{\rm HI}/A_V$} & \colhead{Names} & \colhead{Ref.} & \colhead{IVC} \\
\colhead{} & \colhead{} & \colhead{} & \colhead{(deg)} & \colhead{(deg)} & \colhead{(pc)} & \colhead{(pc)} & \colhead{(km\,s$^{-1}$)} & \colhead{(km\,s$^{-1}$)} & \colhead{(pc)} & \colhead{($10^4\,M_{\odot}$)} & \colhead{\shortstack{($10^{21}$\,cm$^{-2}$\\mag$^{-1}$)}} & \colhead{} & \colhead{} & \colhead{}
}
\startdata
268 & 10 &  & $74.7 \pm 2.2$ & $55.4 \pm 0.8$ & $874 \pm 34$ & $715 \pm 33$ & $-17.4 \pm 0.8$ & $-21.0 \pm 0.9$ & $78.7 \pm 8.4$ & $0.89 \pm 0.21$ & $3.78 \pm 0.46$ & \hlnames{\ldots} & [1] & \ldots \\
292 & 12 &  & $290.7 \pm 0.8$ & $36.1 \pm 0.2$ & $1158 \pm 5$ & $682 \pm 4$ & $3.2 \pm 1.4$ & $12.8 \pm 1.3$ & $97.3 \pm 7.1$ & $5.4 \pm 2.6$ & $9.16 \pm 3.43$ & \hlnames{\ldots} & [1] & \ldots \\
1614 & 3 & 360 & $144.8 \pm 1.4$ & $39.6 \pm 0.5$ & $974 \pm 10$ & $624 \pm 12$ & $-53.4 \pm 5.5$ & $-47.1 \pm 5.5$ & $59.3 \pm 7.3$ & $0.71 \pm 0.16$ & $2.83 \pm 0.52$ & \hlnames{LLIV1$^{[3]}$, LLIV3$^{[3]}$} & [1,3] & \ldots \\
2066 & 11 &  & $322.5 \pm 1.5$ & $43.0 \pm 0.4$ & $899 \pm 20$ & $613 \pm 11$ & $23.8 \pm 1.9$ & $32.5 \pm 1.9$ & $98.8 \pm 11.9$ & $3.9 \pm 1.3$ & $6.32 \pm 2.43$ & \hlnames{\ldots} & \ldots & \ldots \\
309 & 12 &  & $135.3 \pm 3.9$ & $54.3 \pm 2.5$ & $745 \pm 7$ & $606 \pm 18$ & $-47.0$ & $-44.6 \pm 0.3$ & $68.2 \pm 8.4$ & $0.47 \pm 0.11$ & $1.65 \pm 0.07$ & \hlnames{IV21$^{[2,3]}$, \textbf{IVC 135}$^{[1]}$, G135.5+51.3$^{[4]}$ \newline G135.3+54.4$^{[5]}$, G135.6+51.3$^{[5]}$} & [1,2,3,4,5] & \checkmark \\
360 & 11 &  & $151.4 \pm 0.5$ & $37.3 \pm 0.1$ & $996 \pm 18$ & $606 \pm 11$ & $-53.4 \pm 0.9$ & $-47.2 \pm 1.0$ & $94.5 \pm 8.5$ & $1.48 \pm 0.29$ & $2.28 \pm 0.39$ & \hlnames{LLIV1$^{[2]}$, \textbf{LLIV}$^{[2]}$, LLIV3$^{[2]}$ \newline LLIV4$^{[2,3]}$, LLIV2$^{[3]}$} & [1,2,3] & \checkmark \\
527 & 9 &  & $277.9 \pm 0.7$ & $51.6 \pm 0.1$ & $745 \pm 4$ & $584 \pm 4$ & $-31.5 \pm 0.6$ & $-28.9 \pm 0.6$ & $23.6 \pm 3.3$ & $0.153 \pm 0.033$ & $2.77 \pm 0.50$ & \hlnames{\ldots} & [1] & \ldots \\
381 & 12 &  & $278.9 \pm 0.5$ & $53.6 \pm 0.1$ & $725 \pm 2$ & $584 \pm 2$ & $-23.8 \pm 0.7$ & $-20.9 \pm 0.7$ & $75.9 \pm 3.7$ & $1.36 \pm 0.16$ & $2.29 \pm 0.22$ & \hlnames{\textbf{S2}$^{[3]}$, G 283.9+54.9$^{[1]}$, G 288.4+53.2$^{[1]}$} & [1,3] & \ldots \\
357 & 4 &  & $334.6 \pm 1.7$ & $43.7 \pm 1.2$ & $842 \pm 14$ & $582 \pm 16$ & $10.9 \pm 11.4$ & $17.6 \pm 11.6$ & $50.6 \pm 4.7$ & $0.70 \pm 0.32$ & $8.71 \pm 2.24$ & \hlnames{\ldots} & \ldots & \ldots \\
2016 & 3 &  & $42.1 \pm 1.9$ & $49.3 \pm 0.9$ & $768 \pm 24$ & $570 \pm 20$ & $-13.5 \pm 5.9$ & $-20.1 \pm 5.5$ & $65.3 \pm 7.7$ & $2.0 \pm 1.7$ & $38.74 \pm 17.23$ & \hlnames{\ldots} & \ldots & \ldots \\
368 & 4 & 296 & $110.5 \pm 1.3$ & $70.7 \pm 0.4$ & $582 \pm 10$ & $550 \pm 9$ & $-16.1 \pm 4.9$ & $-16.4 \pm 5.0$ & $41.5 \pm 4.0$ & $0.400 \pm 0.035$ & $6.78 \pm 0.96$ & \hlnames{\textbf{IV25}$^{[3]}$, G107.4+70.9$^{[4]}$, G99.3+69.0$^{[4]}$} & [1,3,4] & \ldots \\
332 & 8 &  & $260.2 \pm 0.4$ & $28.9 \pm 0.6$ & $1105 \pm 14$ & $536 \pm 11$ & $2.6 \pm 1.2$ & $-0.5 \pm 1.2$ & $76.8 \pm 6.1$ & $2.06 \pm 0.50$ & $5.29 \pm 1.13$ & \hlnames{\ldots} & \ldots & \ldots \\
331 & 4 &  & $261.0 \pm 0.7$ & $31.2 \pm 0.5$ & $1022 \pm 17$ & $535 \pm 6$ & $0.6 \pm 0.9$ & $-1.5 \pm 1.0$ & $90.0 \pm 11.1$ & $4.21 \pm 0.81$ & $6.26 \pm 0.70$ & \hlnames{\ldots} & \ldots & \ldots \\
296 & 12 &  & $122.5 \pm 10.9$ & $69.3 \pm 1.5$ & $565 \pm 32$ & $527 \pm 24$ & $-7.7 \pm 5.2$ & $-7.6 \pm 5.5$ & $67.5 \pm 10.7$ & $0.91 \pm 0.56$ & $4.67 \pm 0.93$ & \hlnames{G124.1+71.6$^{[4]}$, G107.4+70.9$^{[5]}$} & [4,5] & \ldots \\
2690 & 6 & 296 & $143.5 \pm 0.9$ & $65.8 \pm 0.3$ & $539 \pm 7$ & $492 \pm 6$ & $-5.2 \pm 0.6$ & $-4.7 \pm 0.6$ & $37.1 \pm 5.0$ & $0.123 \pm 0.040$ & $2.89 \pm 0.75$ & \hlnames{\textbf{G149.9+67.4}$^{[4]}$} & [4] & \ldots \\
371 & 12 &  & $204.9 \pm 1.0$ & $60.5 \pm 0.6$ & $555 \pm 4$ & $482 \pm 4$ & $-47.0 \pm 3.7$ & $-47.9 \pm 3.7$ & $46.3 \pm 3.7$ & $0.49 \pm 0.14$ & $4.87 \pm 1.06$ & \hlnames{IV18$^{[3]}$, \textbf{IVC 210}$^{[1]}$, G211+63$^{[5]}$} & [1,3,5] & \checkmark \\
526 & 12 & 461 & $245.9 \pm 0.7$ & $46.5 \pm 0.1$ & $663 \pm 7$ & $481 \pm 6$ & $0.0 \pm 0.6$ & $-1.7 \pm 0.6$ & $36.0 \pm 4.2$ & $0.59 \pm 0.13$ & $3.64 \pm 0.45$ & \hlnames{\ldots} & \ldots & \ldots \\
356 & 12 &  & $338.2 \pm 0.3$ & $-37.3 \pm 0.1$ & $803 \pm 6$ & $-486 \pm 5$ & $7.1$ & $13.8 \pm 0.1$ & $77.3 \pm 4.9$ & $4.33 \pm 0.96$ & $9.98 \pm 1.67$ & \hlnames{\ldots} & \ldots & \ldots \\
293 & 4 &  & $253.8 \pm 8.4$ & $-81.3 \pm 1.6$ & $494 \pm 26$ & $-486 \pm 28$ & $-9.0 \pm 4.0$ & $-8.7 \pm 4.0$ & $56.1 \pm 9.5$ & $1.55 \pm 0.40$ & $15.11 \pm 4.13$ & \hlnames{\ldots} & \ldots & \ldots \\
183 & 6 &  & $260.6 \pm 0.7$ & $-29.7 \pm 0.8$ & $1184 \pm 7$ & $-590 \pm 13$ & $14.8 \pm 3.0$ & $11.9 \pm 3.0$ & $86.0 \pm 4.2$ & $3.24 \pm 0.48$ & $12.35 \pm 3.16$ & \hlnames{\ldots} & [1] & \ldots \\
253 & 3 &  & $26.2 \pm 1.2$ & $-69.3 \pm 0.4$ & $685 \pm 16$ & $-641 \pm 17$ & $-3.2 \pm 0.6$ & $-4.8 \pm 0.6$ & $50.0 \pm 6.8$ & $0.65 \pm 0.18$ & $10.64 \pm 0.88$ & \hlnames{\textbf{G25.1-67.7}$^{[5]}$} & [5] & \ldots \\
\enddata
\tablecomments{Reference key for literature identifications: [1]~\citet{RohserKerp2016}; [2]~\citet{Wakker2001}; [3]~\citet{KuntzDanly1996}; [4]~\citet{GladdersClarke1999}; [5] Magnani et al. Superscripts on cloud names indicate which catalog(s) use that name.  For clouds with multiple names, the bolded name indicates the name we use to refer to the cloud in this work.  A \checkmark\ in the IVC column marks clouds with $|v_{\rm dev}| > 40$\,km\,s$^{-1}$. Rows are ordered by decreasing $z_c$.  Reported properties and uncertainties are the medians and standard deviations of values across draws.  Entries without uncertainties have std = 0 across draws.  See Section \ref{S:catalog} for the full list of properties calculated for each cloud; only a subset is summarized here.}
\end{deluxetable*}

\begin{longrotatetable}
\setlength{\tabcolsep}{3pt}%
\begin{deluxetable*}{lccccccccccc l c}
\tablewidth{0pt}
\tabletypesize{\tiny}
\tablecaption{Properties of highlighted low-altitude clouds (non-HACs) \label{tab:highlights_nonhac}}
\tablehead{
\colhead{ID} & \colhead{$N_{\rm{draws}}$} & \colhead{\shortstack{Parent\\ID}} &\colhead{$\ell_c$} & \colhead{$b_c$} & \colhead{$d_c$} & \colhead{$z_c$} & \colhead{$v_{\rm LSR}$} & \colhead{$v_{\rm dev}$} & \colhead{$R_{\rm eff}$} & \colhead{$M$} & \colhead{$N_{\rm HI}/A_V$} & \colhead{Names} & \colhead{Ref.} \\
\colhead{} & \colhead{} & \colhead{} & \colhead{(deg)} & \colhead{(deg)} & \colhead{(pc)} & \colhead{(pc)} & \colhead{(km\,s$^{-1}$)} & \colhead{(km\,s$^{-1}$)} & \colhead{(pc)} & \colhead{($10^4\,M_{\odot}$)} & \colhead{\shortstack{($10^{21}$\,cm$^{-2}$\\mag$^{-1}$)}} & \colhead{} & \colhead{}
}
\startdata
379 & 12 &  & $12.3 \pm 3.1$ & $35.3 \pm 0.9$ & $713 \pm 30$ & $413 \pm 9$ & $10.9$ & $7.3 \pm 1.2$ & $52.1 \pm 5.4$ & $0.71 \pm 0.17$ & $4.54 \pm 0.64$ & \hlnames{\ldots} & [5] \\
2118 & 9 &  & $114.2 \pm 0.2$ & $30.3 \pm 0.1$ & $744 \pm 6$ & $375 \pm 3$ & $-14.8 \pm 2.0$ & $-10.8 \pm 2.0$ & $31.5 \pm 2.3$ & $0.460 \pm 0.099$ & $6.87 \pm 1.25$ & \hlnames{\ldots} & [1] \\
415 & 3 & 374 & $72.9 \pm 2.1$ & $27.5 \pm 0.5$ & $795 \pm 6$ & $367 \pm 4$ & $8.4 \pm 1.6$ & $1.0 \pm 1.0$ & $56.9 \pm 8.2$ & $1.00 \pm 0.25$ & $3.69 \pm 0.28$ & \hlnames{\textbf{IR4}$^{[5]}$} & [5] \\
460 & 3 &  & $226.4 \pm 0.6$ & $36.9 \pm 0.9$ & $596 \pm 3$ & $355 \pm 8$ & $12.2$ & $8.3 \pm 0.1$ & $69.7 \pm 5.4$ & $1.68 \pm 0.31$ & $1.15 \pm 0.10$ & \hlnames{IR2$^{[5]}$, 235.9+38.2$^{[5]}$} & [5] \\
352 & 12 &  & $92.9 \pm 0.3$ & $38.4 \pm 0.7$ & $516 \pm 7$ & $319 \pm 2$ & $-22.5 \pm 0.6$ & $-24.4 \pm 0.7$ & $38.7 \pm 3.5$ & $0.134 \pm 0.014$ & $1.25 \pm 0.06$ & \hlnames{\textbf{Draco}$^{[1]}$, G90.0+38.8$^{[4]}$, G94.8+37.6$^{[4,5]}$ \newline MBM41$^{[5]}$, MBM42$^{[5]}$, MBM43$^{[5]}$ \newline MBM44$^{[5]}$} & [1,4,5] \\
1791 & 4 &  & $142.1 \pm 0.0$ & $38.7 \pm 0.0$ & $403 \pm 0$ & $252 \pm 0$ & $3.2 \pm 0.6$ & $5.1 \pm 0.6$ & $7.9 \pm 0.2$ & $(4.08 \pm 0.15) \times 10^{-2}$ & $0.75 \pm 0.05$ & \hlnames{\textbf{MBM30}$^{[5]}$} & [5] \\
2134 & 3 & 484 & $45.5 \pm 1.2$ & $36.8 \pm 4.9$ & $406 \pm 19$ & $243 \pm 14$ & $-4.5 \pm 6.1$ & $-8.9 \pm 5.5$ & $46.8 \pm 1.2$ & $2.56 \pm 0.32$ & $4.85 \pm 0.03$ & \hlnames{\textbf{MBM40}$^{[5]}$} & [5] \\
1618 & 8 & 648 & $153.3 \pm 0.1$ & $37.0 \pm 0.0$ & $392 \pm 1$ & $235 \pm 0$ & $-0.6 \pm 0.4$ & $1.0 \pm 0.4$ & $20.9 \pm 1.2$ & $0.254 \pm 0.022$ & $0.70 \pm 0.01$ & \hlnames{MBM31$^{[5]}$, HSVMT24$^{[5]}$, HSVMT27$^{[5]}$ \newline MBM26$^{[5]}$, HSVMT28$^{[5]}$} & [5] \\
648 & 12 &  & $155.7 \pm 0.7$ & $35.4 \pm 0.3$ & $387 \pm 2$ & $225 \pm 2$ & $1.9 \pm 0.6$ & $3.6 \pm 0.6$ & $50.5 \pm 3.2$ & $1.17 \pm 0.20$ & $0.76 \pm 0.08$ & \hlnames{MBM27$^{[5]}$, MBM28$^{[5]}$, MBM29$^{[5]}$} & [5] \\
655 & 12 &  & $140.3 \pm 0.1$ & $48.0 \pm 0.0$ & $297 \pm 0$ & $220 \pm 0$ & $-10.9$ & $-10.4 \pm 0.0$ & $10.0 \pm 0.8$ & $(1.51 \pm 0.10) \times 10^{-2}$ & $1.74 \pm 0.08$ & \hlnames{G139.6+47.6$^{[4,5]}$, G141.1+48.0$^{[4,5]}$} & [4,5] \\
547 & 12 &  & $140.5 \pm 0.6$ & $41.9 \pm 0.1$ & $309 \pm 1$ & $206 \pm 1$ & $4.5$ & $5.4 \pm 0.0$ & $28.0 \pm 1.0$ & $0.230 \pm 0.024$ & $1.06 \pm 0.12$ & \hlnames{\textbf{MBM32}$^{[5]}$} & [5] \\
774 & 12 & 648 & $170.7 \pm 0.1$ & $27.1 \pm 0.1$ & $398 \pm 1$ & $181 \pm 1$ & $-1.9 \pm 0.6$ & $-1.1 \pm 0.6$ & $29.0 \pm 1.6$ & $0.574 \pm 0.044$ & $2.20 \pm 0.04$ & \hlnames{MBM23$^{[5]}$, MBM24$^{[5]}$, MBM25$^{[5]}$} & [5] \\
327 & 12 &  & $82.3 \pm 1.3$ & $42.5 \pm 3.2$ & $265 \pm 38$ & $178 \pm 36$ & $4.5 \pm 0.6$ & $1.5 \pm 0.6$ & $26.5 \pm 10.6$ & $0.14 \pm 0.13$ & $5.91 \pm 0.81$ & \hlnames{\textbf{G81.2+39.2}$^{[4]}$} & [4] \\
545 & 11 &  & $278.2 \pm 4.0$ & $81.9 \pm 0.5$ & $145 \pm 1$ & $143 \pm 1$ & $1.9$ & $2.4 \pm 0.0$ & $14.4 \pm 1.0$ & $(5.78 \pm 0.65) \times 10^{-2}$ & $3.38 \pm 0.24$ & \hlnames{\textbf{G249.0+73.7}$^{[4]}$} & [4] \\
765 & 12 &  & $148.1 \pm 0.5$ & $31.8 \pm 0.2$ & $250 \pm 1$ & $132 \pm 1$ & $1.9 \pm 1.1$ & $2.8 \pm 1.1$ & $12.0 \pm 1.3$ & $(9.4 \pm 1.8) \times 10^{-2}$ & $5.57 \pm 0.72$ & \hlnames{\ldots} & [5] \\
1771 & 12 &  & $340.9 \pm 0.1$ & $23.1 \pm 0.1$ & $155 \pm 0$ & $61 \pm 0$ & $3.2$ & $5.6 \pm 0.0$ & $11.0 \pm 0.8$ & $0.1272 \pm 0.0094$ & $0.80 \pm 0.03$ & \hlnames{\textbf{G344.8+24.0}$^{[5]}$} & [5] \\
905 & 5 &  & $269.3 \pm 0.2$ & $20.4 \pm 0.5$ & $170 \pm 0$ & $59 \pm 1$ & $-1.9 \pm 0.6$ & $1.0 \pm 0.6$ & $32.1 \pm 1.5$ & $1.76 \pm 0.14$ & $3.24 \pm 0.06$ & \hlnames{\textbf{272.9+29.3}$^{[5]}$} & [5] \\
1820 & 3 &  & $162.6 \pm 0.3$ & $-27.5 \pm 0.0$ & $141 \pm 0$ & $-65 \pm 0$ & $10.9$ & $10.9 \pm 0.0$ & $7.5 \pm 0.4$ & $(4.42 \pm 0.57) \times 10^{-2}$ & $1.02 \pm 0.03$ & \hlnames{UT7$^{[5]}$, UT5$^{[5]}$, MBM17$^{[5]}$ \newline UT6$^{[5]}$} & [5] \\
2872 & 4 &  & $159.2 \pm 0.5$ & $-28.5 \pm 0.2$ & $144 \pm 0$ & $-69 \pm 1$ & $10.3 \pm 0.6$ & $10.3 \pm 0.6$ & $7.2 \pm 0.3$ & $(4.19 \pm 0.60) \times 10^{-2}$ & $1.35 \pm 0.19$ & \hlnames{\textbf{UT3}$^{[5]}$} & [5] \\
885 & 11 &  & $200.9 \pm 0.1$ & $-25.4 \pm 0.1$ & $175 \pm 0$ & $-75 \pm 0$ & $9.7$ & $9.1 \pm 0.0$ & $10.5 \pm 0.5$ & $(4.58 \pm 0.37) \times 10^{-2}$ & $2.46 \pm 0.22$ & \hlnames{\textbf{CB28}$^{[5]}$} & [5] \\
1796 & 9 &  & $299.8 \pm 0.2$ & $-24.4 \pm 0.3$ & $212 \pm 1$ & $-88 \pm 1$ & $-9.7$ & $-4.7 \pm 0.0$ & $6.1 \pm 0.7$ & $(2.89 \pm 0.97) \times 10^{-2}$ & $1.06 \pm 0.09$ & \hlnames{299.8-26.2$^{[5]}$, 300.1-24.5$^{[5]}$, 301.2-24.5$^{[5]}$} & [5] \\
1409 & 11 & 819 & $204.0 \pm 0.1$ & $-31.4 \pm 0.2$ & $188 \pm 0$ & $-98 \pm 0$ & $-5.8$ & $-6.3 \pm 0.0$ & $4.9 \pm 0.5$ & $(3.67 \pm 0.84) \times 10^{-3}$ & $0.75 \pm 0.07$ & \hlnames{\ldots} & [5] \\
1488 & 5 & 1197 & $65.4 \pm 0.5$ & $-26.8 \pm 0.2$ & $221 \pm 2$ & $-100 \pm 0$ & $12.2$ & $7.7 \pm 0.0$ & $7.8 \pm 0.9$ & $(2.66 \pm 0.53) \times 10^{-2}$ & $1.46 \pm 0.06$ & \hlnames{\textbf{MBM49}$^{[5]}$} & [5] \\
2755 & 4 &  & $219.4 \pm 0.3$ & $-35.9 \pm 0.4$ & $184 \pm 0$ & $-108 \pm 1$ & $4.5$ & $4.3 \pm 0.0$ & $10.1 \pm 0.6$ & $(4.04 \pm 0.49) \times 10^{-2}$ & $3.79 \pm 0.47$ & \hlnames{\ldots} & [5] \\
1615 & 8 & 1307 & $102.0 \pm 0.1$ & $-28.1 \pm 0.1$ & $289 \pm 1$ & $-136 \pm 1$ & $-3.9 \pm 0.6$ & $-5.2 \pm 0.6$ & $9.3 \pm 0.9$ & $(6.05 \pm 1.00) \times 10^{-2}$ & $1.29 \pm 0.09$ & \hlnames{G102-27$^{[5]}$, MBM56$^{[5]}$} & [5] \\
1290 & 8 &  & $285.4 \pm 1.2$ & $-40.5 \pm 1.0$ & $226 \pm 2$ & $-147 \pm 5$ & $7.1$ & $10.5 \pm 0.1$ & $10.5 \pm 1.2$ & $0.122 \pm 0.035$ & $2.71 \pm 0.32$ & \hlnames{\textbf{295.3-36.2}$^{[5]}$} & [5] \\
1703 & 12 &  & $161.4 \pm 0.1$ & $-30.1 \pm 0.1$ & $296 \pm 1$ & $-149 \pm 1$ & $1.9 \pm 2.5$ & $2.9 \pm 2.5$ & $19.0 \pm 0.9$ & $0.83 \pm 0.23$ & $1.54 \pm 0.40$ & \hlnames{MBM11$^{[5]}$, MBM12$^{[5]}$, MBM13$^{[5]}$ \newline MBM14$^{[5]}$, UT4$^{[5]}$} & [5] \\
1183 & 11 &  & $61.9 \pm 1.8$ & $-38.9 \pm 1.1$ & $249 \pm 3$ & $-156 \pm 1$ & $-5.8$ & $-9.8 \pm 0.2$ & $18.6 \pm 1.8$ & $0.205 \pm 0.061$ & $1.03 \pm 0.10$ & \hlnames{\textbf{HD210121}$^{[5]}$} & [5] \\
1485 & 9 & 1183 & $57.5 \pm 0.3$ & $-42.4 \pm 0.4$ & $241 \pm 2$ & $-162 \pm 0$ & $-5.8$ & $-9.6 \pm 0.1$ & $10.6 \pm 1.6$ & $(6.4 \pm 1.8) \times 10^{-2}$ & $1.04 \pm 0.09$ & \hlnames{\ldots} & [5] \\
1196 & 12 &  & $72.4 \pm 0.4$ & $-44.2 \pm 0.9$ & $234 \pm 3$ & $-163 \pm 2$ & $-5.8$ & $-9.0 \pm 0.1$ & $13.7 \pm 1.6$ & $(9.3 \pm 2.7) \times 10^{-2}$ & $0.92 \pm 0.15$ & \hlnames{\ldots} & [5] \\
788 & 12 &  & $198.3 \pm 0.8$ & $-51.2 \pm 0.5$ & $222 \pm 4$ & $-173 \pm 4$ & $10.9 \pm 0.6$ & $10.7 \pm 0.6$ & $24.4 \pm 2.0$ & $0.400 \pm 0.064$ & $2.68 \pm 0.21$ & \hlnames{G187.1-52.3$^{[5]}$, G189.3-53.3$^{[5]}$} & [5] \\
1021 & 12 &  & $98.6 \pm 0.6$ & $-41.3 \pm 0.5$ & $263 \pm 1$ & $-174 \pm 2$ & $-5.8$ & $-7.5 \pm 0.0$ & $43.5 \pm 2.1$ & $2.45 \pm 0.23$ & $1.10 \pm 0.04$ & \hlnames{G61-34$^{[5]}$, MBM55$^{[5]}$, MBM54$^{[5]}$ \newline MBM01$^{[5]}$, G112-40$^{[5]}$, MBM02$^{[5]}$ \newline MBM03$^{[5]}$, MBM04$^{[5]}$} & [5] \\
199 & 4 & 788 & $184.2 \pm 0.4$ & $-51.8 \pm 0.1$ & $232 \pm 1$ & $-183 \pm 1$ & $10.9 \pm 0.6$ & $10.8 \pm 0.6$ & $13.3 \pm 1.2$ & $0.124 \pm 0.027$ & $3.50 \pm 0.61$ & \hlnames{\textbf{MBM15}$^{[5]}$} & [5] \\
1324 & 3 & 1175 & $107.5 \pm 2.8$ & $-50.0 \pm 0.3$ & $249 \pm 1$ & $-190 \pm 2$ & $-7.1$ & $-8.2 \pm 0.1$ & $12.1 \pm 1.9$ & $0.174 \pm 0.086$ & $2.92 \pm 0.16$ & \hlnames{\ldots} & [5] \\
1175 & 11 &  & $111.7 \pm 3.5$ & $-51.1 \pm 0.2$ & $248 \pm 1$ & $-193 \pm 1$ & $-7.1 \pm 0.4$ & $-8.0 \pm 0.4$ & $17.2 \pm 1.2$ & $0.40 \pm 0.11$ & $3.17 \pm 0.16$ & \hlnames{\ldots} & [5] \\
2342 & 9 & 1324 & $92.8 \pm 1.1$ & $-50.7 \pm 0.2$ & $251 \pm 1$ & $-195 \pm 1$ & $-7.1$ & $-9.0 \pm 0.1$ & $13.6 \pm 1.5$ & $0.114 \pm 0.019$ & $1.36 \pm 0.15$ & \hlnames{\ldots} & [5] \\
2339 & 7 & 1175 & $119.5 \pm 1.1$ & $-53.0 \pm 0.4$ & $248 \pm 1$ & $-198 \pm 2$ & $-8.4$ & $-8.9 \pm 0.0$ & $9.9 \pm 1.4$ & $(6.6 \pm 2.8) \times 10^{-2}$ & $2.28 \pm 0.47$ & \hlnames{\ldots} & [5] \\
1208 & 9 &  & $211.8 \pm 0.2$ & $-28.4 \pm 0.2$ & $426 \pm 1$ & $-203 \pm 2$ & $5.8$ & $2.8 \pm 0.0$ & $18.0 \pm 1.0$ & $0.407 \pm 0.052$ & $2.26 \pm 0.18$ & \hlnames{MBM22$^{[5]}$, MBM21$^{[5]}$} & [5] \\
1519 & 5 &  & $115.7 \pm 0.1$ & $-43.2 \pm 0.1$ & $306 \pm 0$ & $-210 \pm 0$ & $-3.2 \pm 0.0$ & $-3.4 \pm 0.0$ & $7.9 \pm 0.4$ & $(4.94 \pm 0.46) \times 10^{-2}$ & $1.78 \pm 0.11$ & \hlnames{\ldots} & [5] \\
1314 & 10 &  & $150.4 \pm 0.7$ & $-47.2 \pm 0.4$ & $290 \pm 1$ & $-213 \pm 1$ & $-9.7$ & $-9.1 \pm 0.0$ & $13.2 \pm 0.9$ & $(9.5 \pm 1.8) \times 10^{-2}$ & $1.16 \pm 0.15$ & \hlnames{\textbf{MBM05}$^{[5]}$} & [5] \\
1520 & 5 &  & $117.3 \pm 0.3$ & $-44.0 \pm 0.1$ & $308 \pm 0$ & $-214 \pm 1$ & $-3.2 \pm 0.0$ & $-3.3 \pm 0.0$ & $10.0 \pm 0.8$ & $(8.5 \pm 1.3) \times 10^{-2}$ & $1.91 \pm 0.19$ & \hlnames{\ldots} & [5] \\
1056 & 10 &  & $164.8 \pm 0.6$ & $-42.7 \pm 0.1$ & $320 \pm 2$ & $-217 \pm 1$ & $-4.5$ & $-4.0 \pm 0.0$ & $38.9 \pm 1.9$ & $1.87 \pm 0.22$ & $1.20 \pm 0.08$ & \hlnames{G154.7-39.8$^{[5]}$, G161.9-43.3$^{[5]}$, IR1$^{[5]}$} & [5] \\
2469 & 8 &  & $161.4 \pm 0.2$ & $-46.7 \pm 0.1$ & $303 \pm 1$ & $-221 \pm 1$ & $-12.2 \pm 0.6$ & $-11.8 \pm 0.6$ & $7.7 \pm 1.0$ & $(4.05 \pm 0.83) \times 10^{-2}$ & $1.85 \pm 0.24$ & \hlnames{\ldots} & [5] \\
403 & 12 &  & $32.4 \pm 1.5$ & $-18.5 \pm 0.6$ & $700 \pm 12$ & $-222 \pm 10$ & $20.0 \pm 0.5$ & $9.4 \pm 0.7$ & $40.4 \pm 5.2$ & $0.58 \pm 0.18$ & $4.30 \pm 0.51$ & \hlnames{\ldots} & [1] \\
3072 & 4 & 1056 & $168.4 \pm 0.3$ & $-43.0 \pm 0.2$ & $337 \pm 1$ & $-230 \pm 0$ & $-4.5$ & $-4.1 \pm 0.0$ & $29.5 \pm 1.9$ & $0.93 \pm 0.11$ & $1.11 \pm 0.02$ & \hlnames{\textbf{3C75.0}$^{[5]}$} & [5] \\
509 & 12 &  & $88.5 \pm 0.4$ & $-34.6 \pm 0.3$ & $408 \pm 1$ & $-231 \pm 1$ & $-11.6 \pm 2.5$ & $-14.4 \pm 2.5$ & $36.6 \pm 2.8$ & $0.64 \pm 0.11$ & $2.20 \pm 0.18$ & \hlnames{3C454.3$^{[5]}$, MBM53$^{[5]}$} & [5] \\
3035 & 3 &  & $196.7 \pm 0.5$ & $-56.6 \pm 0.2$ & $280 \pm 2$ & $-234 \pm 2$ & $4.5 \pm 0.6$ & $4.2 \pm 0.6$ & $11.7 \pm 1.0$ & $(4.51 \pm 0.30) \times 10^{-2}$ & $2.54 \pm 0.31$ & \hlnames{\ldots} & [5] \\
1283 & 9 & 3072 & $171.6 \pm 0.4$ & $-41.9 \pm 0.3$ & $350 \pm 2$ & $-234 \pm 0$ & $-4.5$ & $-4.2 \pm 0.0$ & $27.0 \pm 1.7$ & $0.656 \pm 0.076$ & $0.99 \pm 0.06$ & \hlnames{\ldots} & [5] \\
703 & 12 & 496 & $229.4 \pm 0.5$ & $-65.9 \pm 0.2$ & $274 \pm 1$ & $-251 \pm 1$ & $0.6 \pm 0.6$ & $0.9 \pm 0.6$ & $9.1 \pm 1.1$ & $(1.38 \pm 0.32) \times 10^{-2}$ & $2.77 \pm 0.23$ & \hlnames{\textbf{225.3-66.3}$^{[5]}$} & [5] \\
965 & 10 &  & $190.3 \pm 0.1$ & $-33.2 \pm 0.1$ & $462 \pm 3$ & $-253 \pm 1$ & $5.8 \pm 0.5$ & $4.5 \pm 0.5$ & $39.5 \pm 2.3$ & $4.31 \pm 0.45$ & $3.93 \pm 0.12$ & \hlnames{MBM19$^{[5]}$, 3C105.0$^{[5]}$, MBM18$^{[5]}$} & [5] \\
1085 & 12 & 965 & $191.0 \pm 0.1$ & $-28.9 \pm 0.9$ & $524 \pm 7$ & $-254 \pm 4$ & $4.5 \pm 0.6$ & $2.8 \pm 0.6$ & $28.2 \pm 2.2$ & $1.24 \pm 0.39$ & $2.95 \pm 0.48$ & \hlnames{\ldots} & [5] \\
504 & 12 &  & $130.1 \pm 0.5$ & $-68.8 \pm 0.1$ & $306 \pm 1$ & $-286 \pm 1$ & $-3.2$ & $-3.6 \pm 0.0$ & $35.0 \pm 1.7$ & $0.797 \pm 0.049$ & $2.53 \pm 0.08$ & \hlnames{G101.9-62.0$^{[5]}$, G135.4-68.7$^{[5]}$} & [5] \\
552 & 11 &  & $16.7 \pm 1.6$ & $-41.4 \pm 1.0$ & $469 \pm 8$ & $-310 \pm 11$ & $-1.9 \pm 0.5$ & $-4.8 \pm 0.6$ & $76.7 \pm 5.5$ & $5.2 \pm 1.1$ & $2.09 \pm 0.30$ & \hlnames{\textbf{MBM45}$^{[5]}$} & [5] \\
1394 & 9 &  & $40.0 \pm 0.0$ & $-35.1 \pm 0.1$ & $551 \pm 2$ & $-317 \pm 1$ & $5.8 \pm 0.4$ & $-1.4 \pm 0.4$ & $15.1 \pm 1.0$ & $0.163 \pm 0.015$ & $1.13 \pm 0.04$ & \hlnames{MBM46$^{[5]}$, MBM48$^{[5]}$, MBM47$^{[5]}$} & [5] \\
414 & 12 &  & $28.8 \pm 0.3$ & $-16.3 \pm 0.5$ & $1170 \pm 6$ & $-330 \pm 11$ & $22.5 \pm 0.8$ & $6.0 \pm 0.8$ & $65.5 \pm 7.1$ & $5.40 \pm 0.95$ & $10.95 \pm 1.29$ & \hlnames{\ldots} & [1] \\
772 & 11 &  & $45.2 \pm 0.0$ & $-32.7 \pm 0.2$ & $652 \pm 2$ & $-352 \pm 1$ & $4.5$ & $-4.5 \pm 0.0$ & $27.2 \pm 1.6$ & $0.777 \pm 0.082$ & $2.68 \pm 0.16$ & \hlnames{\ldots} & [5] \\
343 & 12 &  & $312.0 \pm 0.4$ & $-21.6 \pm 0.1$ & $1074 \pm 6$ & $-396 \pm 2$ & $-16.1 \pm 0.8$ & $0.4 \pm 0.8$ & $82.7 \pm 4.8$ & $2.78 \pm 0.38$ & $6.31 \pm 0.58$ & \hlnames{\ldots} & [1] \\
373 & 3 & 343 & $308.1 \pm 1.9$ & $-23.1 \pm 0.6$ & $1071 \pm 4$ & $-418 \pm 10$ & $-16.1 \pm 0.6$ & $-1.3 \pm 0.6$ & $57.9 \pm 7.1$ & $2.10 \pm 0.52$ & $12.90 \pm 1.29$ & \hlnames{\ldots} & [1] \\
\enddata
\tablecomments{Reference key for literature identifications: [1]~\citet{RohserKerp2016}; [2]~\citet{Wakker2001}; [3]~\citet{KuntzDanly1996}; [4]~\citet{GladdersClarke1999}; [5] Magnani et al. (private communication). Superscripts on cloud names indicate which catalog(s) use that name. For clouds with multiple names, the bolded name indicates the name we use to refer to the cloud in this work.  Rows are ordered by decreasing $z_c$.  Reported properties and uncertainties are the medians and standard deviations of values across draws.  Entries without uncertainties have standard deviation = 0 across draws.  See Section \ref{S:catalog} for the full list of properties calculated for each cloud; only a subset is summarized here.}
\end{deluxetable*}
\end{longrotatetable}

\paragraph{Draco}

Draco is perhaps the most well-studied high-latitude literature IVC, with decades of extensive study revealing a small complex of diffuse neutral hydrogen laced with dense molecular clumps traced by CO \citep[e.g.,][]{GoerigkMebold1983, MagnaniBlitz1985, HerbstmeierHeithausen1993, HeithausenWeiss2001, MivilleDeschenesSalome2017, SchneiderOssenkopfOkada2025}. On the basis of analysis of the stability of fragmented clumps within Draco observed with Herschel-SPIRE, \citet{MivilleDeschenesSalome2017} proposed that Draco was likely formed as the result of colliding flows of gas between the lower Galactic halo and the outer disk.  [CII] was additionally detected in Draco by \citet{SchneiderOssenkopfOkada2024}, suggesting the presence of shocks resulting from the cloud's motion towards the Galactic plane.  The earliest distance constraint to Draco was provided by \citet{GladdersClarke1998}, who used detections/non-detections of Na I absorption towards 4 stars to derive a distance bracket of $463^{+192}_{-136} < d < 618^{+243}_{-174}$ pc.  More recently, using Gaia DR2 parallax measurements to infer stellar distance and foreground extinctions, \citet{ZuckerSpeagle2019} obtained a distance to Draco of 481$^{+50}_{-45} \pm 24$ pc (where the first uncertainty is statistical and the second is systematic).  In this work, we derive a centroid distance $d_c = 516 \pm 7$ pc, centroid z-height $z_c = 319 \pm 2$ pc, and deviation velocity of $v_{\rm{dev}} = -24.4 \pm 0.7$ km/s.  Draco is therefore excluded from our stricter definition of IVCs as clouds with $|v_{\rm{dev}}| > 40 $ km/s, and is also excluded from our HAC definition of $|z| \geq 480$ pc.  This cloud was chosen to visualize our example \HI-matching procedure in Figure \ref{fig:example_ssim}.  

\paragraph{IVC 210}

IVC 210 is another member of the small sample of high-latitude IVCs with associated CO emission \citep{DesertBazell1990,RohserKerp2014,RohserKerp2016b}.  Using a handful of Ca II absorption measurements, \citet{WesseliusFejes1973} estimated a distance bracket for this cloud of 12 pc to 400 pc.  IVC 210 is observed to produce a soft X-ray shadow \citep{SnowdenFreyberg2000}, which \citet{RohserKerp2014} used to estimate a distance towards the cloud of $510 \pm 140$ pc (by extrapolating the relationship between X-ray count rate and the distance for IVC 135, described below, derived by \citet{BenjaminVenn1996} from starlight bracketing).  We derive a centroid distance $d_c = 555 \pm 4$ pc, centroid z-height $z_c = 482 \pm 4$ pc, and deviation velocity of $v_{\rm{dev}} = -47.9 \pm 3.7$ km/s, classifying IVC 210 as both an IVC and an HAC under our observational definitions.  This cloud is also referred to as IV18 by \citetalias{KuntzDanly1996}.

\paragraph{Cloud 296 and IV 25}

This cloud (cloud 296 in our numbered catalog, and associated sub-cloud cloud 2690) contains three of the high-latitude molecular cirrus clouds studied by \citet{HeilesReach1988} and \citetalias{GladdersClarke1999} and Magnani et al.: G124.1+71.6, G107.4+70.9, and G149.9+67.4, along with one additional clump.  For this full complex, we derive a centroid distance $d_c = 565 \pm 32$ pc, altitude $z=527 \pm 24$ pc, and deviation velocity of $v_{\rm{dev}}=-7.6 \pm 5.5$ km/s.  The cloud's LSR velocity is broad and ranges from $-23.8 \pm 4.0$ km/s to $-0.6 \pm 0.6$ km/s.  We classify Cloud 296 as an HAC but not an IVC.  

IV 25 is a less studied clump within the \citetalias{KuntzDanly1996} catalog.  This cloud is embedded within a substructure of Cloud 296 in our hierarchical \perch segmentation, along with two other clumps.  We derive a centroid distance $d_c = 582 \pm 10$ pc, centroid z-height $z_c = 550 \pm 9$ pc, and deviation velocity of $v_{\rm{dev}} = -16.4 \pm 5.0$ km/s, classifying IV 25 solely as an HAC.  We note that this $v_{\rm{dev}}$ is computed from the cloud's peak velocity of $v_{\rm{LSR}} = -16.1 \pm 4.9$, well below the velocity of this cloud in the \citetalias{KuntzDanly1996} catalog; this match is made based on the broad velocity range of this cloud in our catalog, extending from $-35.4 \pm 3.0$ to $-3.2 \pm 1.4$ and overlapping with the \citetalias{KuntzDanly1996} velocity range for IV25.  This cloud is also matched to the lower-velocity G107.4+70.9 and G99.3+69.0 in the \citetalias{GladdersClarke1999} catalog, which are closer to the peak velocity of our cloud.  The distance brackets for this cloud in Fig. \ref{fig:ivc_lit_dist} come from \citetalias{GladdersClarke1999}, and disagree with the cloud's actual distance.

\paragraph{IV Spur S2}

\citetalias{KuntzDanly1996} described an extension from the IV Arch towards the eastern Northern hemisphere, which they referred to as the IV Spur.  They divided the Spur into two clumps: S1 and S2.  We associate one of our clouds with their IV Spur S2.  A cluster of intermediate-velocity CO emission towards three LOS (G283.9+54.9, G288.4+53.2, and G295.0+57.1) was reported by \citet{MagnaniSmith2010}, who noted that these LOS were projected onto the portion of the sky occupied by ``the easternmost extension of the Intermediate Velocity Spur'' and suggested that their CO detections were all part of the same structure.  We confirm \citet{MagnaniSmith2010}'s interpretation that these LOS all trace the same 3D structure known as the IV Spur S2.  \citet{KuntzDanly1996} derived a distance bracket for the IV Spur of 0.3-2.1 kpc from stellar absorption.  We derive a centroid distance $d_c = 725 \pm 2$ pc, centroid z-height $z_c = 584 \pm 2$ pc, and deviation velocity of $v_{\rm{dev}} = -20.9 \pm 0.7$ km/s, classifying the IV Spur S2 as a HAC but not an IVC.  This cloud contains a small embedded substructure, Cloud 527.

\paragraph{LLIV Arch}

The Low-Latitude IV (LLIV) Arch is the third of the three large Northern complexes identified by \citetalias{KuntzDanly1996} (who divided it into six sub-complexes named LLIV1 through LLIV6).  It overlaps in POS position with both the lower-velocity North Celestial Pole Loop and the higher-velocity HVC Complex A, and its origin has long been unclear.  \citetalias{Wakker2001} bracketed the LLIV Arch's distance at  900 - 1800 pc, and estimated the abundance relative to Solar as $1.0 \pm 0.5$ with a depletion pattern consistent with the warm disk; they therefore speculated the LLIV Arch is the returning fragment of a Galactic fountain flow in the interarm region between the Local Arm and the Perseus Arm.  \citet{RichterSavage2001} found evidence for H$_2$ absorption in some LOS sampling the LLIV Arch.  Through analysis of HI emission and absorption data, \citet{VujevaMarchal2023} found evidence for CNM gas within the \citetalias{KuntzDanly1996} sub-clump LLIV1.  We recover a cloud containing the \citetalias{KuntzDanly1996} clumps LLIV1, LLIV2, LLIV3, and LLIV4; we do not detect LLIV5 or LLIV6, which are offset in POS position by a significant amount from the main complex.  We additionally detect an embedded sub-clump (cloud 1614) within our primary LLIV match (cloud 360) matched solely to LLIV1 and LLIV3.  We derive a centroid distance for our detected cloud of $d_c = 966 \pm 18$ pc, centroid z-height $z_c = 606 \pm 11$ pc, and deviation velocity of $v_{\rm{dev}} = -47.2 \pm 1.0$ km/s, classifying the LLIV Arch as both a HAC and an IVC.
  
\paragraph{IVC 135}

IVC 135 is another member of the class of molecular IVCs with CO detections \citep{HeilesReach1988, ReachKoo1994, WeissHeithausen1999, HeithausenWeiss2001}.  This cloud is also referred to as IV21 in the \citetalias{KuntzDanly1996} catalog.  IVC 135 is X-ray-shadowed and has long been speculated to be related to the IV Arch \citep{BenjaminVenn1996}.  \citet{HernandezWakker2013} derived an unusually low dust-to-gas ratio and a sub-solar metallicity ($\sim0.4 \ Z_\odot$) for the cloud, which they argued implies an extragalactic or circumgalactic origin.  \citet{LenzKerp2015} challenged this interpretation, deriving a high dust emissivity for IVC 135 similar to values found for low-velocity disk gas, implying a solar metallicity and probable Galactic origin.  \citet{HeilesReach1988} divided the cloud into three parts: G135.5+51.3, G137.3+53.9, and G135.3+54.5.  \citet{BenjaminVenn1996} analyzed optical spectra for 9 stars in the direction of G135.3+54.5, and, using distance brackets from the presence or absence of Na I absorption, derived a distance of $d = 355 \pm 95$ pc based on a detection at the cloud's velocity in the foreground to BD +63 985 (located at $(\ell,b)=(133.74^\circ, 53.42^\circ)$).  Also using Na I with a sample of 3 stars, \citetalias{GladdersClarke1999} derived a distance bracket for the adjacent G135.5+51.3 of $746^{+1747}_{-433} -1419^{+485}_{-382} $ pc. We derive a centroid distance $d_c = 745 \pm 7$ pc, centroid z-height $z_c = 606 \pm 18$ pc, and deviation velocity of $v_{\rm{dev}} = -44.6 \pm 0.3$ km/s, classifying IVC 135 as both an IVC and an HAC.  \citet{LenzKerp2015} suggested that an adjacent (on the POS) cloud within the HVC Complex C may be interacting with IVC 135; we find no evidence for an associated nearby HVC in 3D space (but note that a dust-poor HVC would likely not be detected at current 3D dust map sensitivity limits).  

\paragraph{Cloud 268} This unnamed cloud is detected in the \citetalias{RohserKerp2016} survey and is adjacent on the POS to Cloud 296, but located at a higher-altitude; we derive a centroid $z=715$ pc, $v_{\rm{LSR}}$ between -27.7 km/s to -12.2 km/s, and deviation velocity $v_{\rm{dev}}=-21.0$ km/s.  We therefore classify this cloud as an HAC.  This cloud contains a clump visible on the edge of the IV Arch (see next subsection), but not cataloged by \citetalias{KuntzDanly1996}.

\subsubsection{The IV Arch}

\begin{figure*}
    \centering
\includegraphics[width=\textwidth]{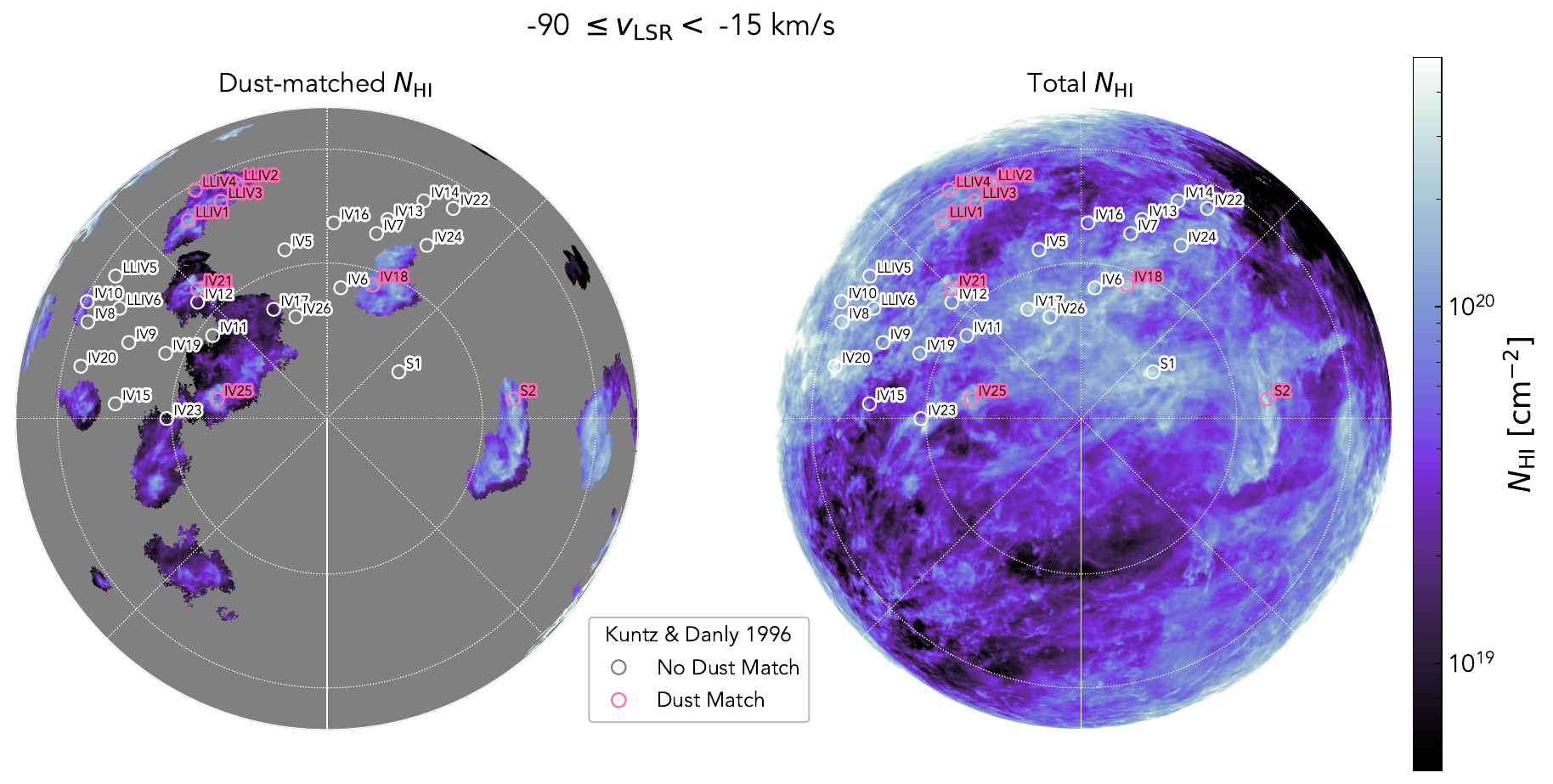}
\includegraphics[width=\textwidth]{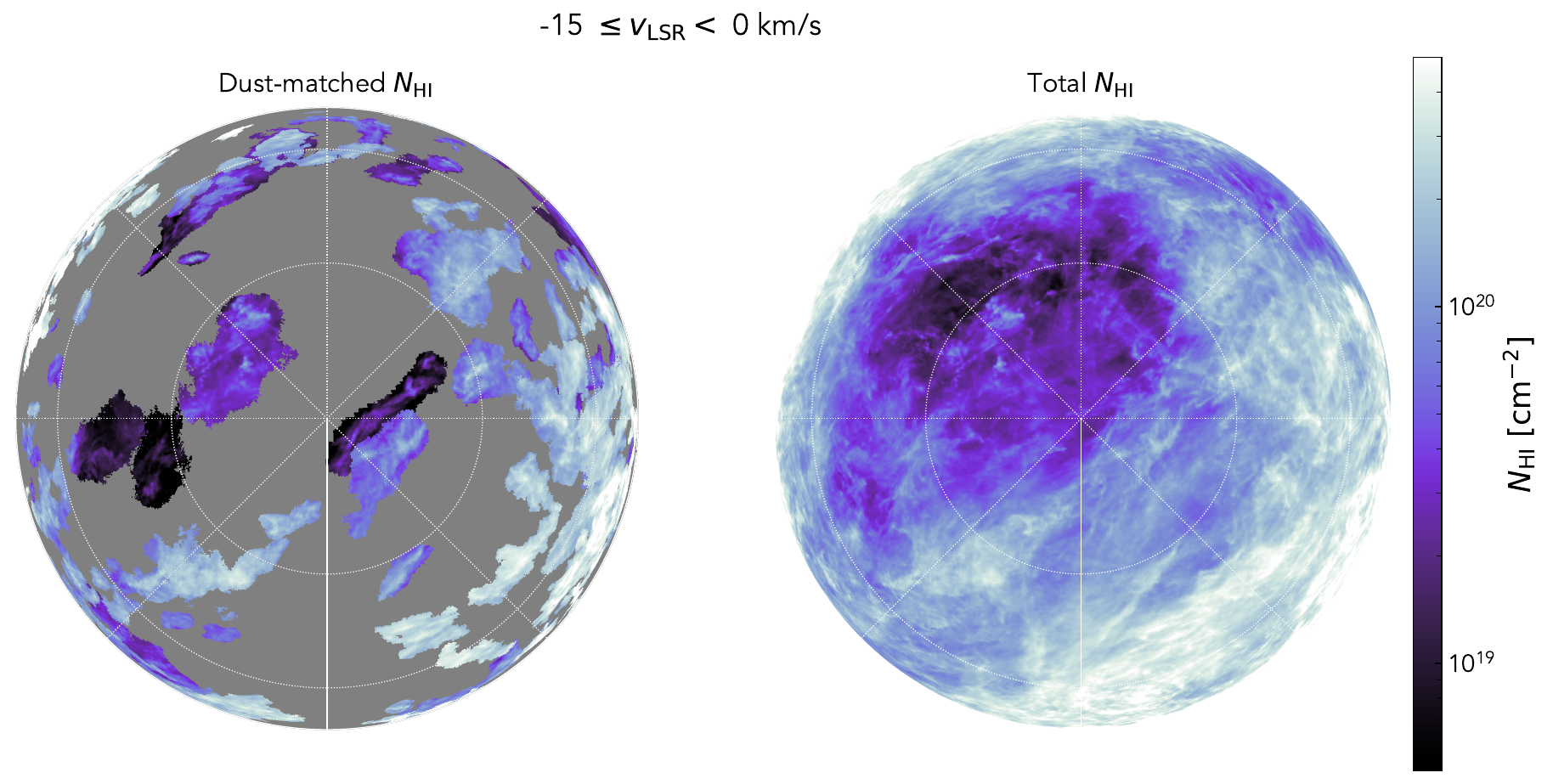}
 \caption{\textit{Top:} Integrated \HI{} column density $N_{HI}$ over the velocity range $-90 \leq v_{\rm{LSR}} < -15$ km/s in the Northern galactic hemisphere, for \HI{} emission matched to 3D dust clouds (\textit{left}, calculated over the dust clouds' maximum extent areas) and for all \HI{} emission (\textit{right}).  The labeled open circles mark the clumps in this velocity range defined by \citet{KuntzDanly1996} that constitute the Intermediate Velocity Arch.  \citet{KuntzDanly1996} clumps that we have matched to a 3D dust cloud are marked in pink, while those with no 3D dust match are marked in white.  \textit{Bottom:} as top, but for the lower-velocity range $-15 \leq v_{\rm{LSR}} < 0$ km/s.}
    \label{fig:ivarch}
\end{figure*}

Sweeping over an enormous fraction of the Galactic Northern sky, the history of studies of the IV Arch is long (dating back to \citealt{WesseliusFejes1973}).  This portion of the sky displays a variety of LVCs, IVCs, and HVCs, making disentangling various features' distances, origins, and eventual fates difficult.  \citetalias{Wakker2001} found it useful to separate the IV Arch into a higher-velocity ($v_{\rm{LSR}} <-60$ km/s) component (consisting of \citetalias{KuntzDanly1996} cores IV5-IV17) and a lower-velocity ($v_{\rm{LSR}} > -60$ km/s) component (consisting of \citetalias{KuntzDanly1996} cores IV18-IV26), which overlap on the POS.  Figure \ref{fig:ivarch} compares the integrated $N_{HI}$ for the IV Arch over the range $-90 \leq v_{\rm{LSR}} < -15$ km/s to disk-confused emission at $-15 \leq v_{\rm{LSR}} < 0$ km/s.   

We associate the clouds IVC 135 (IV21), IVC 210 (IV18), and IV25 with this lower-velocity component of the IV Arch.  Their combined extent on the POS (see Fig. \ref{fig:ivarch}) produces a clear match to portions of the established morphology and kinematics of the lower-velocity component of the IV Arch.  Cloud 296 (which contains IV25) and Cloud 268 appear to be a low-velocity, similar-altitude ``bridge'' connecting these three clouds in both 3D space and on the POS.  We argue that these clouds (IVC 135, IVC 210, IV 25/Cloud 296, Cloud 268) constitute the lower-velocity component of the IV Arch.  For the lower-velocity IV Arch at $\ell > 150^\circ$, \citetalias{Wakker2001} derived a $z$-bracket of 0.4-1.7 kpc with an abundance of roughly Solar and a depletion pattern consistent with the Halo; for $\ell < 150^\circ$ (excluding IVC 135), they derived a $z-$bracket of 0.4-0.8 kpc.  Our four clouds span centroid $z$-heights ranging between 482-606 pc, consistent with the \citetalias{Wakker2001} brackets.    

We fail to recover several portions of the IV Arch complex.  We do not identify any of the \citetalias{KuntzDanly1996} clumps in the higher-velocity IV Arch component, for which \citetalias{Wakker2001} derived a $z$-bracket of 0.7-1.7 kpc and an abundance consistent with Solar.  We additionally do not recover the \citetalias{KuntzDanly1996} IV Spur S1 cloud, which appears to serve as a bridge between the IV Arch and the IV Spur S2 on the POS.  We do not observe any compelling potential visual matches to the IV Spur S1 in the set of \perch dust clouds that failed our \HI{}-matching quality cuts (see Appendix \ref{ap:failed_matches}), nor to the higher-velocity part of the IV Arch.  We therefore tentatively suggest that this higher-velocity subcomplex and the IV Spur S1 either a) have distances greater than the outer limits of the GXP dust map (1.25 kpc) and/or b) have sufficiently low dust content such that they are not detectable with current dust map sensitivity and/or c) were not identified as discrete clouds by \perch.

The origin of the IV Arch complex has been the subject of intense speculation for many years, with explanations proposed ranging from a high-$z$ supernova to remnants from a burst superbubble to infalling gas from a satellite galaxy \citep[see e.g.,][]{KuntzDanly1996}.  While near-Solar metallicities appear to rule out the stripped gas hypothesis, uncertain distances (and, by extension, uncertain masses) to the clouds making up the IV Arch have made reaching a consensus viewpoint on its origins and fate difficult.  This work's joint detection of the lower-velocity IV Arch, the LLIV Arch, and a portion of the IV Spur marks the first time the IV Arch complex has been mapped in 4D (three spatial dimensions and one velocity dimension) and so, by extension, the first time that robust masses (along with other key physical quantities such as altitude, size, density, kinematic deviation from rotation, etc.) can be derived.   

We derive a combined total mass for the Northern IV complex (the four lower-velocity IV Arch clouds excluding IV25 to avoid double-counting with Cloud 296, the LLIV Arch, and the IV Spur S2) of $(5.09 \pm 0.95) \times 10^4 \ M_\odot$.  This is significantly lower than the \citetalias{Wakker2001} total mass estimate for these features of $(2.5 - 22.2) \times 10^5 \ M_\odot$ (an estimate which includes the mass of the IV Spur S1).  We attribute the majority of this difference to the clouds being on the nearer end of their \citetalias{Wakker2001} distance brackets (since masses scale with distances squared); however, note that the mass contributed by the non-recovered portions of the IV Arch would partially reconcile this discrepancy.  This implied lower mass estimate should be considered when developing physical models for potential origins of the IV Arch complex.  We emphasize that, in creating a cloud catalog traced by both 3D dust and neutral hydrogen, we are only sensitive to dust-traced neutral gas mass; \citet{WerkRubin2019} analyzed seven blue horizontal branch stars on the edge of the IV Arch and found evidence for a substantial, coherent envelope of ionized gas moving at the same velocity as the neutral IV gas, implying a significant mass infall rate additionally carried by the warm ionized medium.

We are agnostic on the question of whether Draco is part of the IV Arch.  Although it shares a similar negative deviation velocity to clouds that we associate with the IV Arch, it is located at significantly lower altitudes with a notable separation in 3D space.  If Draco is indeed part of the IV Arch, we speculate that the [CII] detection in Draco by \citet{SchneiderOssenkopfOkada2024} marks the leading edge of the encounter between the larger IV Arch and the disk.

\citet{ONeillZucker2024} observed that the IV Arch occupies a similar position on the POS as the Northern ``Chimney'' extension of the Local Bubble, and speculated that a connection between the burst Local Bubble and the IV Arch may exist \citep[an idea also proposed by e.g.,][]{DickeyLockman1990,LallementWelsh2003, WelshSallmen2004}.  Detailed dynamical modeling of the 3D velocities of the IV Arch, IV Spur, and LLIV Arch, informed by this work's 4D maps of these structures, may prove useful in probing support for a causal relationship between these structures.

\subsubsection{Other Clouds of Interest}

Several lower-altitude clouds cross-matched to the literature IVC and cirrus catalogs are located on the shell of the Local Bubble as defined by \citet{ONeillZucker2024}.  G81.2+39.2 and Cloud 655 (containing both G139.6+47.6 and G141.1+48.0) are cirrus clouds studied by \citet{ReachKoo1994} that we find are located along the nearby, low-latitude portions of the Bubble's Chimney feature.  G249.0+73.7 is located on the Bubble's Northern closed cap, and is also referred to in the literature as Markkanen’s cloud \citep{Markkanen1979} and the North Galactic Pole Rift (NGPR); the connection of this cloud to the Local Bubble was previously noted by \citet{ONeillZucker2024} and \citet{SnowdenKoutroumpa2015}.

Clouds detected within the Magnani catalog are: 225.3-66.3, 235.9+38.2, 272.9+29.3, 295.3-36.2, 299.8-26.2, 300.1-24.5, 301.2-24.5, 3C105.0, 3C454.3, 3C75.0, CB28, G101.9-62.0, G102-27, G107.4+70.9, G112-40, G135.3+54.4, G135.4-68.7, G135.6+51.3, G139.6+47.6, G141.1+48.0, G154.7-39.8, G161.9-43.3, G187.1-52.3, G189.3-53.3, G211+63, G25.1-67.7, G344.8+24.0, G61-34, G94.8+37.6, HD210121, HSVMT24, HSVMT27, HSVMT28, IR1, IR2, IR4, MBM01, MBM02, MBM03, MBM04, MBM05, MBM11, MBM12, MBM13, MBM14, MBM15, MBM17, MBM18, MBM19, MBM21, MBM22, MBM23, MBM24, MBM25, MBM26, MBM27, MBM28, MBM29, MBM30, MBM31, MBM32, MBM40, MBM41, MBM42, MBM43, MBM44, MBM45, MBM46, MBM47, MBM48, MBM49, MBM53, MBM54, MBM55, MBM56, UT3, UT4, UT5, UT6, UT7.  We present all derived results for these clouds in our extended data tables, but defer further discussion of their detailed properties to future work.

We now describe several clouds of interest (some visualized in Figure \ref{fig:ivcs_pos} and/or interactive figures; see in particular the interactive figure at \url{https://theo-oneill.github.io/HACs_and_IVCs/moments/index.html}, which allows highlighting of specific clouds) that, to our knowledge, are unnamed in the literature.  We concentrate our descriptions on the positive deviation velocity clouds in our sample, owing to the well-known lack of positive-velocity IVCs in the literature.  Southern IVCs are also historically scarce (in comparison to the abundance of IVCs in the Northern sky concentrated around the greater IV Arch complex); e.g., \citetalias{RohserKerp2016} observed a strong North-South imbalance in their sample of MIVCs (with 3.6$\times$ as many MIVCs in the North) as well as a discrepancy in the inflowing-outflowing symmetry in the North vs. South (with the North being 98\% inflowing MIVCs and the South being 60\% inflowing).

\paragraph{Cloud 183} A Southern positive-velocity cloud with a striking head-tail morphology, with denser material concentrated towards the Galactic midplane.  This cloud was erroneously split into two clusters by our clustering algorithm, with one cluster (6 draws) passing our quality cuts and containing a trailing low-density extension towards the South, and the other smaller cluster (4 draws) failing our quality cuts.  The cloud's distance and height (centroid distance $d_c = 1184 \pm 7$ pc, centroid $z_c = -590 \pm 13$ pc) classifies it as an HAC, and its positive but moderate $v_{\rm{dev}}=11.9 \pm 3.0$ classifies it as an LVC.  This cloud is truncated by the edge of the GXP dust map at $1250$ pc.  Matching detections in the \citetalias{RohserKerp2016} MIVC catalog are concentrated towards the cloud's dense head.  This cloud's cometary morphology is reminiscent of head-tail morphologies in clouds such as the HVC Smith Cloud \citep{Smith1963} and the IVC PP Arch (W01, discussed further in \S\ref{S:nondetec}), which have both been interpreted as the result of interactions with the Galactic disk \citep[e.g.,][]{LockmanBenjamin2008,FukuiKoga2021,SheltonWilliams2022}.

\paragraph{Cloud 292} A massive Northern positive-velocity cloud ($z=682 \pm 4$ pc), consisting of an elongated clump and containing detections in the \citetalias{RohserKerp2016} MIVC catalog.  It has a moderate positive deviation velocity of $v_{\rm{dev}}=12.8 \pm 1.3$ km/s (based on comparison to the central velocity of $v_{\rm{LSR}}=3.2 \pm 1.4$ km/s), although morphological support for velocity extends to higher positive velocities ($\rm{max}(v_{\rm{LSR}})=28.3 \pm 5.7$ km/s), implying a potentially underestimated deviation from rotation.  The cloud is truncated by the edge of the GXP dust map at $d=1250$ pc, which may explain its high $N_{HI}/A_V$ measurement.  

\paragraph{Cloud 2066} A second massive Northern ($z=613 \pm 11$ pc) positive-velocity cloud, composed of a variety of clumps arranged in a jagged pattern parallel to the midplane.  The cloud has the highest positive deviation velocity in our sample ($v_{\rm{dev}}=32.5 \pm 1.9$ km/s).  This cloud was not detected in any of our compared catalogs.

\subsubsection{Foreground Clouds to non-detected IVCs}\label{S:nondetec}

A number of clouds in our catalog are located along the same sightlines as other IVCs described in the literature.  Although we do not detect these IVCs, we remark on these interloper clouds here in the hopes of making corrections for the local foreground more straightforward for future studies of these known IVCs.  

\citetalias{Wakker2001} introduced an IVC referred to as the Pegasus-Pisces Arch (PP Arch), stretching from (l,b)=(90,-40) to (l,b)=(130,-60) with velocities between $-85 \lesssim v_{\rm{LSR}} \lesssim -45$ km/s; \citet{FukuiKoga2021} studied a cloud (IVC 86-36) located within the dense head of this structure.  \citetalias{Wakker2001} derived an upper limit on distance for the northern ``knot'' within the PP arch of $<$2.7 kpc, and an upper limit on distance for the southern part of the arch of $<1.1$ kpc.  This latter limit is well within the limits of the GXP dust map ($d < 1.25$ kpc); however, we do not detect this cloud within our catalog.  Our catalog's Cloud 509 is therefore apparently located in the foreground to the PP Arch, with a velocity of $v_{\rm{LSR}}=-11.6 \pm 2.5$ km/s and distance of $d=408 \pm 1$ pc, that is clearly visible in the velocity channel map shown by \citet{FukuiKoga2021} in their study of IVC 86-36.  

\citetalias{Wakker2001} also introduced an IVC Complex gp centered near (l,b)=(50, -25), overlapping on the POS with the HVC complex GP \citep{WakkervanWoerden1991} (a complex that includes the Smith cloud, \citealt{Smith1963}).  \citetalias{Wakker2001} estimated the distance of Complex gp to be between 0.8-4.3 kpc based on stellar bracketing.  We do not detect clouds consistent with this complex's velocities of $+60 < v_{\rm{LSR}} < +90$ km/s, but do detect a number of lower-positive-velocity clouds in this area of the sky.

\section{Conclusions}\label{S:conclusions}

In this work, we identify clouds in a parsec-resolution 3D dust map of the Solar Neighborhood \citep{EdenhoferZucker2024}.  We match the morphology of these clouds on the plane-of-the-sky to the morphology of \HI 21 cm emission in the \HIPI survey \citep{HI4PICollaborationBenBekhti2016} to constrain the kinematics of our cloud sample.  Key takeaways from this work are as follows.

\begin{enumerate}
    \item We release a catalog of 1,695 clouds within 1.25 kpc of the Sun, mapped with 3D dust and identified with the topological structure-finding method \perch. Many members of the catalog correspond to well-known molecular clouds in the Solar Neighborhood. Cloud properties derived from 3D dust include factors such as distance, altitude $z$, and size. (\S\ref{S:methods}) 
    
    \item We find high-confidence morphological matches between dust and \HI{} for 519 of these clouds, and for this subset release additional properties derived from \HI{} including radial velocity, line width, and column density-to-extinction ratio (\S\ref{S:methods}).   The full cloud decomposition and catalog is available at \url{https://doi.org/10.5281/zenodo.20349203}.

    \item Three of our \HI{}-matched dust clouds have velocities consistent with being IVCs, deviating from Galactic rotation by greater than 40 km/s.  We detect no HVCs.  The detected IVCs range in $|z|$-height between 482 pc to 606 pc, and are all located in the Galactic northern hemisphere (\S\ref{S:kinematic}).  We define an additional extended sample of clouds previously referred to as IVCs in the literature, that deviate from rotation by a smaller amount.  Our total sample of clouds identified with 3D dust for the first time includes all well-known high-latitude molecular IVCs (Draco, IVC 135, IVC 210, and the IV Spur S2), making up the molecular component of the well-known IV Arch cloud complex, as well as the Low-latitude IV Arch (LLIV Arch) cloud (\S\ref{S:discuss}). 

    \item We observe a significant asymmetry in the number of high-altitude dust-traced clouds in the Northern vs. Southern Galactic hemispheres, with the asymmetry beginning for clouds with heights of $|z| > 100$ pc.  We define a sample of 17 high-altitude clouds (HACs) with $|z|$-heights greater than or equal to the minimum IVC altitude ($|z| \geq 480$ pc); at this altitude, there are $2.9 \pm 0.2$ times more clouds in the North than the South.  IVCs make up only 18\% (3/17) of the total number of clouds located at these high altitudes, suggesting that a significant fraction of clouds at the disk-halo interface are not traced by intermediate-radial-velocity gas (\S\ref{S:vertical}).

    \item We observe significant correlations between cloud properties and altitude.  Interpretation of these correlations is complicated by correlations between low-altitude cloud properties and distance, suggesting that subtle systematics may affect structures reconstructed at greater distances in 3D dust maps (\S\ref{S:corr_props}).  Despite those systematics, robust negative correlations are found between cloud altitude and the maximum neutral volume density within a cloud, the estimated external pressure confining clouds, and the estimated internal pressure.
    
    \item We additionally observe a robust positive correlation between cloud altitude and the neutral-gas-to-dust ratio $N_{HI}/A_V$, implying variations in cloud properties, environment, and/or origin with altitude (\S\ref{S:dgr}).  We derive a median gas-to-dust ratio of $3.3 \times 10^{21}$ mag$^{-1}$ cm$^{-2}$ across our full cloud sample.  We observe that our three IVCs have \dgr consistent with lower-altitude disk clouds, and that our low-velocity HACs have \dgr on average twice as high in the South as in the North.  We emphasize that nearly two dex of variation in the \dgr ratio is observed within our sample of local clouds, suggesting that the assumption of a constant gas-to-dust ratio in the Solar Neighborhood is not well-founded.

    \item  Our cloud sample generally appears gravitationally unbound (in the absence of sufficient external confining pressure), and falls in the range of expected pressures and densities typical of the cold and unstable neutral medium phases of the ISM (\S\ref{S:phases}).    

\end{enumerate}

Future work will evaluate whether it is possible to connect recent feedback events in the Solar Neighborhood to the IVCs and HACs identified in this work.  Detailed kinematic modeling and/or chemical observations of our sample of IVCs and HACs would likely prove useful in evaluating these questions.

Galactic fountain flows through the ISM and CGM play a fundamental role in driving the evolution of galaxies.  This work places the Solar Neighborhood within that context using 3D dust maps that allow us to directly constrain the distances, morphologies, and properties of high-altitude clouds.  Our results underscore the importance of low-radial-velocity, high-altitude gas in tracing the full extent of the disk-halo interface --- as has been speculated and occasionally observed for many years \citep[e.g.,][]{PeekHeiles2009,BishWerk2021}.  As 3D dust maps reach an increasing fraction of our Galaxy, we anticipate further insights into IVCs, HACs, and related structures that trace and shape the exchange of mass between the disk and halo.

\begin{acknowledgements}
We thank Gina Panopoulou, Doug Finkbeiner, Charlie Lada, and Alyssa Goodman for insightful discussions.  We thank Loris Magnani for providing a catalog of high-latitude CO detections.  T.J.O. acknowledges that this material is based upon work supported by the National Science Foundation Graduate Research Fellowship under Grant No. DGE 2140743. A.K.S. acknowledges that support for this work was provided by NASA through the NASA Hubble Fellowship grant HST-HF2-51564.001-A awarded by the Space Telescope Science Institute, which is operated by the Association of Universities for Research in Astronomy, Inc., for NASA, under contract NAS5-26555.  The authors acknowledge Interstellar Institute's program "ii7" and the Paris-Saclay University's Institut Pascal for hosting discussions that nourished the development of the ideas behind this work. C.Z. and T.J.O. acknowledge support by the NSF CAREER award \#2442546 (CAREER: Charting the Formation, Transformation, and Destinies of Gas and Young Stars in the Solar Neighborhood).  We acknowledge the use of Claude and Claude Code (Anthropic, model Claude Opus 4.7 and earlier models) in debugging and refining our pipeline and figure-generation scripts, as well as in suggesting minor corrections (spelling, grammar, punctuation, and clarity) to the text of this manuscript. 
\end{acknowledgements}

\software{Astropy \citep{astropy_2013,astropy_2018,AstropyCollaborationPriceWhelan2022}; Claude Code (Anthropic); Cmasher \citep{cmasher2020}; Healpy \citep{Zonca2019_healpy}; Matplotlib \citep{matplotlib_Hunter2007}; Numpy \citep{harris2020_numpy}; Pandas \citep{pandas_mckinney-proc-scipy-2010}.}

\paragraph{Data Availability} – All data products are available at \url{https://doi.org/10.5281/zenodo.20349203}.

\newpage
\appendix
\restartappendixnumbering

\section{Extended Methods: Inter-Draw Clustering}\label{ap:cluster}

In this appendix, we describe the technical workflow of our inter-draw cloud clustering method.  We describe the calculation of cloud similarity metrics (\S\ref{S:similarity}), the construction of a training set defining clouds that are the ``same'' vs. ``different'' structures between draws (\S\ref{S:train}), our clustering method (\S\ref{S:cv_cluster}), and our post-clustering definition of embedded substructures within clouds (\S\ref{S:embedded}).

\subsection{Similarity Metrics}\label{S:similarity}

Across all 12 draws, a total of $N_{tot} = 15,328$ structures were identified.  We must now define metrics indicating when a structure is likely the same vs. different between draws. 

To do so, we calculate three metrics of the similarity of two structures $\mathcal{S}_i$ and $\mathcal{S}_j$ identified in draws $x$ and $y$, respectively.  The first is the Jaccard index measuring the fraction of pixels on the POS shared between the two structures,
\begin{equation}
    f_{\rm{overlap},ij} = \frac{N_{\rm{pix}}(\mathcal{A}_i \cap \mathcal{A}_j)}{N_{\rm{pix}}(\mathcal{A}_i) + N_{\rm{pix}}(\mathcal{A}_j) - N_{\rm{pix}}(\mathcal{A}_i \cap \mathcal{A}_j)}.
    \label{eqn:f_overlap}
\end{equation}
where the notation $N_{\rm{pix}}(S)$ denotes the number of \Hpx pixels in the mask defined by $S$.  This yields a sparse $N_{tot} \times N_{tot}$ matrix $F_{\rm{overlap}}$ ranging between 0 and 1.  We only populate $F_{\rm{overlap}}$ for structures that are identified in different draws ($x \neq y$), to avoid matching structures in the same draw to each other; the exception is the diagonal matching each structure to itself ($i=j$), which we set to 1.

The second and third metrics are the root mean square error (RMSE) between integrated maps of $\mathcal{S}_i$ and $\mathcal{S}_j$.  To calculate these metrics, we first scale the POS ($\mathcal{A}_i$) and LOS ($\mathcal{R}_i$) maps of each structure to a 0--1 range,
\begin{equation}
    \hat{\mathcal{A}}_i(p) = \frac{\mathcal{A}_i(p) - \rm{min}(\mathcal{A}_i)}{\rm{max}(\mathcal{A}_i) - \rm{min}(\mathcal{A}_i)}
\end{equation}
\begin{equation}
    \hat{\mathcal{R}}_i(r) = \frac{\mathcal{R}_i(r) - \rm{min}(\mathcal{R}_i)}{\rm{max}(\mathcal{R}_i) - \rm{min}(\mathcal{R}_i)}
\end{equation}
We then calculate the RMSE between scaled maps,
\begin{equation}
\rm{RMSE}(\mathcal{A})_{ij} = \sqrt{\frac{1}{N_{\rm{pix}}(\mathcal{A}_i \vee \mathcal{A}_j)}\sum_{p=1}^{N_{\rm{pix}}(\mathcal{A}_i \vee \mathcal{A}_j)} (\hat{\mathcal{A}}_i(p) - \hat{\mathcal{A}}_j(p))^2}    
\end{equation}
\begin{equation}
\rm{RMSE}(\mathcal{R})_{ij} = \sqrt{\frac{1}{N_{dr}(\mathcal{R}_i \vee \mathcal{R}_j)}\sum_{r=1}^{N_{dr}(\mathcal{R}_i \vee \mathcal{R}_j)} (\hat{\mathcal{R}}_i(r) - \hat{\mathcal{R}}_j(r))^2}    
\end{equation}
for all structure pairs with $f_{\rm{overlap},ij} > 0$. 

\begin{figure}
    \centering
\includegraphics[width=0.48\textwidth]{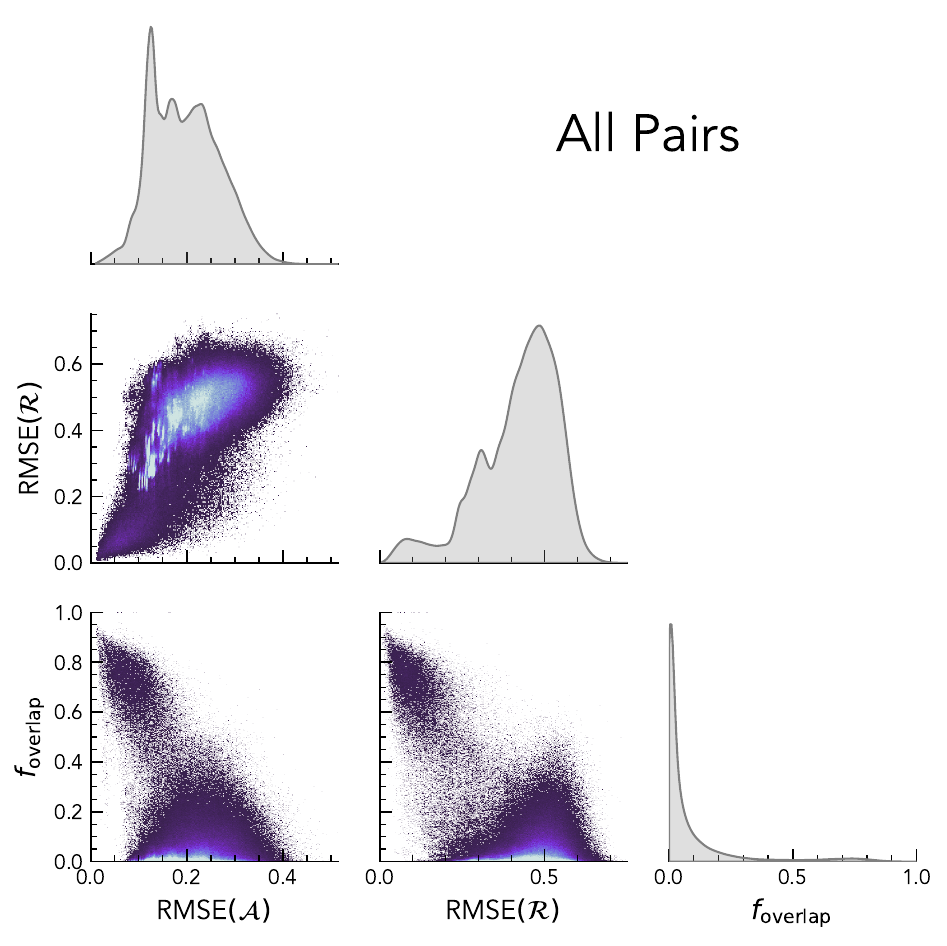}
\includegraphics[width=0.48\textwidth]{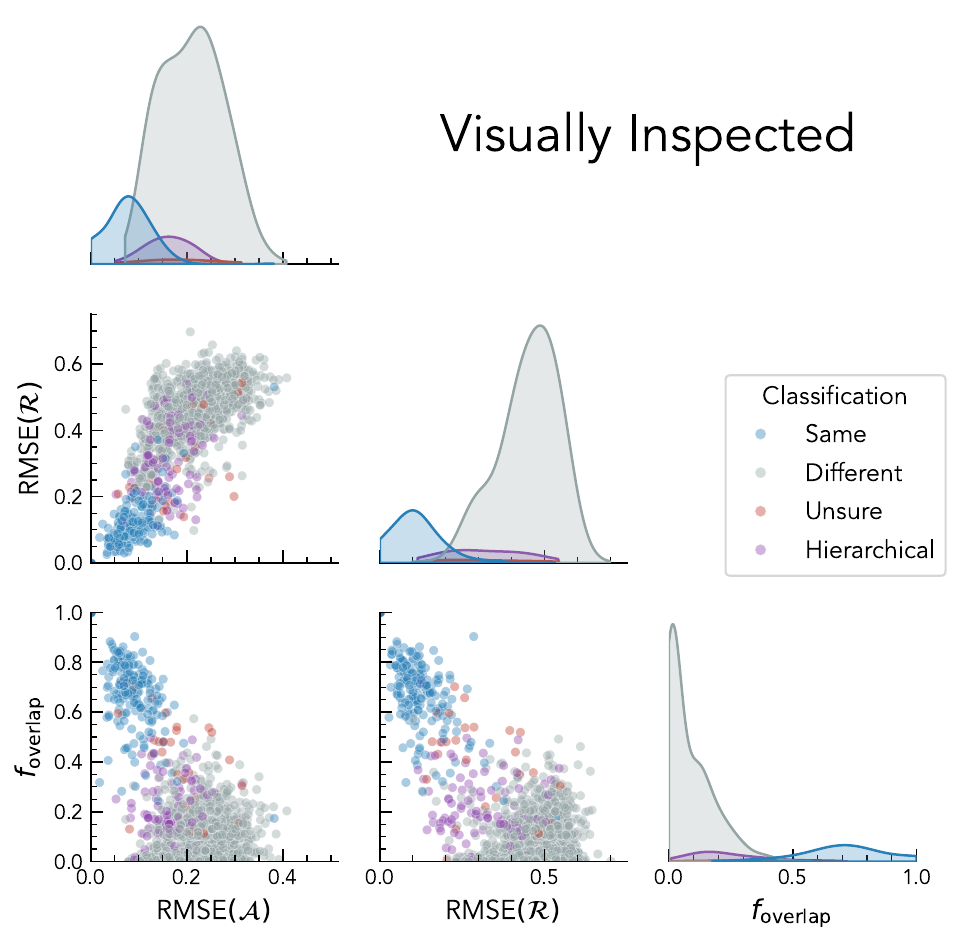}
    \caption{\textit{Top:} Corner plot comparing, for all pairs of overlapping clouds between draws of the dust map, the distributions of cloud-cloud similarity metrics: similarity along the LOS (RMSE$(\mathcal{R})$) (where $\mathcal{R}$ is radially-integrated extinction as defined in Eqn. \ref{eqn:integrated_Ri}), similarity on the POS (RMSE$(\mathcal{A})$) (where $\mathcal{A}$ is POS-integrated extinction as defined in Eqn. \ref{eqn:integrated_Ai}), and fraction of pixels overlapping on the POS ($f_{\rm{overlap}}$).  \textit{Bottom:} As top, but for the subset of visually inspected and labeled cloud pairs.  Pairs labeled as the same cloud are shown in blue, different in grey, unsure in red, and hierarchical variants in purple.}
    \label{fig:similarity_corner}
\end{figure}

We have now defined 3 metrics comparing the similarity of structures between draws: in POS overlap fraction ($F_{\rm{overlap}}$), in POS similarity ($\rm{RMSE}(\mathcal{A})$), and in LOS similarity ($\rm{RMSE}(\mathcal{R})$).  Figure \ref{fig:similarity_corner} presents the distributions of our metrics within all overlapping pairs.

\subsection{Training Set Construction}\label{S:train}

We then construct a set of visually-inspected pairs of clouds that are the same vs. different between draws.  To do so, we randomly select 1,000 pairs of structures with $f_{\rm{overlap},ij} > 0$.  We supplement this with an additional 500 structures with $f_{\rm{overlap},ij} > 0.1$ in order to achieve greater balance in our training set between structures that are the same vs. different (since different structures tend to have lower overlap fractions, and represent the vast majority of our pair population).  This set of 1,500 pairs represents 0.1\% of the total number of pairs with $f_{\rm{overlap},ij} > 0$.  

We then visually inspect $\mathcal{A}_i$ and $\mathcal{A}_j$ for each pair, and label it as either being the [1] the same structure or [0] different structures.  Additional options included are being [-1] unsure if it is the same structure, or [-2] a hierarchical variant where one structure is contained within the other.  In total, 77.7\% of inspected pairs were labeled as being different, 14.1\% as the same, 1.7\% as unsure, and 6.5\% as hierarchical.  

Figure \ref{fig:similarity_corner} additionally presents the distributions of our metrics within the labeled set of clouds.  Pairs labeled as being the same tend to have high $F_{\rm{overlap}}$ and low $\rm{RMSE}(\mathcal{A})$) and $\rm{RMSE}(\mathcal{R})$.  Pairs labeled as different tend to have low $F_{\rm{overlap}}$ and high $\rm{RMSE}(\mathcal{A})$) and $\rm{RMSE}(\mathcal{R})$, while unsure and hierarchical pairs exist in the transition region between the same and different classes.  

\subsection{Clustering}\label{S:cv_cluster}

We use these labeled pairs to construct a weighted similarity metric incorporating our three individual metrics using logistic regression.  We first relabel unsure clouds to be included in the different class (class = 0), and hierarchical clouds to be included in the same class (class = 1).  We then split our pairs into a 70\% training set and 30\% test set with assignments stratified by class.  

Within our training set, we perform stratified K-fold cross validation (with $K=10$) to learn the optimal regularization strength $C$ for a logistic regression model with L2 regularization (implemented in {\tt sklearn}).  Within each fold (consisting of a training and validation subset), we perform a grid search cross-validation (also with $K=10$) on the training subset to determine the best $C$.  We construct a logistic regression model using the best $C$,
\begin{equation}
    p(\rm{same})_{ij} = \sigma(c_1 \ \rm{RMSE}(\mathcal{A})_{ij} + c_2 \ \rm{RMSE}(\mathcal{R})_{ij} + c_3 \ f_{\rm{overlap},ij} + c_4)
\end{equation}
where $\sigma$ is the sigmoid function.  Using the best-fit $c_i$'s derived from this model, we now have a weighted similarity metric with which we can derive cross-draw structure clusters.  

To do so, we construct a distance matrix encoding the distance between all pairs in our structure sample, $d_{ij} = p(\rm{different})_{ij} = 1 - p(\rm{same})_{ij}$, where $0 \leq d_{ij} \leq 1$.  We fill non-overlapping and same-draw structures ($f_{\rm{overlap},ij}=0$) with a large invalid value (1,000) so that they cannot be clustered together, and ensure that the diagonal of the distance matrix ($i=j$) is equal to 0 to indicate that there is no distance between a structure and itself.

We then construct a complete-linkage dendrogram from our distance matrix using {\tt scipy.cluster.hierarchy.linkage}.  Complete-linkage clustering assigns distances between two clusters as the maximum distance between an individual member of one cluster to an individual member of the other cluster; unlike other clustering strategies, this allows us to prevent duplicate-draw cluster members.  We now need to select a connectivity threshold from which to define flat clusters of structures from the linkage hierarchy (using {\tt scipy.cluster.hierarchy.fcluster}), with distances between clusters no greater than the selected connectivity threshold.  To do so, we test a variety of potential thresholds from which to define clusters, and for each threshold calculate the accuracy and $F_1$ score for both the training and validation subsets,
\begin{equation}
    \rm{accuracy} = \frac{\rm{TP + TN}}{\rm{TP + TN + FP + FN}}
\end{equation}
\begin{equation}
    \rm{F_1} = \frac{2 \ \rm{TP}}{2\ \rm{TP}+\rm{FP}+\rm{FN}}
\end{equation}
where TP is the number of true positives, TN is the number of true negatives, FP is the number of false positives, and FN is the number of false negatives.  We select the threshold that maximizes the $F_1$ score for the validation subset.  We use the $F_1$ score because our labeled pairs are imbalanced between the same and different classes, and wish to avoid over-inflating our confidence in the performance of our model or neglecting the importance of identifying the ``same'' class.  We record the optimal connection threshold and regularization strength resulting from the fold.

\begin{figure}
    \centering
\includegraphics[width=0.48\textwidth]{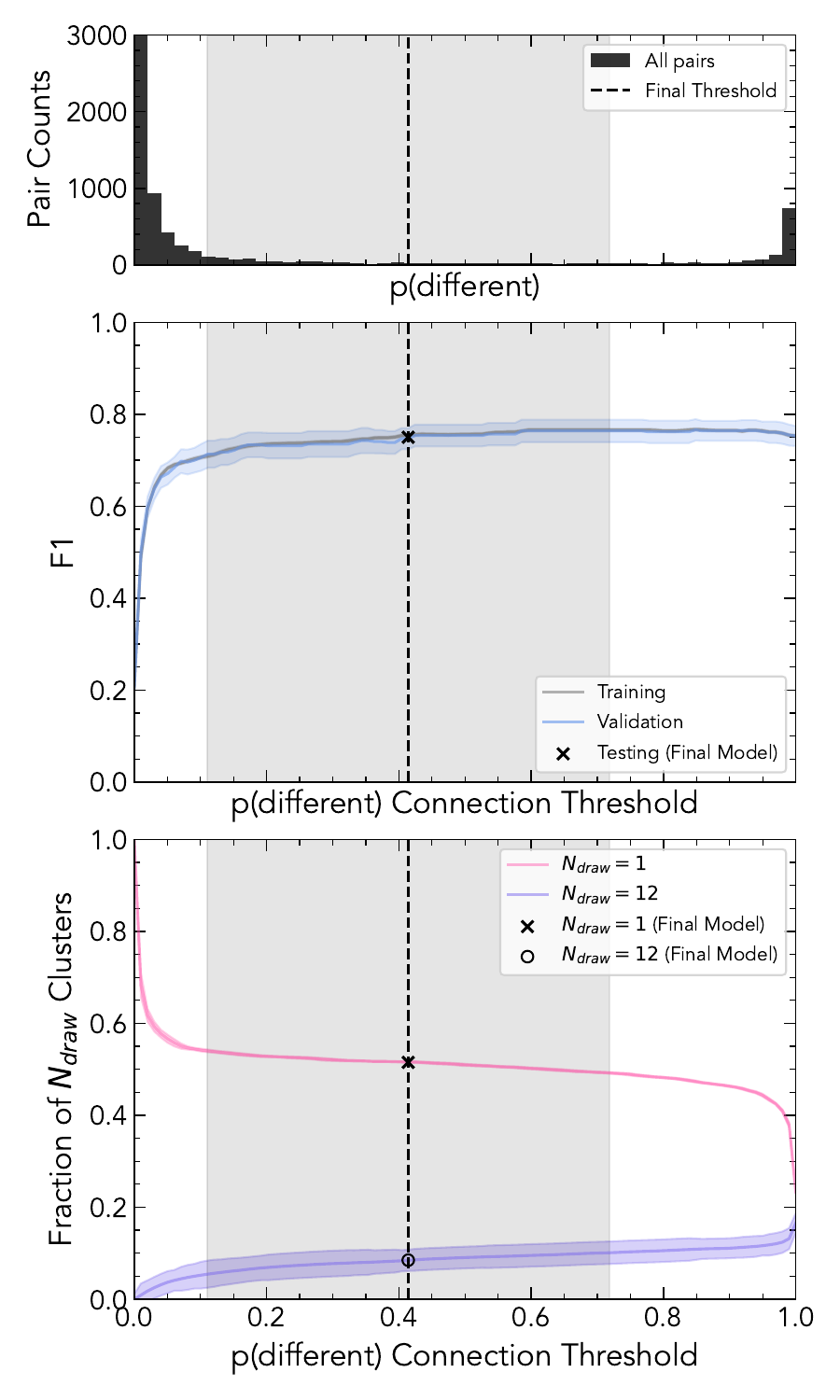}
    \caption{\textit{Top:} Distribution of distance metric between pairs of overlapping structures, p(different), for all cloud pairs.  Final selected distance threshold, used to determine cluster membership, is shown by the black vertical dashed line, with standard deviation across the cross-validated pipeline shown by the black shaded region.  \textit{Center:} F1-score as a function of p(different) connection threshold for the training set (grey curve) and validation set (blue curve), with standard deviations shown by the shaded regions.  The black x marks the F1 score obtained by applying our final model parameters to our held-out testing set.  \textit{Bottom:} Fraction of clusters with $N_{\rm{draw}}=$1 (pink curve) and $N_{\rm{draw}}=12$ (purple curve) as a function of p(different) connection threshold.  The black x and black open circle respectively show the $N_{\rm{draw}}=$1 and $N_{\rm{draw}}=12$ fractions for our final model.}
    \label{fig:clustering_thresh}
\end{figure}

We repeat this process for each of our $k=10$ folds, and then construct a final logistic regression model using the median optimal regularization strength $C=61.68$.  This final model has parameters $c_1=-8.542, c_2=-8.972, c_3=7.304, c_4=1.681$.  The distribution of p(different) obtained from this model is shown in Figure \ref{fig:clustering_thresh}, with most pairs being relatively similar (low p(different)) and a small tail having unambiguous differences (high p(different)).  Figure \ref{fig:clustering_thresh} also shows the mean and standard deviation of F1 scores for the training subsets and validation subsets as a function of p(different) connection threshold, along with the performance of the final model.  The figure additionally shows the fraction of 1-draw-structure and 12-draw-structure clusters resulting from the tested thresholds.  These latter metrics are useful in determining if the pairs have been over- vs. under- clustered.  We observe that most curves are relatively slow-changing between a threshold of p(different)$\simeq0.3-0.9$, suggesting that our choice of specific threshold within this range is relatively unimportant to the ultimate clustering results. 

We then cluster our pairs with the median optimal connection threshold across folds of $0.414$ (with a standard deviation of $0.304$).  At this threshold, the held-out 30\% test set achieves a clustering accuracy of 91.6\% and F1 score of 75.0\% relative to our visually-inspected pair labels.  The high standard deviation of our optimal connection threshold reinforces our interpretation that a wide range of possible connection thresholds are valid choices yielding very similar final clustering results. 

As a final check, we ensure that our choice of seed to make our initial training/test split has no significant impact on our downstream clusters by testing 10 random seeds through our full pipeline and calculating the Adjusted Rand Index (ARI) between final cluster assignments \citep{hubert1985comparing} as implemented in {\tt sklearn},
\begin{equation}
    ARI = \frac{\rm{RI - E(RI)}}{\rm{max(RI)-E(RI)}}
\end{equation}
where $RI$ is the Rand Index,
\begin{equation}
RI = \frac{TP + TN}{TP + FP + FN +T N}
\end{equation}
and $E(RI)$ is the expected value of the $RI$.  The ARI quantifies the similarity of two clustering assignments and has a maximum score of 1, indicating complete agreement between the two clusterings.  The median ARI between our pairs of random seeds is 0.984, indicating that the choice of this seed has essentially no meaningful effect on our cluster choices.

\begin{figure}
    \centering
\includegraphics[width=0.48\textwidth]{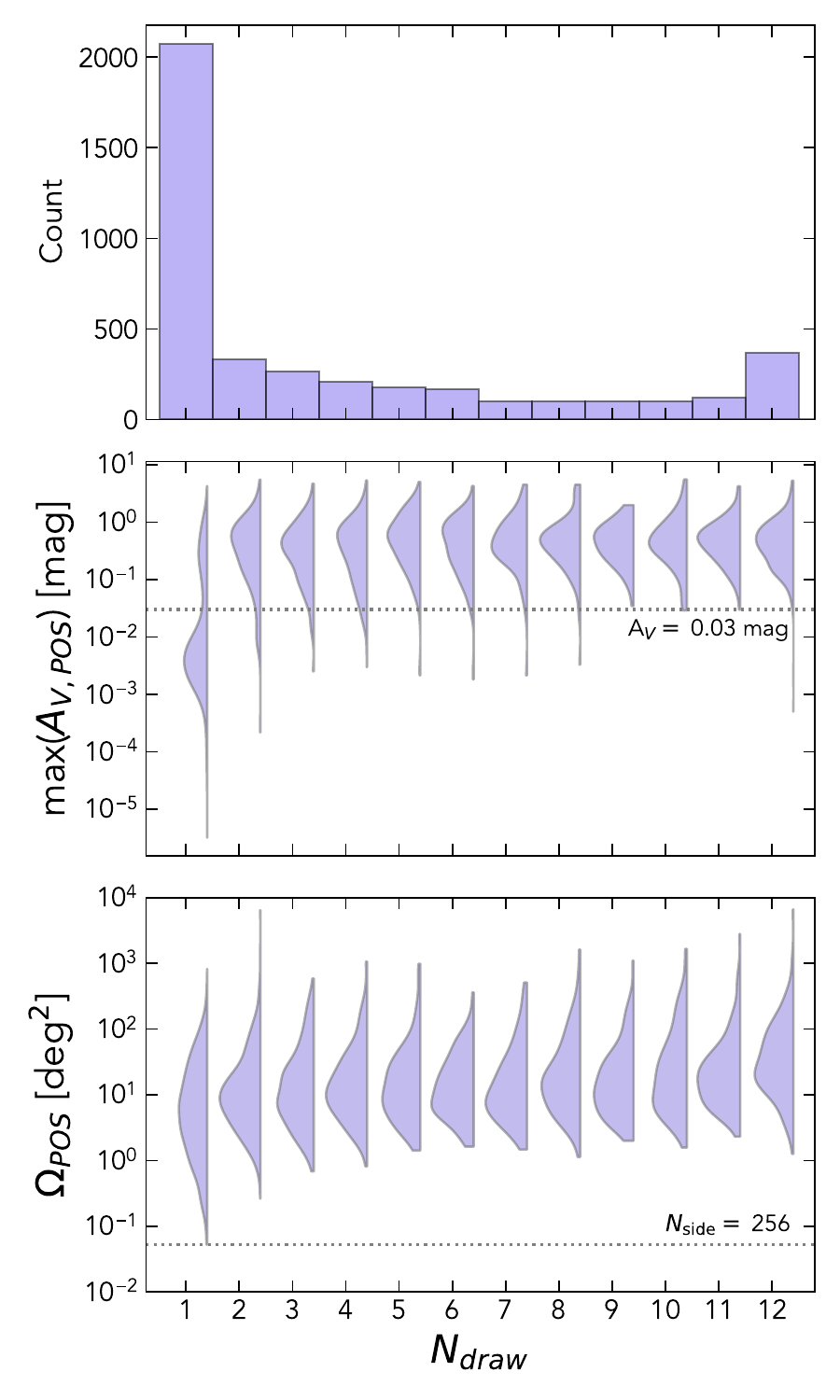}
    \caption{\textit{Top:} Histogram of the distribution of clusters with $N_{\rm{draw}}$ members, i.e., with structures appearing in $N_{\rm{draw}}$ draws of the GXP dust map.  \textit{Center:} Kernel density estimations (KDEs) of the distributions of the (median across draws) maximum $A_V$ extinction on the POS, max$(A_{\rm{V,POS}})$, vs. $N_{\rm{draw}}$.  The black horizontal dotted line shows our estimated sensitivity of the GXP dust map, $A_V = 0.03$ mag.  \textit{Bottom:} As center, but for (median across draws) area on the POS $\Omega_{\rm{POS}}$.  The black horizontal dotted line shows the area on sky at the pixelation scale of the dust map (\Hpx \nside{256}).}
    \label{fig:cluster_n_draws}
\end{figure}

Having performed this check, we now describe our initial clustering results under the model parameters described above.  A total of 4,195 clusters are initially identified.  We recall that, as described in \S\ref{S:perch}, the median number of significant structures identified within each individual draw of the dust map was 1276.5 structures.  The majority of our initially-identified clusters ($51.5\%$) contain only one structure identified in a single draw ($N_{\rm{draw}}=1$).  We attribute this apparent over-fragmenting primarily to two factors: the hierarchical structure-finding implicit to \perch combined with our requirement that each cluster contain at most one structure from each draw, and the generous threshold used to define significant structures in individual draws.    

For each cluster, we find its nearest-neighbor cluster by finding the cluster with the smallest $p(\rm{different})$ between overlapping cross-cluster structure pairs; this is essentially the single-linkage distance between clusters, instead of complete-linkage.  We then classify the cluster as falling into one of the following categories as to why it did not merge with its nearest neighbor:
\begin{itemize}
\item Isolated: the cluster has no overlapping structures in other clusters, and so has no nearest-neighbor cluster (0.55\% of clusters)
\item Merge candidate: the single-linkage distance is less than our fiducial connection threshold of 0.414 (12.09\% of clusters)
\item Separated: the single-linkage distance is greater than 0.414 (22.34\% of clusters)
\item Blocked: the clusters both have structures from the same draw(s) of the dust map
\end{itemize}
We further subdivide the blocked category into:
\begin{itemize}
    \item Blocked by hierarchy: all of the same-draw structures exist in the same \perch hierarchy tree (parent/child/descendants in common) (27.20\% of clusters)
    \item Blocked, mixed: some of the same-draw structures are hierarchically-related (3.60\% of clusters)
    \item Blocked, independent: all of the same-draw structures are not related in the \perch hierarchy (34.23\% of clusters)
\end{itemize}
The large fraction of clusters completely or partially blocked from merging by same-draw hierarchical structures (30.8\% in total across hierarchy and mixed categories), strongly suggests that future work should take into account hierarchical relationships between structures as part of the clustering process, rather than as a post-processing diagnostic as used in this initial work.  

Among the dominant $N_{\rm{draw}}=1$ category, though, we find that same-draw hierarchical blocks are less of an issue (representing 10.1\% of single-draw clusters), with the vast majority of these solo clusters being instead comprised of low-significance, noise-driven structures.  We gain this intuition by estimating the minimum dust extinction to which the GXP dust map is sensitive as 30 mmag.  We estimate this limit by first imposing a quality cut on the ZGR23 extinction catalog that was the input to the GXP map, requiring that the reduced chi-squared be less than 1.5. Then, we find that the lowest one percentile of the formal extinction uncertainties on the ZGR catalog is 15 mmag $A(V)$, which corresponds to the sight lines with the highest extinctions. This uncertainty reflects the internally-estimated systematic uncertainty associated with model misspecification relative to the data. However, \citet{EdenhoferZucker2024} found that the formal extinction uncertainties from the ZGR catalog had to be inflated by $\sim2\times$ over a wide range of extinction in order to obtain a dust map with residuals consistent with a normal distribution (see their Figure D.1). As such, we multiply the lower limit on the formal extinction uncertainty from ZGR by $2\times$ to obtain the reasonable estimate of dust sensitivity limit of 30 mmag $A(V)$. It would be difficult to imagine that the GXP map is sensitive to dust extinctions below this value due to the systematic uncertainties in the input dust catalog, especially at high Galactic latitudes where the source density is lowest. 

Among our $N_{\rm{draw}}=1$ clusters, 74.1\% have maximum $A_V$ on the POS less than 30 mmag.  The breakdown of these structures becomes clearer by category: the classes of clusters that are isolated, independently blocked, or appropriately separated are all $>80\%$ composed of $A_V < 0.03$ mag structures.  In contrast, structures that are blocked by hierarchy are only 15\% low-$A_V$, while merge candidates are 33\% low-$A_V$.  This suggests that $N_{\rm{draw}}=1$ structures in the isolated, independently blocked, and appropriately separated classes are low-significance structures identified by \perch and correctly segregated into isolated clusters.  

We then turn our attention to addressing the merge-candidate clusters.  These merge-candidate structures were blocked from merging into their best-match cluster due to complete linkage's max-pair-distance thresholding requirement, despite having no draw conflict and potentially being similar to most members of a given cluster.  We perform a merging pass where we allow each structure to have the opportunity to merge into another cluster if a) they share no common draws and b) the merge would lower the mean $p(\rm{different})$ by at least $\geq 0.01$.  We assign $N_{\rm{draw}}=1$ clusters an initial neutral $p(\rm{different})=0.5$ to allow them the ability to participate in this merging.  We allow each structure to move at most one time, in order to prevent cyclical migrations, and then perform iterations until no further migrations occur (converging after 6 passes).  292 of the total 15,328 structures (1.9\%) are migrated to a better-matching cluster by the final pass.  This process slightly reduces our total number of clusters to 4,098 clusters, with a slight reduction in the number of $N_{\rm{draw}}=1$ clusters and slight increase in $N_{\rm{draw}}=12$ clusters.  

The final distribution of cluster sizes post-merging is shown in Figure \ref{fig:cluster_n_draws}.  $N_{\rm{draw}}=1$ clusters are still dominant, largely as a result of the previously-discussed low-extinction structures and hierarchical blocks.  Figure \ref{fig:cluster_n_draws} also shows the distributions of  maximum extinction on the POS and structure area on the POS as a function of $N_{\rm{draw}}$.  We see that the $N_{\rm{draw}} = 1$ cluster distributions have extended tails with very low extinctions (well-below $A_V = 0.03$ mag, our estimated reliability limit of the GXP dust map described above) and smaller average areas on the POS (truncating at the pixelation scale of a \Hpx \nside{256} map).  Since the draws of the GXP dust map are not independent (as described in \S\ref{S:dustmap}), we require that clusters have $N_{\rm{draw}} \geq 3$ appearances across draws (i.e., the structure represented by the cluster appears in at least 25\% of draws) to be included in our final cluster catalog.  This yields a truncated 1,695 clusters for subsequent analysis. 

Note that cluster IDs are initially assigned by our clustering method in the order in which independent subtrees are encountered within the dendrogram in a depth-first search, and then (after our one-way migration step) relabeled to be contiguous within our range of the total number of post-migration clusters.  Our cluster IDs are therefore unique identifiers, but contain no intrinsic information about size, position, or any other physical property.  Note, however, that since IDs were initially assigned in a depth-first manner, sibling subtrees with a parent that merges just above our selected connectivity threshold and were not merged in our migration step will have consecutive cluster IDs.  As a result, the handful of clusters in our catalog that by-eye appear to be oversplit may tend to have adjacent ID numbers (e.g., Clouds 253 and 245, Clouds 331 and 332).  This is especially relevant for a small number of clouds which appear to be over-split with disjoint draws, causing them to be undetected as embedded children.

\subsection{Definition of Embedded Clouds}\label{S:embedded}

Within each draw of the GXP dust map, \perch defines a hierarchical parent/child relationship between structures.  When working with our cross-draw clusters of clouds, we must redefine a final hierarchical structure across all draws that allows us to identify repetitive sub-structures within clouds, in order to avoid downstream biases incurred by ``double-counting'' duplicate clouds.

To do so, for each cluster in our $N_{\rm{draw}}\geq 3$ subset, we aggregate \perch hierarchical ancestry across draws.  We label each cluster with one of the following categories:
\begin{itemize}
    \item as being the \textit{morphologically-similar child} of another cluster if 1) at least 50\% of shared draws between the two clusters share an ancestry tree, 2) the $p(\rm{different})$ between members of the two clusters is $\leq 0.5$, and 3) the cluster has no morphologically-duplicated children of its own.  This category is 13.7\% of our cluster catalog.
    \item as being the \textit{parent} of a morphologically-similar cluster, with no parents of its own higher in the ancestry tree.  This represents 10.0\% of our cluster catalog.
    \item as being in the \textit{middle} of a cluster hierarchy, having both morphologically-similar parent(s) and child(ren).  This represents 5.4\% of our catalog.
    \item as being a \textit{standalone} cluster if it has no morphologically-similar parents or children and is not a superstructure (defined below).  This represents 70.3\% of our catalog.
    \item as being a \textit{superstructure} if it has $\geq 50$ \perch-descendant clusters (ignoring morphological similarity).  This represents 0.5\% of our catalog.
\end{itemize}      

We attempt \HI{}-matching on clusters of all categories, but on occasion throughout this work exclude morphologically-similar children (including ``middle'' children) from analysis that requires a notion of a ``unique'' cloud (e.g., our North-South cloud asymmetry analysis in \S\ref{S:vertical}).

\restartappendixnumbering
\section{Extended Methods: Morphological Matching}\label{ap:morphological}

\subsection{Adaptive Smoothing Scale}\label{ap:fwhm}

\begin{figure*}
    \centering
\includegraphics[width=\textwidth]{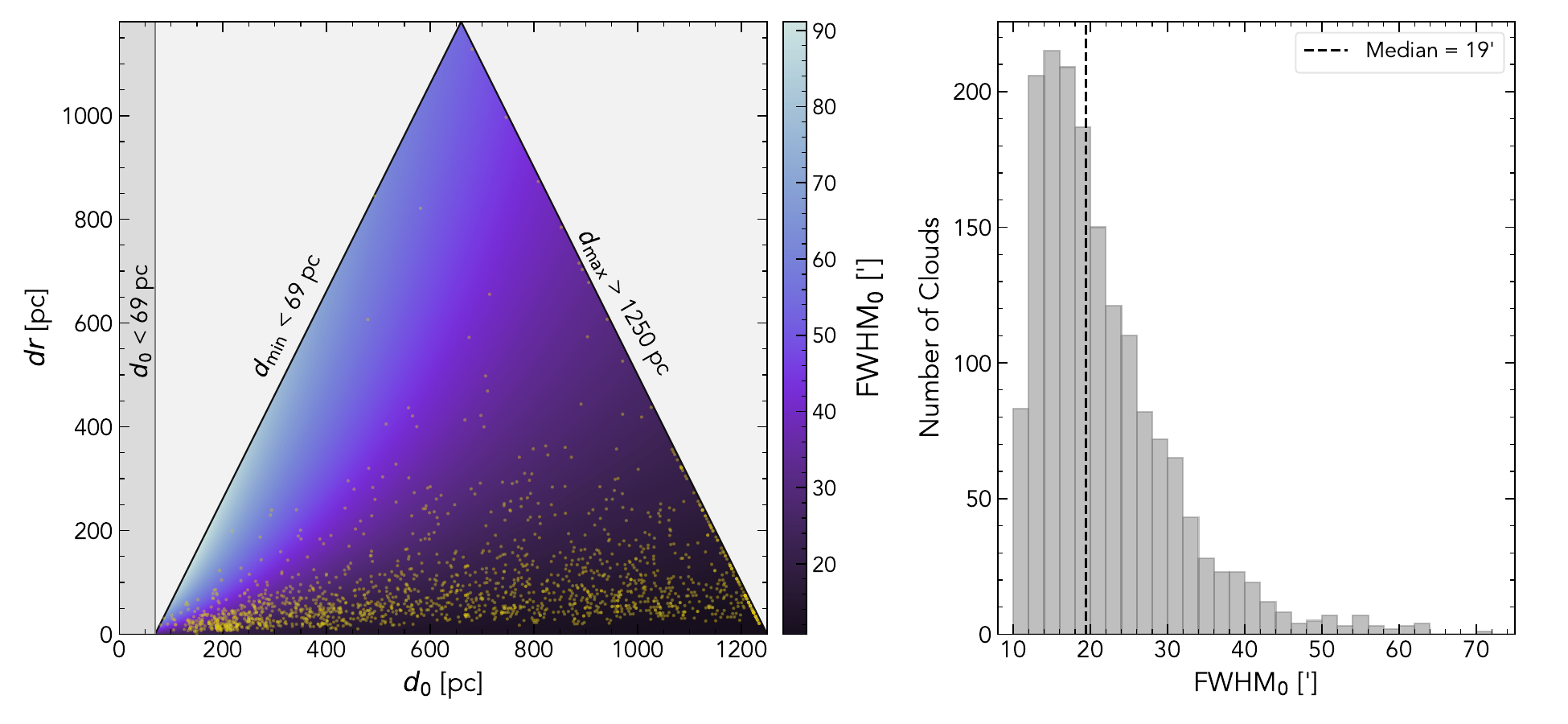}
    \caption{\textit{Left:} Effective FWHM within the GXP dust map, FWHM$_0$, as a function of cloud central distance $d_0$ and depth along the LOS $dr$.  Regions of this parameter space that are inaccessible within the dust map distance limits ($69 < d < 1250$ pc) are shaded in gray.  The \perch-identified clouds are marked by the yellow circles.  \textit{Right:} Histogram of the effective FWHM for the sample of \perch clouds.  The median FWHM$_0$ is marked by a black line.}
    \label{fig:dust_kernel}
\end{figure*}

In order to perform our morphological matching, we require that the HI4PI HI data and our integrated maps of clouds extinctions, $\mathcal{A}_i$, have the same angular resolution.  As described in \S\ref{S:hi4pi}, the native angular resolution of HI4PI is 16.2'.  However, the effective angular resolution of the GXP dust map is not constant throughout the volume of the map.  This is the result of the prior kernel used to construct the map as part of the Gaussian process-based inference.  The prior kernel, $\mathcal{K}$, is defined in 3D space; the resulting angular resolution of the map is consequently not uniform with distance.  

We must therefore estimate the effective angular resolution within each cloud's volume in order to faithfully correct for the effects of angular resolution differences between the HI and 3D dust.  Although the shape of the prior kernel itself is not Gaussian, we then use these estimated resolutions to smooth the \HIPI data with a Gaussian beam with the same FWHM as the prior kernel.  This allows us to approximate the angular resolution of each cloud on the POS.  

We estimate the angular resolution for each cloud by calculating the full-width at half maximum (FWHM) of $\mathcal{K}$ at all distances within the volume of the map assigned to the cloud by \perch.  A given cloud exists between some minimum and maximum distance, $d_{\rm{min}}$ and $d_{\rm{max}}$.  For each draw of a cloud, we define its geometric average distance,
\begin{equation}
    d_0 = \frac{1}{2}(d_{\rm{max}} + d_{\rm{min}}),
    \label{eqn:d_0}
\end{equation}
and LOS depth,
\begin{equation}
    dr = d_{\rm{max}}-d_{\rm{min}}.
    \label{eqn:dr}
\end{equation} 
We integrate the 3D kernel over 128 discrete points centered at $d_0$ and spanning between $d_0 - dr/2$ and $d_0 + dr/2$ (i.e., between $d_{\rm{min}}$ and $d_{\rm{max}}$).  This yields the effective 2D angular kernel for the prior on this draw of the cloud.  We then calculate the median FWHM across draws of the integrated kernel, FWHM$_{0}$.  

Figure \ref{fig:dust_kernel} shows the range of FWHM$_{0}$ over a grid of $69 \leq d_0 \leq 1250$ pc and $0 \leq dr \leq 1250$ pc.  The masked regions of the grid representing combinations of $d_0$ and $dr$ that are not sampled by the map (pairs with $d_{\rm{min}} < 69$ pc or $d_{\rm{max}} > 1250$ pc) create the triangular shape of this parameter space.  Within the volume of allowable $d_0$ and $dr$, possible FWHM$_{0}$ range between 10\farcm8 and 91\farcm2.  Note that the \Hpx \nside{256} grid has pixels of angular size 13\farcm7.

For each cloud identified in this work, we calculate its FWHM$_{0}$ based on the cloud's mean $d_{\rm{min}}$ and mean $d_{\rm{max}}$ across draws.  The positions of the cloud clusters in ($d_0, dr$) space are marked by the yellow points on the FWHM$_{\rm{prior}}$ grid.  The distribution of their effective FWHM$_{0}$ are additionally summarized in the histogram in Figure \ref{fig:dust_kernel}, ranging between 10\farcm9 and 71\farcm8 with a median of 19\farcm4.

\subsection{Definition of the SSIM}\label{ap:ssim}

The SSIM is defined as the product of three terms comparing local regions of two images, $x$ and $y$,
\begin{equation}
    \rm{SSIM}(x,y) = [l (x,y)]^\alpha \cdot [c(x,y)]^\beta \cdot [s(x,y)]^\gamma
\end{equation}
where $l$(x,y) represents similarity in \textit{luminance}, $c$(x,y) represents similarity in \textit{contrast}, and $s$(x,y) represents similarity in \textit{structure}.

The luminance term evaluates the similarity of the images' mean intensities as a proxy for similarity in overall perceptual brightness, and is mathematically represented as,
\begin{equation}
    l(x,y) = \frac{2 \mu_x \mu_y + C_1}{\mu_x^2 + \mu_y^2 + C_1}
\end{equation}
where $\mu_x$ and $\mu_y$ are the means of pixels in Images $x$ and $y$, respectively, and $C_1$ is a small constant introduced to stabilize $l(x,y)$ in the event of small means.  For images $x$ and $y$ that range between $\rm{Im} \in [0, L]$ , the minimum value of $l(x,y)$ will occur when $\mu_x = 0$ and $\mu_y = L$ (or vice versa) \citep{NilssonAkenineMoller2020}, and the maximum when $\mu_x=\mu_y$.  This term is then bounded to values $l(x,y) \in \left[\frac{C_1}{L^2 + C_1}, 1 \right] \rightarrow [0,1]$ when $C_1 \rightarrow 0$.

The contrast term considers the similarity of the images' standard deviations, and is defined as,
\begin{equation}
    c(x,y) = \frac{2 \sigma_x \sigma_y + C_2}{\sigma_x^2 + \sigma_y^2 + C_2}
\end{equation}
where $\sigma_x$ and $\sigma_y$ are the standard deviations of pixels in Images $x$ and $y$, respectively, and $C_2$ is a small stabilizing constant.  For images $x$ and $y$ that range between $\rm{Im} \in [0, L]$ , it can be shown that the variance $\sigma^2 \leq (L/2)^2$.  The minimum value of $c(x,y)$ will then occur when $\sigma^2_x = 0$ and $\sigma^2_y = (L/2)^2$ (or vice versa) \citep{NilssonAkenineMoller2020}, and the maximum when $\sigma_x=\sigma_y$.  This term is then bounded to values $c(x,y) \in \left[\frac{C_2}{(L/2)^2 + C_2}, 1 \right] \rightarrow [0,1]$ when $C_2 \rightarrow 0$.

The structure term encodes the similarity in local variations in the images, and is defined as,
\begin{equation}
    s(x,y) = \frac{\sigma_{xy} + C_3}{\sigma_x \sigma_y + C_3}
\end{equation}
where $\sigma_{xy}$ is the covariance between pixels in Images $x$ and $y$, respectively, and $C_3$ is a small stabilizing constant.  For images $\rm{Im} \in [0, L]$ , it can be shown that the covariance $|\sigma_{xy}| \leq (L/2)^2$.  The minimum value of $s(x,y)$ will then occur when $\sigma_{xy} = -(L/2)^2$ and $\sigma^2_x = \sigma^2_y = (L/2)^2$ \citep{NilssonAkenineMoller2020}, and the maximum when $\sigma_{xy}=\sqrt{\sigma_x^2 \sigma_y^2}$.  This term is then bounded to values $s(x,y) \in \left[\frac{-(L/2)^2 + C_3}{(L/2)^2 + C_3}, 1 \right] \rightarrow [-1,1]$ when $C_3 \rightarrow 0$.  Note than when $C_3 = 0$, $s(x,y)$ is simply the Pearson correlation coefficient.

In practice, as suggested by \citet{WangBovik2004}, most users of the SSIM set $\alpha = \beta = \gamma = 1$ and $C_3 = C_2/2$.  In this limit, the SSIM simplifies to,
\begin{equation}
    \rm{SSIM}(x,y) = \frac{(2 \mu_x \mu_y + C_1)(2 \sigma_{xy} + C_2)}{(\mu_x^2 + \mu_y^2 + C_1)(\sigma_x^2 + \sigma_y^2 + C_2)}
\end{equation}
The stabilizing constants are typically chosen as $C_1 = (K_1 L)^2$ and $C_2 = (K_2 L)^2$, with $K_1 = 0.01$ and $K_2 = 0.03$; we adopt these default values.  When $C_1 \rightarrow 0$ and $C_2 \rightarrow 0$, the SSIM $\in [-1, 1]$.  Higher values of the SSIM correspond to greater morphological similarity. 

The SSIM is evaluated locally within regions of images, typically with a sliding window or within a Gaussian beam, and then pooled (typically via computing the mean) to yield an overall measure of the images' similarity.  In our implementation on a \Hpx{} grid, we define local regions as the set of pixels within a disk within an angular radius of 5$\sigma_{\rm{cloud}}$ (where $\sigma_{\rm{cloud}} = \rm{FWHM}_0 / (2 \sqrt{2 \log 2})$, see Appendix \ref{ap:fwhm} for the definition of $\rm{FWHM}_0$) , queried using the {\tt healpy} function {\tt query\_disc}.  We then calculate the angular separation $\varphi$ between the central \Hpx{} pixel and each neighbor using the {\tt astropy} function {\tt separation}.  We derive Gaussian weights based on angular separation as $w_i = \exp(-\varphi^2 / 2\sigma^2)$, where $\sigma$ is the width of the Gaussian beam.  We find that $\sigma =  \sigma_{\rm{cloud}}$ strikes an appropriate balance between a large enough smoothing scale to overcome noise while preserving local signals (and, confirm that our results are essentially identical if a choice of $\sigma = 2 \sigma_{\rm{cloud}}$ were instead used).

We then calculate the metrics entering into the SSIM using these weights, e.g.,
\begin{equation}
    \mu_x = \frac{\Sigma_i x_i w_i}{\Sigma_i w_i}
\end{equation}
\begin{equation}
    \sigma_x^2 = \frac{\Sigma_i w_i (x_i - \mu_x)^2}{(\Sigma_i w_i)- (\Sigma_i w_i^2)/(\Sigma_i w_i)}
\end{equation}
\begin{equation}
    \sigma_{xy} = \frac{\Sigma_i w_i (x_i - \mu_x)(y_i - \mu_y)}{(\Sigma_i w_i)- (\Sigma_i w_i^2)/(\Sigma_i w_i)}
\end{equation}
where $\sigma_x$ and $\sigma_{xy}$ are calculated using the unbiased forms of these weighted statistics.

We calculate the local SSIM for each \Hpx{} pixel represented in the draw's masked POS extent.  We then pool this local SSIM map into a single SSIM summary value by calculating the dust-weighted average:
\begin{equation}
    \rm{SSIM}(v) = \frac{\Sigma_k^{N_{\rm{pix}}} \ SSIM(H\mathrm{I}(v, p_k),\mathcal{A}_i(p_k)) \cdot \mathcal{A}_{i}(p_k)}{\Sigma_k^{N_{\rm{pix}}}  \mathcal{A}_i(p_k)}
\end{equation}
where $A_i$ is defined in Equation \ref{eqn:integrated_Ai}.

\subsection{Match Significance}\label{ap:sig_matches}

In order to understand the null distribution of morphological match statistics for each cloud, we create 9 rotated versions of $\mathcal{A}_i$ located at the same Galactic latitude and rotated in increments of 36$\degree$ in Galactic longitude.  

For the original, non-rotated $\mathcal{A}_i$ and for each rotated comparison, we calculate the following statistics on the curve of SSIM(v):
\begin{itemize}

\item The height of the primary SSIM peak, $P_1$

\item The ratio of $P_1$ to the next-highest SSIM peak ($P_2$), $P_1/P_2$ (with the boundaries of the $P_1$ peak and the location and height of $P_2$ defined with a prominence-based method as described in \S\ref{S:scan}).

\item The significance of $P_1$ relative to the noise level in the high-velocity regime of SSIM(v), ($\delta_1 = [P_1 - \mu_{hv}]/\sigma_{hv}$), where $\mu_{hv}$ and $\sigma_{hv}$ are estimated from the high-velocity wings on the opposite side of the velocity at which $P_1$ is located ($-\rm{sign}(v_{HI}) \times (200 < v < 300)$ km/s)

\item The maximum column density of HI emission within the integrated SSIM peak, $\rm{max}(N(HI))$, from the adaptively-smoothed \HI cube

\item The maximum single-channel column density in the high-velocity regime, $\rm{max}(N(HI)_{hv})$, from the original-resolution \HI cube

\item The standard deviation (STD) in the velocity of the primary SSIM peak ($v_{HI}$) between draws, $\sigma_{v_{HI}}$

\end{itemize}

Note that the statistic $\delta_1$ assumes that the SSIM signal in the high-velocity regime is noise-driven.  This assumption is not valid for those LOS with significant HI emission in the high-velocity wings, which generally corresponds to extragalactic structures or HVCs. To flag these LOS, we identify original maps or rotations with maximum single-channel $\rm{max}(N(HI)_{hv})$ greater than the theoretical $5\sigma$ detection limit of the \HIPI survey, $\rm{max}(N({HI})_{hv}) > 2.3 \times 10^{18}$ cm$^{-2}$ (see \S\ref{S:data}).  We exclude flagged rotations from being used to understand the null distribution of that statistic and, if the original map itself is flagged, we exclude this statistic wholesale from being involved in deciding if the detection is significant.

To determine significant matches, we require that the values of $P_1$, $P_1/P_2$, and $\delta_1$ for the original map $\mathcal{A}_i$ are all greater than the maximum value of those statistics within the rotated maps.  We additionally require that the HI column density $N_{HI}$ of the match is greater than the detection limit of the \HIPI survey, and also that the standard deviation of match velocity between draws $\sigma_{v_{HI}}<20$ km/s (in order to ensure the velocity match is well-constrained between draws; this removes 16 clouds that pass all other cuts applied, most of which by-eye inspection reveals to contain a single mis-clustered dust cloud).  After applying these cuts, we inspected our lowest $N_{HI}$ passing cloud (cluster 1373), which we observed to have an anomalously-broad width in SSIM space, anomalously-high-velocity for being at very low altitudes, very low $N_{HI}$ in individual channels, and a very poor match by-eye between dust and \HI{}.  We therefore tighten our HI column density cut to require $N_{HI} > 10^{19.7}$ cm$^{-2}$ in order to exclude this obviously erroneous match.

\begin{figure*}
    \centering
\includegraphics[width=\textwidth]{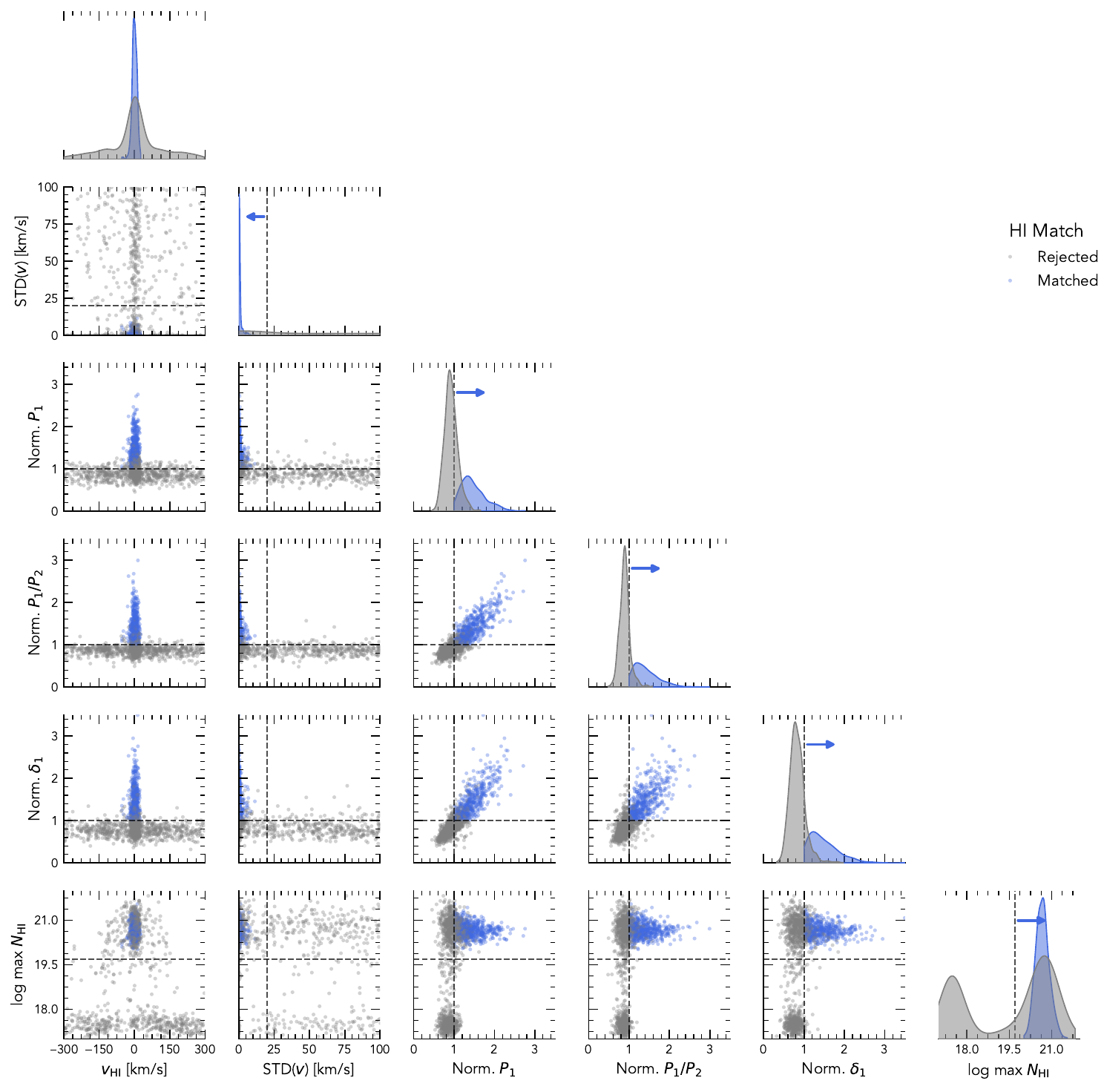}
    \caption{Corner plot demonstrating the relationship between \HI{}-matched velocity, $v_{HI}$, and metrics of match significance: 1) the standard deviation of $v_{HI}$ across draws, $\sigma_v$, 2) the height of the SSIM peak $P_1$ normalized by the maximum value of $P_1$ among the longitude-rotated null set, 3) the height of $P_1$ relative to the second-highest peak $P_2$ normalized by the maximum value of $P_1/P_2$ in the null set, 4) the significance of $P_1$ relative to the noise in the SSIM curve in the high-velocity regime, $\delta_1$, normalized by the maximum value of $\delta_1$ in the null set, and 5) the maximum \HI{} column density max(N$_{HI}$).  Clouds that pass all cuts populate the blue scatter points, while rejected clouds populate the grey scatter points.  The values of the cuts applied for significance are shown by the black dashed lines, with the blue arrows in the diagonal one-dimensional KDEs showing the direction of the cut.}
    \label{fig:hisig_corner}
\end{figure*}

519 clouds (30.6\% of the $N_{\rm{draw}} \geq 3$ cluster sample) pass all cuts.   In Figure \ref{fig:hisig_corner}, we plot the distributions of our statistics divided by the maximum rotated values for clouds passing vs. failing our cuts.  We observe that there is a clear bimodality in the distribution $\rm{max}(N_{HI})$ within the set of all clouds matched to \HI (with the separation between modes falling roughly at the \HIPI detection limit threshold). We note that the distribution of $v_{HI}$ is much broader for the low-column-density group, suggesting that random noise is responsible for the match, while the higher column density group is peaked around $v_{HI} \sim 0$ km/s corresponding to local Galactic structures.  We discuss the failed matches further in Appendix \ref{ap:failed_matches}.

We find that our embedded-cloud classes (defined in \S\ref{S:embedded}) yield significantly different \HI{}-matching success rates.  Clouds in a morphologically-similar hierarchy (parents, middle, children) have the highest matching rates (45.9\%, 38.5\%, 36.9\%, respectively).  Standalone clouds (the largest category by far) have dramatically lower success (26.8\%).  The nine defined superstructures all fail to match (0\%).  We attribute the superstructure matching failures to their large and complex structures (unlikely to be well-described by a single, isolated velocity peak) and typically low-latitudes (leading to disk confusion).  We interpret the higher success of parents as suggesting that these structures are perhaps more stable across draws.

\subsection{Failed Dust-\HI{} Matches}\label{ap:failed_matches}

\begin{figure*}
    \centering
\includegraphics[width=\textwidth]{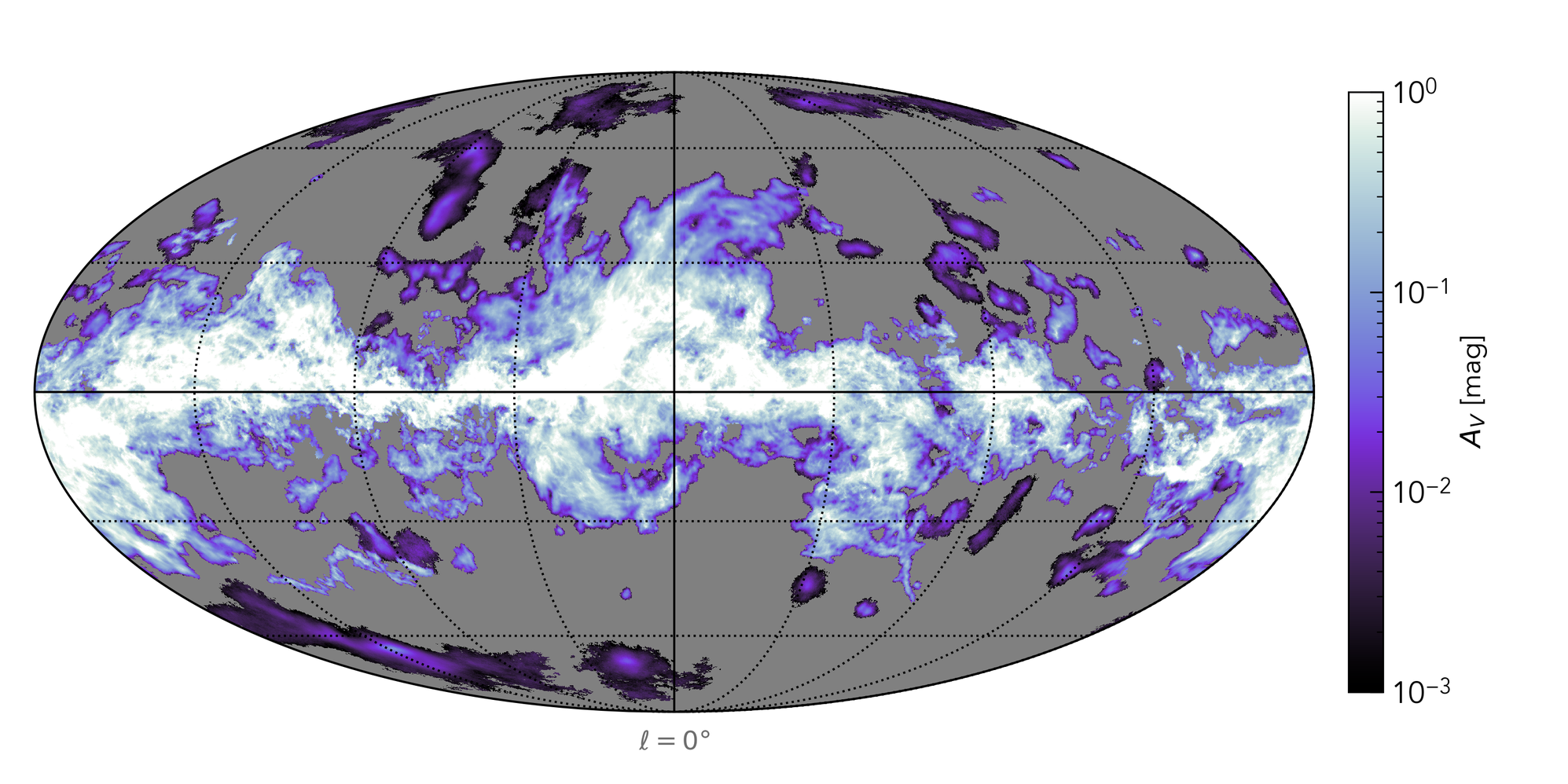}
\includegraphics[width=\textwidth]{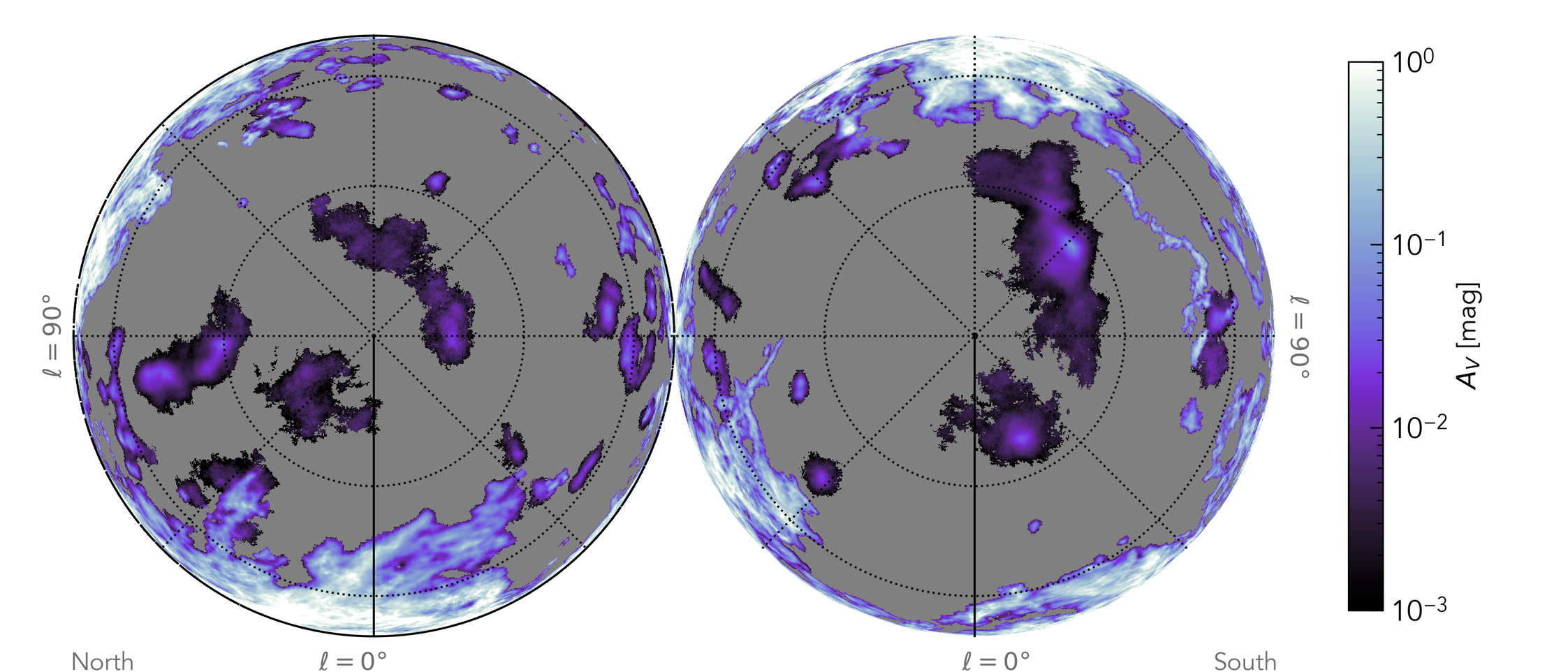}
    \caption{Integrated extinction $A_V$ for all dust clouds with no successful \HI{} match, in Mollweide (top) and polar (bottom) projections.}
    \label{fig:failed_match}
\end{figure*}

Here we present maps (Figure \ref{fig:failed_match}) of those 1,176 dust clouds (in our sample of $N_{\rm{draw}}\geq 3$ clusters) that failed to have high-quality morphological matches with \HI{}.  The majority of failed-match clouds are located at low latitudes.  Since \perch is a hierarchical structure-finding method, many failed-match clouds share substructures with successfully-matched clouds.  Additionally, we observe that many of the failed-match clouds appear to have good visual correlation between their dust morphologies and their rejected-\HI{} matches (e.g., Clouds 507, 665).  The candidate \HI{} matches to our failed match sample can be viewed in the interactive figure hosted at \url{https://theo-oneill.github.io/HACs_and_IVCs/moments/}.

We experimented extensively with definitions of significance cuts, and found that any cut that would retrieve some of these potentially-falsely-rejected matches would also let in a similar number of spurious, correctly-rejected matches.  We therefore chose in this work to prioritize the purity of our catalog over completeness; future work should investigate potential improvements to our morphological matching methodology in order to improve the \HI{}-matching fraction without decreasing accuracy.

\restartappendixnumbering
\section{Constraints on Turbulent Velocity Dispersion}\label{ap:turbulence}

\begin{figure*}
    \centering
    \includegraphics[width=\textwidth]{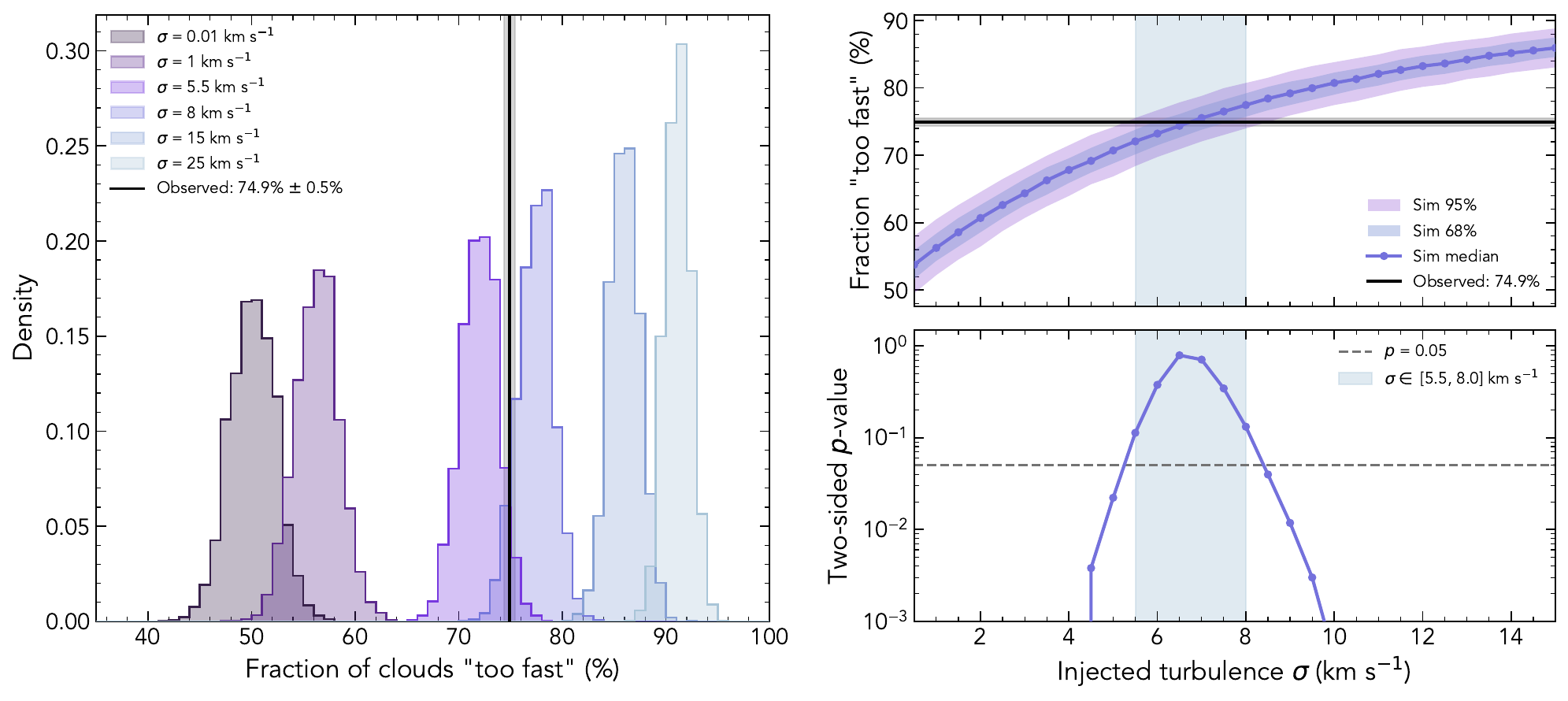}
    \caption{\textit{Left:} Histograms of the distributions of synthetic ``too-fast'' cloud fractions $F_{\rm{syn}}$ for various amplitudes of injected isotropic Gaussian turbulence: $\sigma = 0.01$ km s$^{-1}$, 1 km s$^{-1}$, 5.5 km s$^{-1}$, 8 km s$^{-1}$, 15 km s$^{-1}$, and 25 km s$^{-1}$, with histogram color increasing continuously from dark purple to light blue in that order.  Our observed ``too-fast'' fraction $f_{\rm{obs}} = 74.9\% \pm 0.5\%$ is shown by the black vertical line and gray shaded region, with the mean and uncertainties calculated by bootstrap resampling.  \textit{Top right:} Variation in synthetic $F_{\rm{syn}}$ as a function of injected turbulence $\sigma$.  The median synthetic value across tested $\sigma$ is shown by the purple curve with circular markers, and 68\% and 95\% ranges of the synthetic distributions are shown by the blue and purple shaded regions, respectively.  The observed $f_{\rm{obs}}$ is marked by the black horizontal line.  \textit{Bottom right:} Two sided $p$-value obtained by comparing $F_{\rm{syn}}$ to $f_{\rm{obs}}$, as a function of $\sigma$.  Our selected significance threshold of $p=0.05$ is shown by the gray horizontal dashed line, and the range of non-rejected $\sigma$ ($p>0.05$) is shown by the blue shaded region.}
    \label{fig:turbulence_constraint}
\end{figure*}

In this appendix, we describe our numerical experiment injecting turbulence into a synthetic cloud sample (described in \S\ref{S:kinematic}) in greater detail.  We observed a striking imbalance in the fraction of our clouds that are moving ``too fast'' as compared to Galactic rotation ($f_{\rm{obs}} = 74.9\% \pm 0.5\%$ of clouds in our \HI{}-matched sample, with uncertainties obtained by bootstrap resampling in \S\ref{S:kinematic}).  In this experiment, we assess if the injection of isotropic Gaussian turbulence onto our assumed rotation curve could reproduce this excess.

To do so, for each of our \HI{}-matched clouds, we obtain its median line-of-sight unit vector ($\hat{r}$) and predicted radial velocity under the \citet{ReidMenten2019} rotation curve ($v_{\rm{rot}}$).  We then draw a velocity perturbation from a 3D Gaussian distribution for each cloud, $v_{\rm{turb}} \sim \mathcal{N}(0,\sigma^2)$, and project it onto $\hat{r}$ to obtain $v_{\rm{turb}, \rm{LOS}}$.  We then define a synthetic $v_{\rm{LSR}}=v_{\rm{rot}}+v_{\rm{turb,LOS}}$ and $v_{\rm{dev}}=v_{\rm{turb,LOS}}$.  Finally, we calculate the fraction of clouds in this realization that are moving ``too fast'' (as defined in \S\ref{S:kinematic}).  We repeat this procedure 10,000 times for each tested value of $\sigma$, in order to generate a synthetic distribution of ``too fast'' fractions $F_{\rm{syn}}$.

We perform this test for $\sigma$ ranging between $0.5-15$ km s$^{-1}$ in $0.5$ km s$^{-1}$ intervals.  For each $\sigma$, we compute the two-sided $p$-value as $p = 2 \ \rm{min}(P[F_{\rm{syn}} \geq f_{\rm{obs}}],P[F_{\rm{syn}} \leq f_{\rm{obs}}])$.  We reject values of $\sigma$ with $p \leq 0.05$ as being inconsistent with our observed ``too fast'' fraction at the 95\% level.  

The results of this test are shown in Figure \ref{fig:turbulence_constraint}.  We fail to reject $\sigma$ between $5.5 - 8.0$ km s$^{-1}$.  This suggests that our observed cloud sample may be consistent with a model of clouds with kinematics affected by rotation and moderate amplitudes of isotropic Gaussian turbulence.

\restartappendixnumbering
\section{Correlations between Cloud Properties and Altitude}\label{ap:cloud_corr}

\begin{figure*}
    \centering
\includegraphics[width=\textwidth]{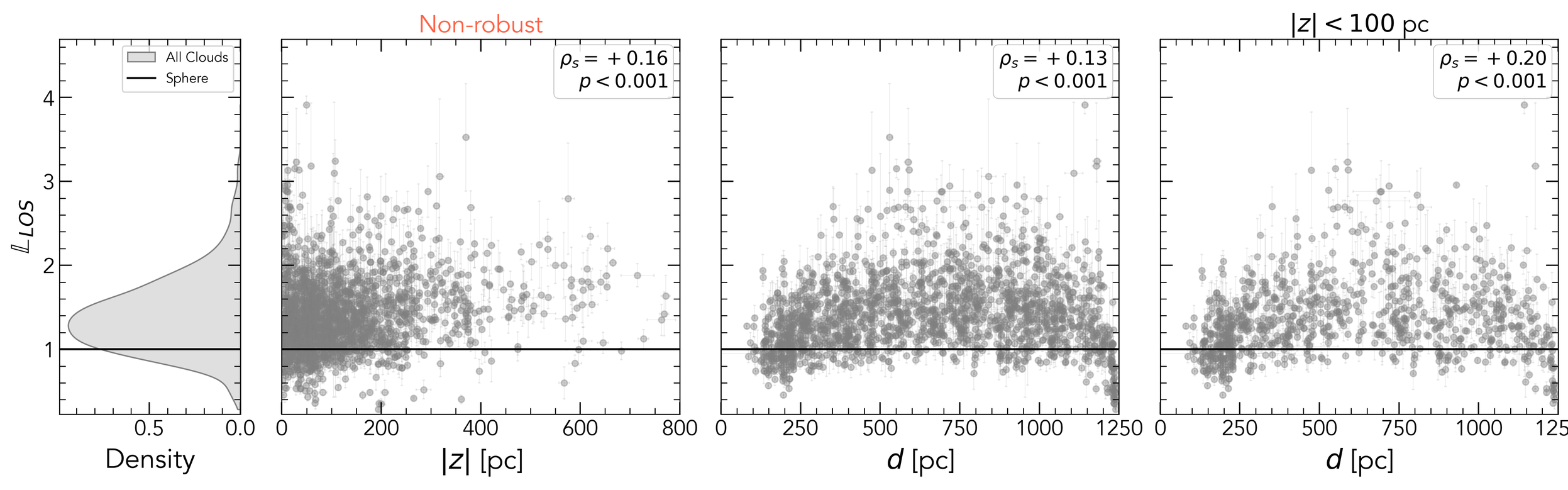}
\includegraphics[width=\textwidth]{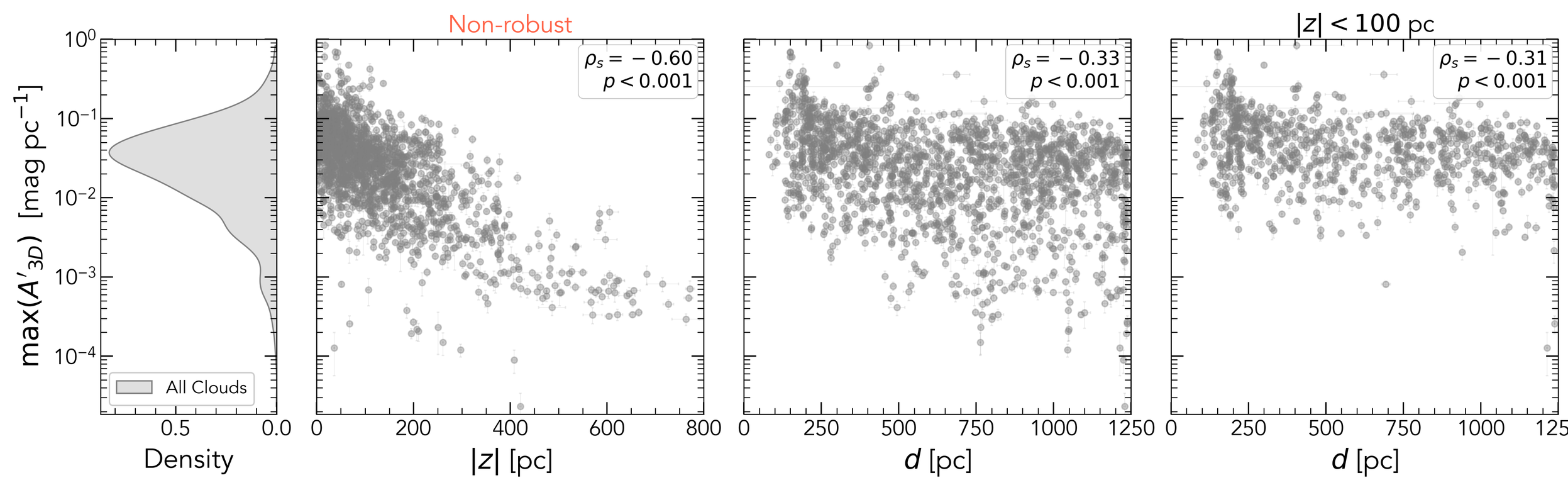}
\includegraphics[width=\textwidth]{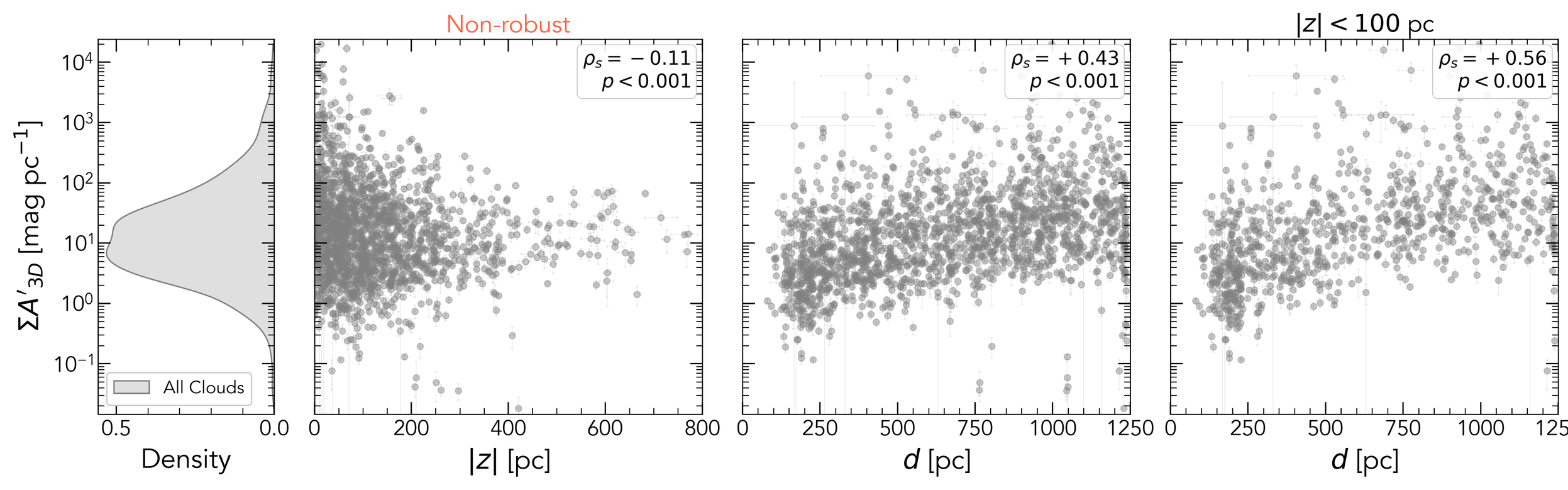}
\includegraphics[width=\textwidth]{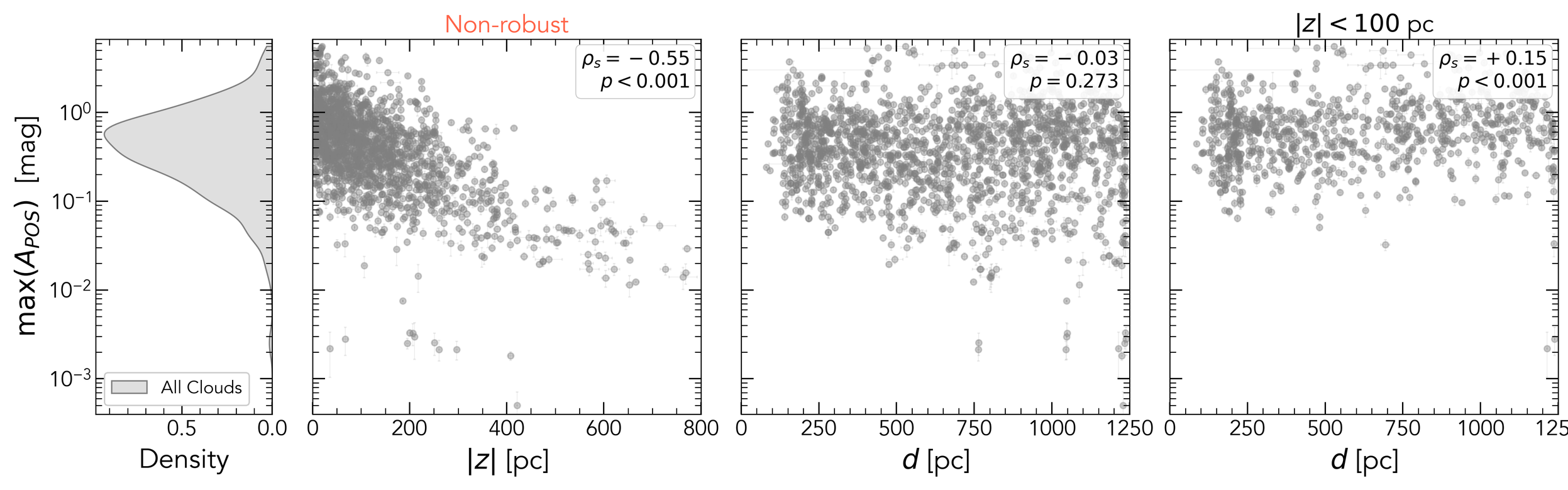}
    \caption{As Figure \ref{fig:ztrends_ex}, but for cloud properties derived solely from 3D dust, including, from top: elongation along the LOS $\mathbb{L}_{\rm{LOS}}$, maximum differential extinction in 3D space max($A'_{3D}$), total integrated differential extinction in 3D space $\sum A'_{3D}$, and maximum extinction integrated on the POS max($A_{POS}$). }
    \label{fig:ztrends_all}
\end{figure*}

\begin{figure*}
    \centering
\includegraphics[width=\textwidth]{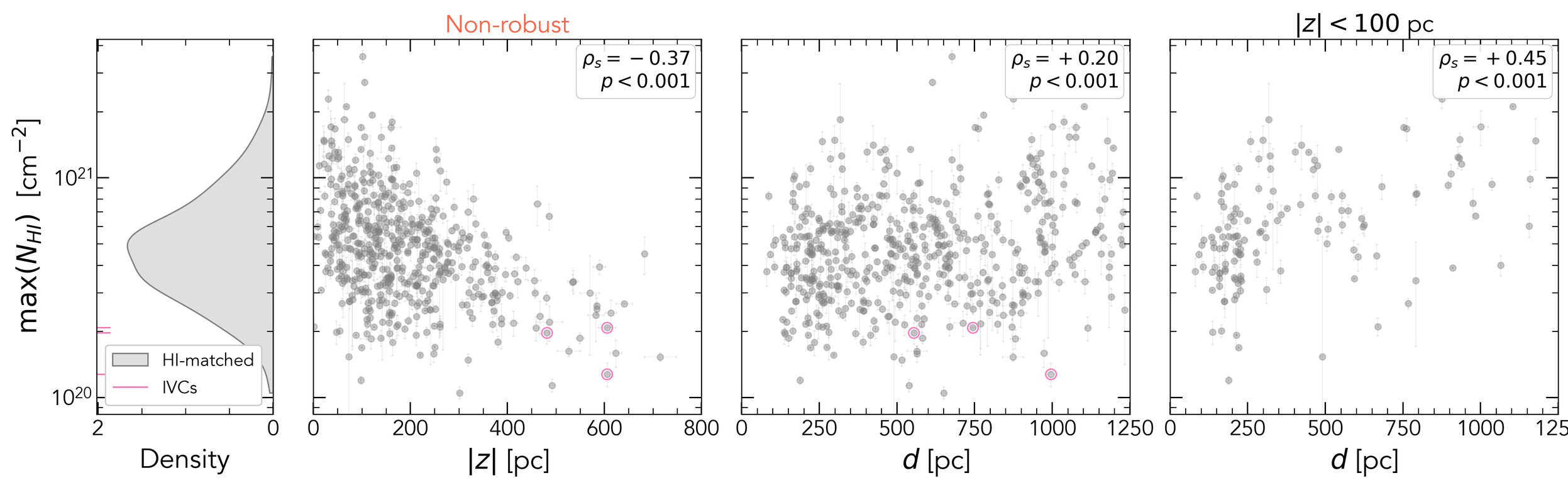}
\includegraphics[width=\textwidth]{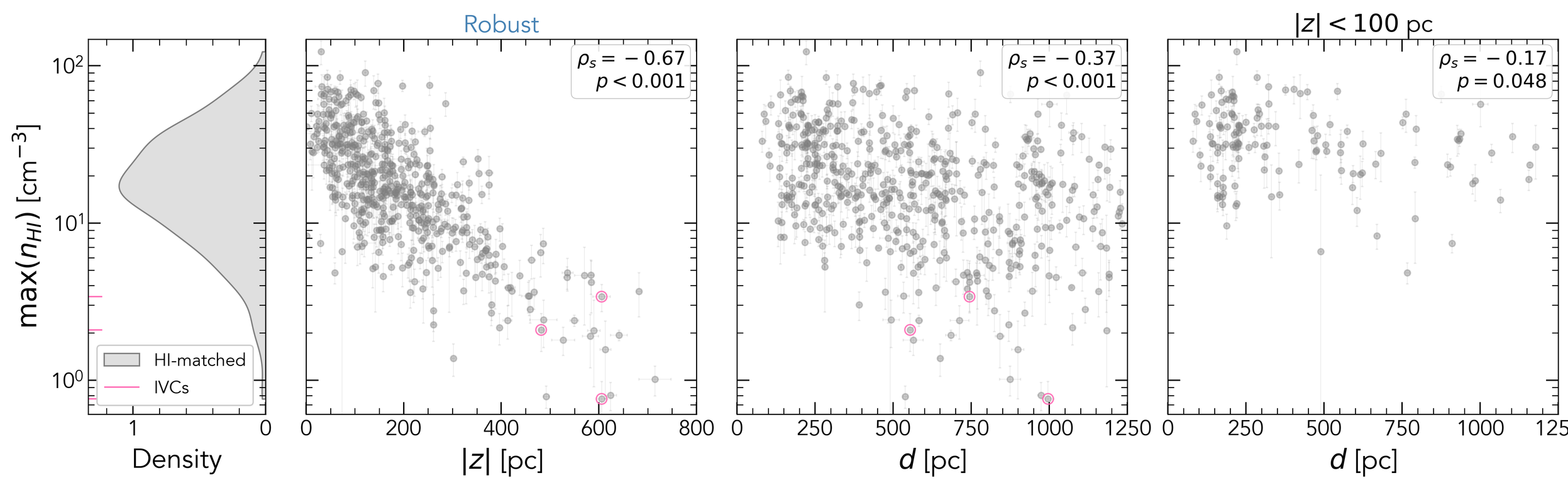}
\includegraphics[width=\textwidth]{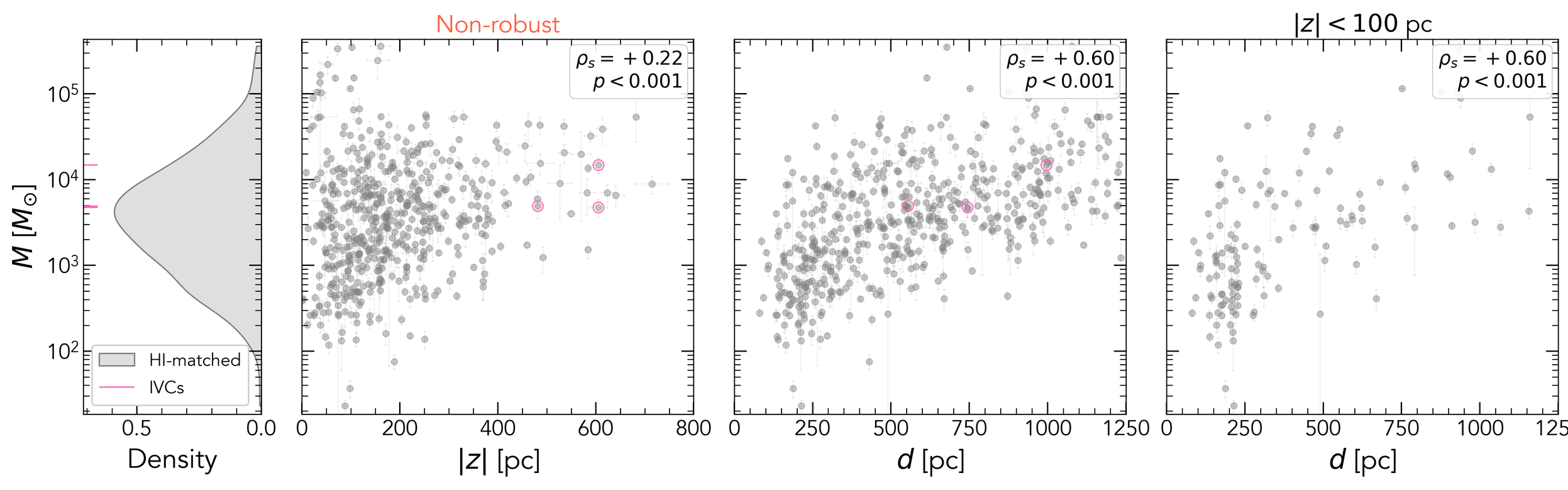}
    \caption{As Figure \ref{fig:ztrends_all}, but for cloud properties derived from the combination of 3D dust and \HI{}, including, from top: maximum \HI{} column density on the POS, max($N_{HI}$), maximum \HI{} volume density in 3D space, max($n_{HI}$), and neutral mass $M$.}
    \label{fig:ztrends_pass}
\end{figure*}

\begin{figure*}
    \centering
\includegraphics[width=\textwidth]{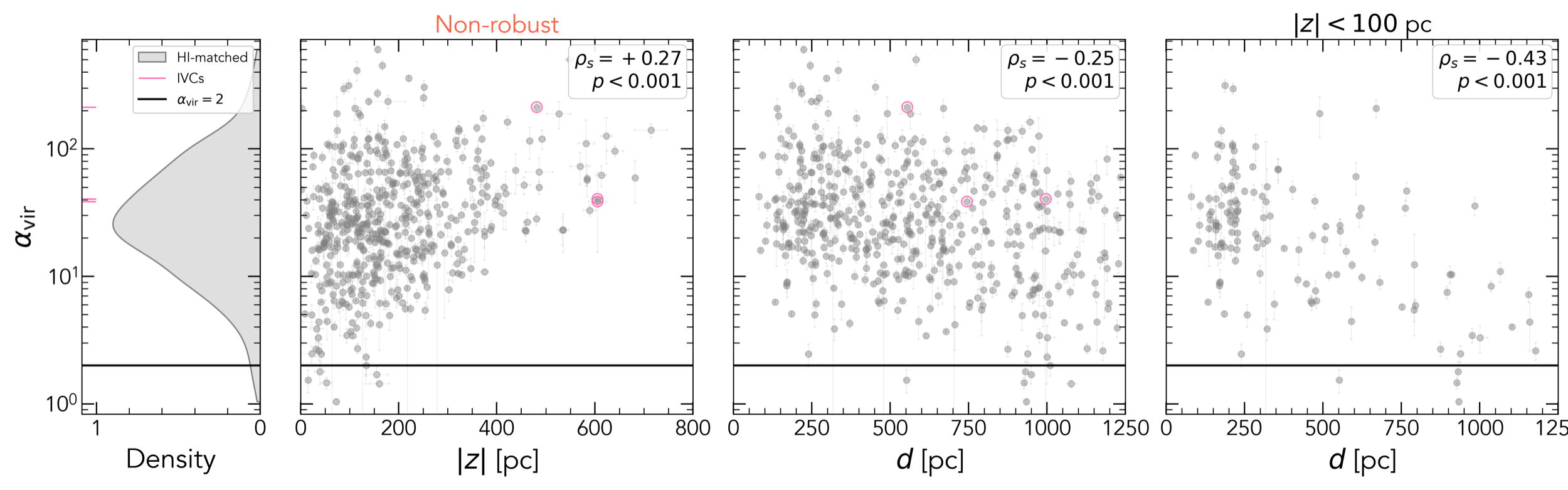}
\includegraphics[width=\textwidth]{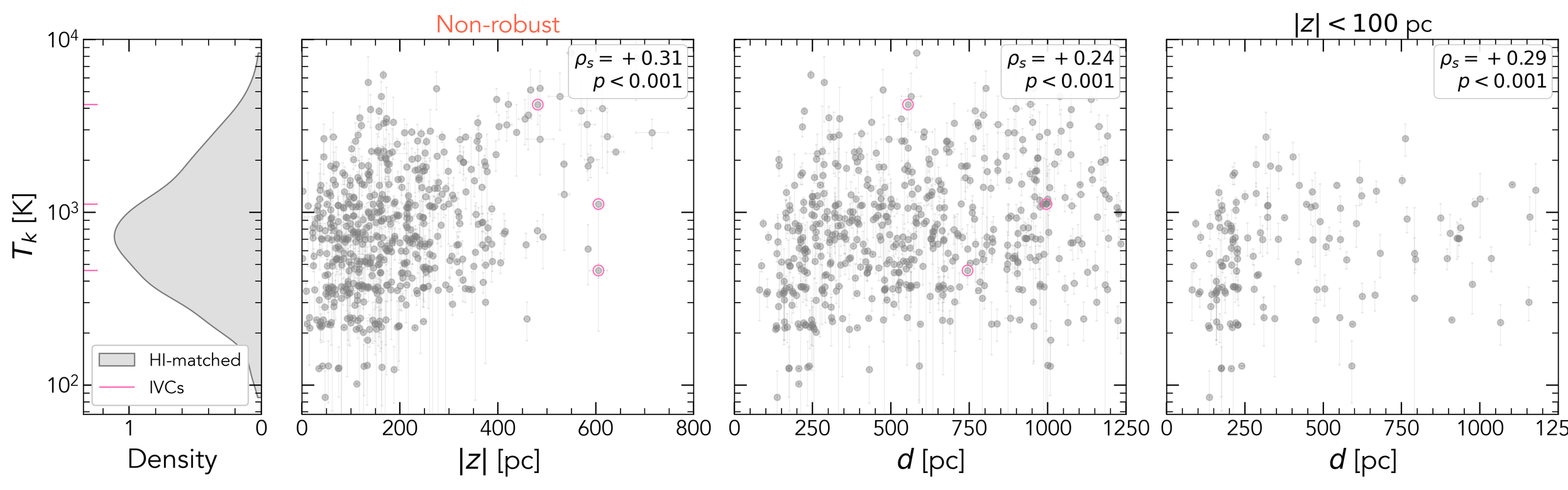}
\includegraphics[width=\textwidth]{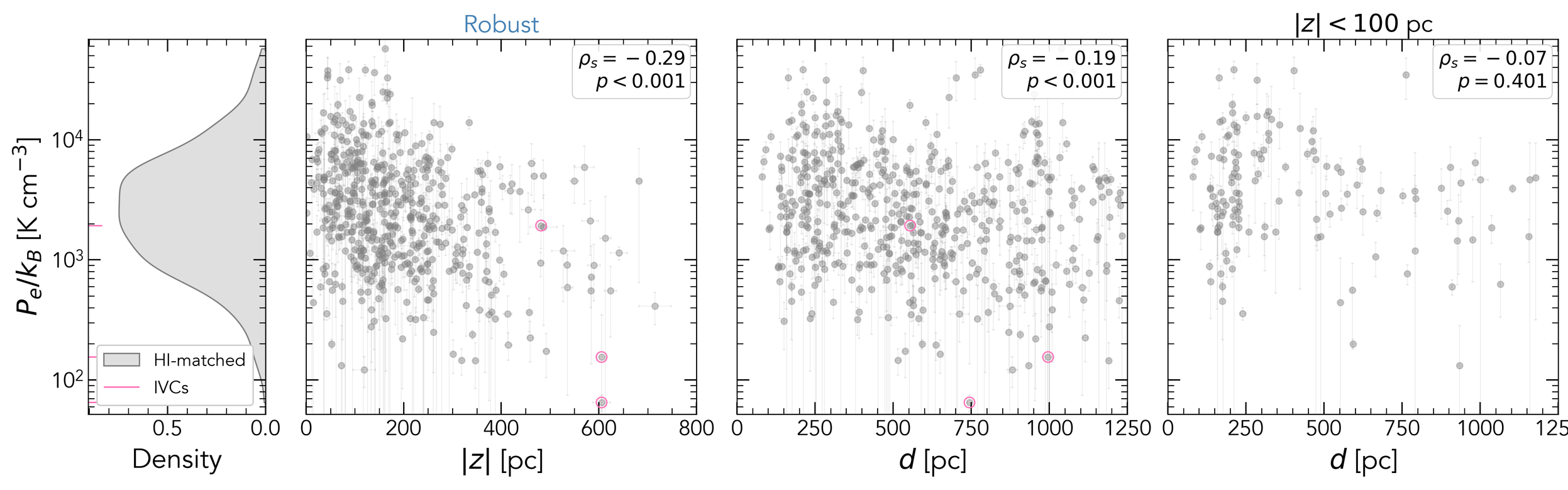}
\includegraphics[width=\textwidth]{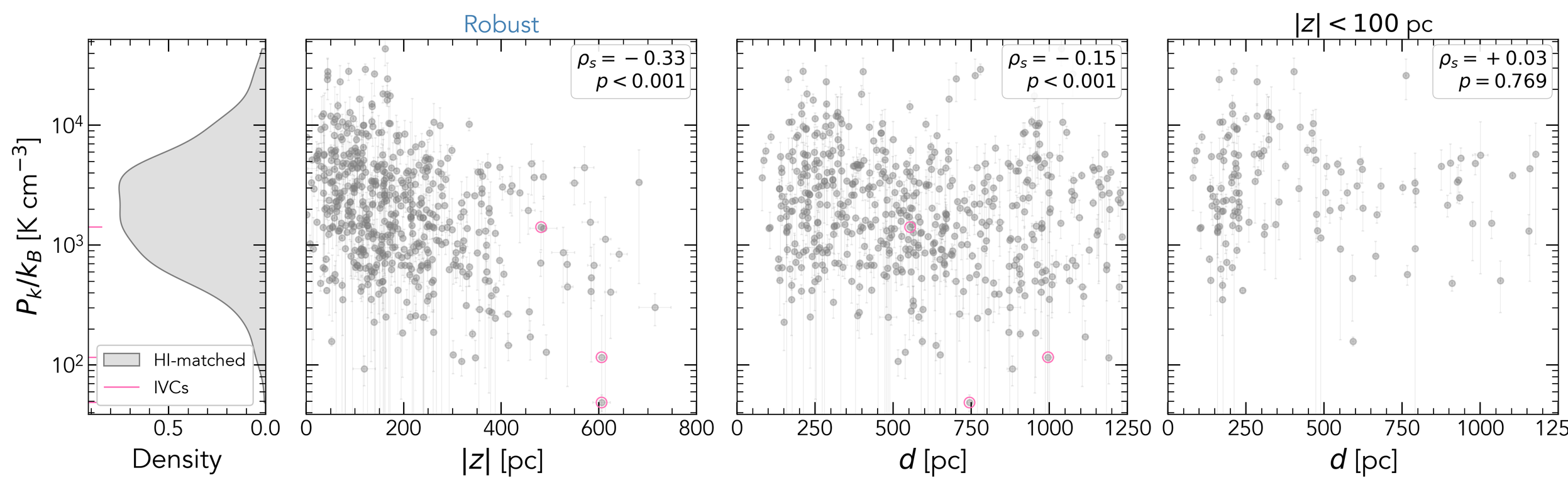}
    \caption{As Figure \ref{fig:ztrends_pass}, for additional cloud properties derived from the combination of 3D dust and \HI{}, including, from top: virial parameter $\alpha_{\rm{vir}}$, kinetic temperature $T_k$, external pressure necessary for virialization $P_e/k_B$, and internal pressure $P/k_B$.}
    \label{fig:ztrends_pass2}
\end{figure*}

Here we present additional properties inspected for robust trends with $|z|$, as described in \S\ref{S:trends}.  Figure \ref{fig:ztrends_all} presents properties solely dependent on the 3D dust map, while Figures \ref{fig:ztrends_pass} and \ref{fig:ztrends_pass2} presents properties additionally dependent on \HI{}.

\bibliographystyle{yahapj}
\bibliography{refs.bib}

\end{document}